\pdfoutput=1

\documentclass[11pt,twoside,a4paper,cmspaper,final,collab]{cms-tdr}
\def\svnVersion{324156}\def\cmsCernNo{2013-037}\def\cmsCernDate{\today}\def\cmsMessage{Published in the Journal of High Energy Physics as \href{http://dx.doi.org/10.1007/JHEP01(2016)096}{\doi{10.1007/JHEP01(2016)096.}}}

\begin{document}\cmsNoteHeader{TOP-14-021}

\hyphenation{had-ron-i-za-tion}
\hyphenation{cal-or-i-me-ter}
\hyphenation{de-vices}
\RCS$Revision: 321849 $
\RCS$HeadURL: svn+ssh://svn.cern.ch/reps/tdr2/papers/TOP-14-021/trunk/TOP-14-021.tex $
\RCS$Id: TOP-14-021.tex 321849 2016-02-02 09:40:59Z abrinke1 $
\newlength\cmsFigWidth
\ifthenelse{\boolean{cms@external}}{\setlength\cmsFigWidth{0.85\columnwidth}}{\setlength\cmsFigWidth{0.4\textwidth}}
\ifthenelse{\boolean{cms@external}}{\providecommand{\cmsLeft}{top\xspace}}{\providecommand{\cmsLeft}{left\xspace}}
\ifthenelse{\boolean{cms@external}}{\providecommand{\cmsRight}{bottom\xspace}}{\providecommand{\cmsRight}{right\xspace}}
\providecommand{\NA}{---\xspace}

\providecommand{\ptmiss}{\ensuremath{p_{\mathrm{T}}^{\text{miss}}}\xspace}
\newcommand{\ttH}{\ttbar\PH}
\newcommand{\ttW}{\ttbar\PW}
\newcommand{\ttZ}{\ttbar\PZ}
\newcommand{\MHT}{\ensuremath{H_\mathrm{T}^\text{miss}}\xspace}
\newcommand{\MT}{\ensuremath{M_{\mathrm{T}}}\xspace}
\newcommand{\cuB}{\ensuremath{\bar{c}_{\mathrm{uB}}}\xspace}
\newcommand{\cpHQ}{\ensuremath{\bar{c}'_{\mathrm{HQ}}}\xspace}
\newcommand{\cHQ}{\ensuremath{\bar{c}_{\mathrm{HQ}}}\xspace}
\newcommand{\cHu}{\ensuremath{\bar{c}_{\mathrm{Hu}}}\xspace}
\newcommand{\cThreeW}{\ensuremath{\bar{c}_{\mathrm{3W}}}\xspace}

\cmsNoteHeader{TOP-14-021}
\title{Observation of top quark pairs produced in association with a vector boson in pp collisions at \texorpdfstring{$\sqrt{s} = 8\TeV$}{sqrt(s) = 8 TeV}}

\date{\today}

\abstract{
Measurements of the cross sections for top quark pairs produced in association with
a $\PW$ or $\PZ$ boson are presented, using 8\TeV pp collision data corresponding
to an integrated luminosity of 19.5\fbinv, collected by the CMS experiment at the
LHC. Final states are selected in which the associated $\PW$ boson decays to a charged lepton
and a neutrino or the $\PZ$ boson decays to two charged leptons. Signal events are identified by matching
reconstructed objects in the detector to specific final state particles from
$\ttW$ or $\ttZ$ decays.
The $\ttW$ cross section is measured to be
$382^{+117}_{-102}\unit{fb}$ with a significance
of 4.8 standard deviations from the background-only hypothesis. The
$\ttZ$ cross section is measured to be
$242^{+65}_{-55}\unit{fb}$ with a
significance of 6.4 standard deviations from the background-only hypothesis.
These measurements are used to set bounds on five anomalous dimension-six operators
that would affect the $\ttW$ and
$\ttZ$ cross sections.
}

\hypersetup{%
pdfauthor={CMS Collaboration},%
pdftitle={Observation of top quark pairs produced in association with a vector boson in pp collisions at sqrt(s) = 8 TeV},
pdfsubject={CMS},%
pdfkeywords={CMS, physics, top, W, Z, electroweak}}

\maketitle
\section{Introduction}
\label{sec:intro}

\begin{figure}[htb]
  \centering
  $\vcenter{\hbox{\includegraphics[width=0.38\textwidth]{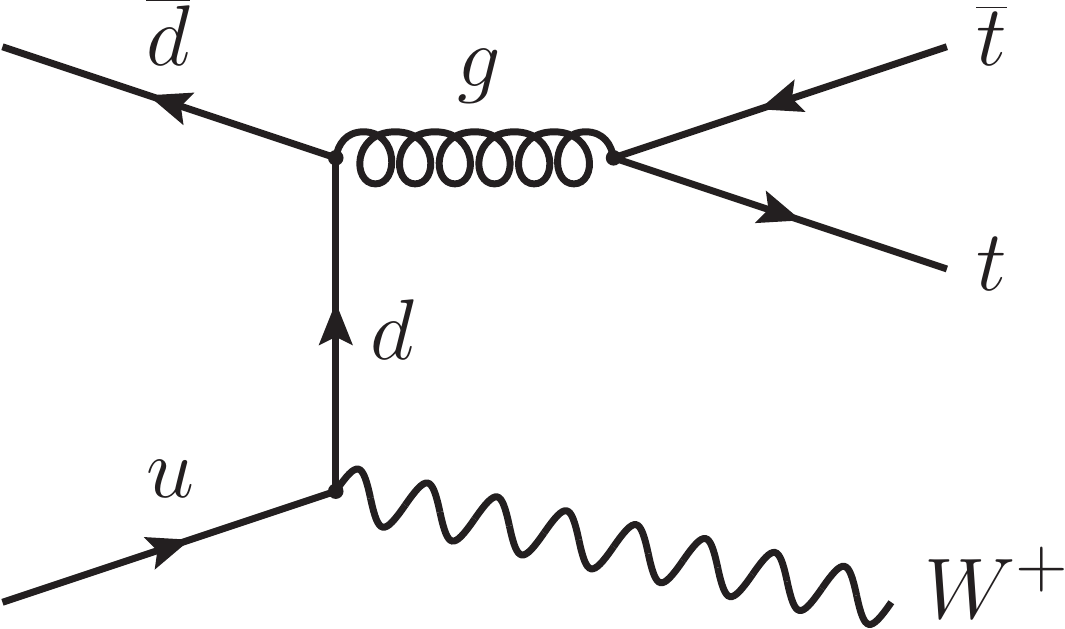}}}$
  \hspace{0.4 in}
  $\vcenter{\hbox{\includegraphics[width=0.40\textwidth]{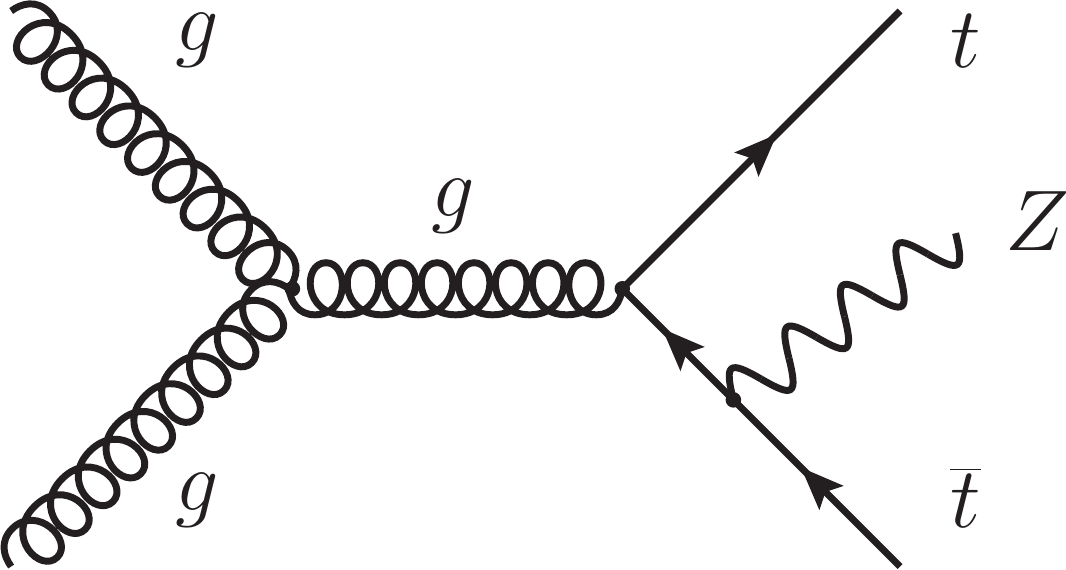}}}$
  \caption{Dominant leading order Feynman diagrams for $\ttW^{+}$ and $\ttZ$ production
           at the LHC. The charge conjugate process of $\ttW^{+}$
		   produces $\ttW^{-}$.}
  \label{fig:ttV_feynman}
\end{figure}

Since the LHC at CERN achieved proton-proton collisions at
center-of-mass energies of 7 and 8\TeV, it has become possible to study signatures
at significantly higher mass scales than ever before. The two heaviest sets of particles
produced in standard model (SM) processes that could be observed using the data already
collected are top quark pairs produced in association with a $\PW$ or $\PZ$ boson ($\ttW$ and
$\ttZ$), which have expected cross sections of $\sigma(\ttW) = 203^{+20}_{-22}\unit{fb}$ and
$\sigma(\ttZ) = 206^{+19}_{-24}\unit{fb}$ in the SM in 8\TeV collisions~\cite{Garzelli:2012bn}. The dominant production
mechanisms for $\ttW$ and $\ttZ$ in pp collisions are shown in Fig.~\ref{fig:ttV_feynman}.
The $\ttZ$ production cross section provides the most accessible direct measurement
of the top quark coupling to the $\PZ$ boson. Both $\sigma(\ttW)$ and $\sigma(\ttZ)$ would be
altered in a variety of new physics models that can be parameterized
by dimension-six operators added to the SM Lagrangian.

The $\ttZ$ cross section was first measured by the CMS experiment in 7\TeV collisions, with
a precision of about 50\%~\cite{PRL-110-172002}.
Measurements in events containing three or four leptons in 8\TeV collisions at CMS~\cite{CMS-TOP-12-036}
have constrained $\sigma(\ttZ)$ to within 45\% of its SM value, and yielded evidence
of $\ttZ$ production at 3.1 standard deviations from the background-only hypothesis. The CMS collaboration
also used same-sign dilepton events to constrain $\sigma(\ttW)$ to within 70\% of the SM prediction,
with a significance of 1.6 standard deviations from the background-only hypothesis.
Most recently, the ATLAS experiment used events containing two to four leptons to measure
$\sigma(\ttW) = 369^{+100}_{-91}\unit{fb}$ at 5.0 standard deviations from the background-only
hypothesis, and $\sigma(\ttZ) = 176^{+58}_{-52}\unit{fb}$ with a significance of 4.2 standard
deviations from the background-only hypothesis~\cite{ATLAS-ttV-8TeV}.

We present the first observation of $\ttZ$ production and measurements of
the $\ttW$ and $\ttZ$ cross sections using a full reconstruction of the top
quarks and the $\PW$ or $\PZ$ boson from their decay products. We target events in which the
associated $\PW$ boson decays to a charged lepton and a neutrino ($\PW  \to \ell\nu$)
or the $\PZ$ boson decays to two charged leptons ($\PZ \to \ell\ell$). In this paper,
``lepton'' ($\ell$) refers to an electron, a muon, or a $\tau$ lepton decaying into other leptons. The top quark pair may decay into final
states with hadronic jets ($\ttbar  \to \PQb\PQq\PAQq\ \PAQb\PQq\PAQq$), a lepton
plus jets ($\ttbar  \to \PQb\ell\nu\ \PAQb\PQq\PAQq$), or two leptons ($\ttbar
 \to \PQb\ell\nu\ \PAQb\ell\nu$). The $\ttZ$ process is measured in channels
with two, three, or four leptons, with exactly one pair of same-flavor opposite-sign
leptons with an invariant mass close to the $\PZ$ boson mass~\cite{PDG2014}.  The $\ttW$ process is measured
in channels with two same-sign leptons or three leptons, where no lepton
pair is consistent with coming from a $\PZ$ boson decay.  Additional $\PQb$-tagged jets and
light flavor jets are required to enable full or partial reconstruction
of the top quark and $\PW$ boson decays.

Channels defined by lepton charge and multiplicity are further subdivided by
lepton flavor and the number of jets, in order to provide an initial
separation between signal and background (Section~\ref{sec:selection}).
Background processes with leptons from $\PW$ and $\PZ$ boson decays are estimated
using Monte Carlo simulations that are validated in separate control
regions (Section~\ref{sec:bkg_prompt}).  Processes with leptons from other
sources are estimated directly from the data, using events in which one or more
leptons fail to satisfy a strict set of selection criteria
(Sections~\ref{sec:bkg_NP} and~\ref{sec:bkg_QF}). In each channel, we attempt
a full or partial reconstruction of the $\ttW$ or $\ttZ$ system with a linear
discriminant that matches leptons and jets to their parent particles using mass,
charge, and \PQb tagging information (Section~\ref{sec:reconstruction}). Additional
kinematic variables from leptons and jets are combined with output from the
linear discriminant in a multivariate analysis that is used to make the final
measurement of the $\ttW$ and $\ttZ$ cross sections (Sections~\ref{sec:signal}
and~\ref{sec:results_sm}). Finally, the measured cross sections are used
to constrain the coupling of the top quark to the $\PZ$ boson, and to set bounds
on five anomalous dimension-six operators (Section~\ref{sec:results_extended}).

\section{The CMS detector}
\label{sec:detector}

The central feature of the CMS apparatus is a superconducting solenoid of
6\unit{m} internal diameter, providing a magnetic field of 3.8\unit{T}.
Within the solenoid volume are a silicon pixel and strip
tracker, a lead tungstate crystal electromagnetic calorimeter (ECAL), and
a brass and scintillator hadron calorimeter (HCAL), each composed of a
barrel and two endcap sections. Extensive forward calorimetry complements
the coverage provided by the barrel and endcap detectors. Muons are detected
in gas-ionization muon chambers embedded in the steel flux-return yoke outside
the solenoid.

A global event description is obtained using the CMS particle-flow (PF)
algorithm~\cite{CMS-PAS-PFT-09-001, CMS-PAS-PFT-10-001}, which
combines information from all CMS sub-detectors to
reconstruct and identify individual particles in collision events.
The particles are placed into mutually exclusive classes:
charged hadrons, neutral hadrons, photons, muons, and
electrons.  The primary collision vertex is identified as the
reconstructed vertex with the highest value of $\sum \pt^{2}$, where \pt
is the momentum component transverse to the beams, and the sum is
over all the charged particles used to reconstruct the vertex.
The energy of photons is directly obtained
from the ECAL measurement, corrected for zero-suppression effects. The
energy of electrons is determined from a combination of the electron
momentum at the primary interaction vertex as determined by the tracker,
the energy of the corresponding ECAL cluster, and the energy sum of all
bremsstrahlung photons spatially compatible with originating from the
electron track. The energy of muons is obtained from the curvature of
the corresponding track and hits in the muon chambers. The energy of
charged hadrons is determined from a combination of their momentum
measured in the tracker and the matching ECAL and HCAL energy deposits,
corrected for zero-suppression effects and for the response function of
the calorimeters to hadronic showers. Finally, the energy of neutral hadrons
is obtained from the corresponding corrected ECAL and HCAL energy.

A more detailed description of the CMS detector, together with a definition
of the coordinate system used and the relevant kinematic variables, can
be found in Ref.~\cite{Chatrchyan:2008zzk}.

\section{Data and simulated samples}
\label{sec:samples}

This search is performed with an integrated luminosity of $19.5 \pm 0.5 \fbinv$ of proton-proton collisions
at $\sqrt{s} = 8 \TeV$, collected in 2012~\cite{CMS-PAS-LUM-13-001}.
Dilepton triggers were used to collect data for all channels. The
dilepton triggers require any combination of electrons and muons, where
one lepton has $\pt > 17\GeV$ and another has $\pt > 8\GeV$. A trielectron
trigger with minimum $\pt$ thresholds of 15, 8, and 5\GeV was also used
for channels with three or more leptons. These triggers approach their maximum efficiency
for leptons with \pt values at least 2~\GeV higher than the thresholds.

Expected signal events and some of the background processes are modeled with simulation.
The signal processes $\ttW$ and $\ttZ$, as well as background processes producing a
single $\PZ$ boson, $\PW \PZ$, $\PZ \PZ$, $\PW^{\pm}\PW^{\pm}$, $\PW \PW \PW$, $\PW \PW \PZ$, \ttbar, $\ttbar\gamma$, $\ttbar\gamma^{*}$,
$\ttbar\PW\PW$, and the associated production of a $\PZ$ boson with a single top
quark ($\PQt \PQb \PZ$), are all generated with the \MADGRAPH 5.1.3~\cite{MadGraph5} tree-level
matrix element generator, combined with \PYTHIA 6.4~\cite{PYTHIA} for the parton
shower and hadronization.  The associated production of a Higgs boson with a top
quark pair ($\ttH$) is modeled using the \PYTHIA generator assuming a Higgs boson mass
of 125\GeV. Samples that include top quark production are generated with a top
quark mass of 172.5\GeV. The CTEQ6L1 parton distribution function
(PDF) set~\cite{Pumplin:2002vw} is used for all samples.

The CMS detector response is simulated using \GEANTfour software~\cite{Agostinelli:2002hh}.  Both data and simulated events
are required to pass the same trigger requirements and are
reconstructed with identical algorithms.  Effects from additional
proton-proton collisions in the same bunch crossing (pileup) in the
simulation are modeled by adding simulated inclusive proton-proton
interactions (generated with \PYTHIA) to the generated hard collision, with
the pileup interaction multiplicity in simulation reflecting the profile
inferred from data. Correction factors are applied to individual
objects and events to bring object properties and efficiencies in simulation
into better agreement with data, as described in Section~\ref{sec:objects}.

\section{Object reconstruction and identification}
\label{sec:objects}

Certain types of particles reconstructed with the PF algorithm are particularly
useful in identifying and reconstructing $\ttW$ and $\ttZ$ events.
These objects are electrons, muons, charged and neutral hadrons clustered
into jets, and the imbalance in \ptvec arising from
neutrinos in the event.

Electrons with $\pt > 10\GeV$ are reconstructed over the full
pseudorapidity range of the tracker, $\abs{\eta} < 2.5$.  The reconstruction
combines information from clusters of energy deposits in the ECAL and
the electron trajectory reconstructed in the inner
tracker~\cite{Khachatryan:2015hwa,CMS_DP_2013-003}.
A multivariate analysis technique combines observables sensitive to the
amount of bremsstrahlung, spatial and momentum matching between the
track and associated ECAL clusters, and shower shape
observables, to distinguish genuine electrons from charged hadrons~\cite{Khachatryan:2015hwa}.

Muons with $\abs{\eta} < 2.4$ and $\pt > 10\GeV$ are reconstructed using
information from both the silicon tracker and the muon spectrometer~\cite{Chatrchyan:2012xi}.
Track candidates must have a minimum number of tracker hits, be compatible with
hits in the muon chambers, and match the associated energy deposits
in the calorimeters, to be selected as PF muons~\cite{CMS-PAS-PFT-10-003}.

The $\tau$ leptons decay before reaching the ECAL, and are not identified
in this analysis.  Their decay products are instead identified as hadrons, which may be
clustered into jets, or as electrons or muons, depending on whether the $\tau$ lepton decays to
hadrons or leptons.

Prompt leptons (electrons or muons from a $\PW$, $\PZ$, or Higgs boson, or the decay of a $\tau$
lepton) are distinguished from non-prompt leptons (misidentified jets or leptons from
hadron decays) in part by assessing their isolation from surrounding hadronic activity.
Lepton isolation is calculated by summing the $\pt$ of other particles
in a cone of radius $\Delta R = \sqrt{ \smash[b]{(\Delta \eta)^2 + (\Delta \phi)^2} } = 0.4$ around
the lepton direction, where $\Delta \eta$ and $\Delta \phi$ are the
pseudorapidity and azimuthal angle difference (in radians) from the lepton direction. Contributions from
charged particles not originating from the primary collision vertex are subtracted
from the isolation sum, multiplied by a factor of 1.5 to account for
the neutral pileup contribution~\cite{CMS-PAS-PFT-10-002}. The relative isolation of the lepton
is defined as the ratio of the corrected isolation sum to the lepton $\pt$.

Prompt leptons are also identified by having low impact parameter (IP) and impact
parameter significance ($S_{\mathrm{IP}}$) values, where the impact parameter is
the minimum three-dimensional distance between the lepton trajectory and the
primary vertex, and its significance is the ratio of the IP value to its uncertainty.
(These values tend to be higher for electrons and muons from the decay of $\tau$ leptons,
which have a nonnegligible lifetime.)
Furthermore, the properties of the nearest jet enclosing the lepton (within $\Delta R < 0.5$)
can be used to identify non-prompt leptons.  The ratio of the lepton $\pt$ to the $\pt$
of this enclosing jet tends to be lower for non-prompt leptons. Also, an enclosing jet
identified as coming from a bottom quark indicates that the lepton is likely non-prompt
and originates from a $\PQb$-hadron decay.

Three levels of lepton selection are defined: preselected, loose, and tight.
The preselection includes leptons in data sidebands used to compute non-prompt backgrounds,
the loose criteria select signal leptons in channels dominated by prompt lepton
events, and the tight selection is used when the largest backgrounds contain
non-prompt leptons.  Loose leptons form a subset of the preselected leptons, and
tight leptons form a subset of the loose leptons. The selection requirements are
described below and summarized in Table~\ref{tab:lep_cuts}.

The preselection removes leptons with an enclosing jet identified as a bottom jet,
as described below, and imposes very loose requirements on the distance from the lepton
trajectory to the primary vertex in the $z$ direction and in the $x$-$y$ plane,
and on the $S_{\mathrm{IP}}$ value. Preselected leptons must also have a relative
isolation less than 0.4.  The preselection has $\approx$100\% efficiency for prompt
leptons, and accepts a substantial number of non-prompt leptons.  Loose leptons must
lie below certain thresholds on the relative isolation calculated using only charged particles
(0.15 for electrons and 0.20 for muons), and loose muons pass a tighter requirement on $S_{\mathrm{IP}}$.
The loose selection retains 93--99\% of prompt muons and 89--96\% of prompt electrons,
depending on $\pt$ and $\eta$, and rejects ${\approx}50\%$ of non-prompt leptons
that pass the preselection.  Tight leptons must pass several selection criteria:
the charged relative isolation must be less than 0.05 for electrons and 0.10
for muons; the ratio of lepton to enclosing jet $\pt$ must be more than 0.6;
and for electrons, the IP must be less than 0.15\unit{mm}. The tight selection efficiency
is ${\approx}90\%$ for prompt muons and ${\approx}80\%$ for prompt electrons, with
efficiency ranges of 68--98\% for muons and 49--93\% for electrons, depending on
$\pt$ and $\eta$. The tight selection rejects ${\approx}80\%$ of non-prompt muons and
${\approx}85\%$ of non-prompt electrons that pass the preselection.

In order to reject leptons with misreconstructed charge, the preselected,
loose, and tight leptons in some channels must pass additional charge
identification (ID) requirements.  Electrons must pass a veto on
electrons from photon conversions and have no missing hits in the inner tracker,
and muons must have more than five inner tracker hits. Electrons
must also have the same charge assignment from the tracker and from the
relative location of ECAL energy deposits from the electron itself and from
its bremsstrahlung radiation. This charge ID selection efficiency ranges from
85 to 100\% for tight electrons with correctly
identified charge, depending on $\pt$ and $\eta$, while more than 97\% of
electrons with misreconstructed charge are rejected. The charge ID selection
has 99\% efficiency for tight muons with correctly identified charge and
rejects ${\approx}100\%$ of muons with misreconstructed charge.
Lepton selection efficiencies are measured using same-flavor (SF) lepton
pairs with an invariant mass near the $\PZ$ boson mass. The charge ID
selection requirements are summarized in Table~\ref{tab:lep_cuts}.

\begin{table}[htb]\small
\renewcommand{\arraystretch}{1.2}
\centering
\topcaption{Summary of preselected, loose, tight, and charge ID lepton selection requirements.
            The charge ID requirements are applied in addition to the preselected, loose, or tight lepton criteria.}
\newcolumntype{.}{D{,}{ }{3}}
\begin{tabular}{l|..|..|..|cc}
\hline
\multicolumn{1}{l|}{Lepton selection criteria} & \multicolumn{2}{c|}{Preselected} &
\multicolumn{2}{c|}{Loose} & \multicolumn{2}{c|}{Tight} & \multicolumn{2}{c}{Charge ID} \\
\hline
Lepton flavor & \multicolumn{1}{c}{$\Pe$} & \multicolumn{1}{c|}{$\PGm$} & \multicolumn{1}{c}{$\Pe$} & \multicolumn{1}{c|}{$\PGm$} & \multicolumn{1}{c}{$\Pe$} & \multicolumn{1}{c|}{$\PGm$} & $\Pe$ & $\PGm$ \\
\hline
$\pt$~(\GeVns{})                     & {>},10 & {>},10 & {>},10 & {>},10 & {>},10 & {>},10 & & \\
$\abs{\eta}$                        & {<},2.5 & {<},2.4 & {<},2.5  & {<},2.4  & {<},2.5  & {<},2.4  & & \\
Relative isolation              & {<},0.4 & {<},0.4 & {<},0.4  & {<},0.4  & {<},0.4  & {<},0.4  & & \\
Charged relative isolation      &         &         & {<},0.15 & {<},0.20 & {<},0.05 & {<},0.15 & & \\
Ratio of lepton $\pt$ to jet $\pt$ &      &         &          &          & {>},0.6  & {>},0.6  & & \\
$x$-$y$ distance to vertex~(mm) & {<},5   & {<},5   & {<},5    & {<},5    & {<},5    & {<},5    & & \\
$z$ distance to vertex~(mm)     & {<},10  & {<},10  & {<},10   & {<},10   & {<},10   & {<},10   & & \\
$\abs{\mathrm{IP}}$~(mm)                   &         &         &          &          & {<},0.15 &          & & \\
$S_{\mathrm{IP}}$               & {<},10  & {<},10  & {<},10   & {<},4    & {<},10   & {<},4    & & \\
Inner tracker hits              &       & &       & &       & &      & ${>}5$  \\
Missing inner tracker hits      & {<},2 & & {<},2 & & {<},2 & & 0    &       \\
Tracker charge $-$ ECAL charge    &       & &       & &       & & 0    &       \\
Electron conversion veto        &       & &       & &       & & Pass &
\\
\hline
\end{tabular}
\label{tab:lep_cuts}
\end{table}

Charged and neutral PF particles are clustered into jets using
the anti-\kt algorithm with a distance parameter of
0.5~\cite{Cacciari:2005hq,Cacciari:2008gp}.  Selected jets must be
separated by $\Delta R > 0.5$ from the selected leptons, and have
$\pt>25\GeV$ and $\abs{\eta}<2.4$.
Charged PF particles not associated with the primary event vertex
are removed from jet clustering, and additional
requirements remove jets arising entirely from pileup
vertices~\cite{CMS-PAS-JME-13-005}.
A neutral component is removed by applying a residual energy correction
following the area-based procedure described in
Refs.~\cite{Cacciari:2008gn,Cacciari:2007fd}, to account for pileup activity.
Fake jets from instrumental effects are rejected by requiring each jet to
have at least two PF constituents and at least 1\% of its energy from
ECAL and HCAL deposits.

The combined secondary vertex (CSV) algorithm~\cite{CMS:2012hd,CMS-PAS-BTV-13-001}
is used to identify (or ``tag'') jets originating from a bottom quark. The CSV algorithm utilizes
information about the impact parameter of tracks and reconstructed secondary
vertices within the jets to assign each jet a discriminator,
with higher values indicating a likely $\PQb$-quark origin.
For a selection with the medium working point of the CSV discriminator, the \PQb tagging
efficiency is around 70\% (20\%) for
jets originating from a bottom quark (charm quark), and the chance of mistagging
jets from light quarks or gluons is about
1\%. For the loose working point, the efficiency to tag jets from \PQb quarks
(c quarks) is approximately 85\% (40\%), and the probability to tag
jets from light quarks or gluons is about 10\%.  These efficiencies
and mistag probabilities vary with the $\pt$ and $\eta$ of the jets.

The missing transverse momentum vector, arising from the presence
of undetected neutrinos in the event, is calculated
as the negative vector sum of the \ptvec of all PF candidates
in the event. This vector is denoted as \ptvecmiss, and its magnitude as
\ptmiss. Since pileup interactions can cause missing transverse momentum not associated
with the primary interaction, the magnitude of the negative vector sum of the \ptvec
of only selected jets and leptons (\MHT) is also used. The \MHT~variable has worse resolution
than \ptmiss, but it is more robust as it does not rely on low-$\pt$ objects
in the event.

The simulation is corrected with data-to-simulation scale factors in order
to match the performance of reconstructed objects
in data.  Simulated events with leptons are corrected for trigger
efficiency, as well as for lepton identification and isolation
efficiency. Scale and resolution corrections accounting for residual
differences between data and simulation are applied to the muon and
electron momenta. All lepton corrections are derived from samples
with a $\PZ$ boson or $\JPsi$ decaying into two leptons.  Jet
energy corrections based on simulation and on $\Pgg+$jets, $\PZ+$jets,
and dijet data are applied as a function of the
jet $\pt$ and $\eta$~\cite{cmsJEC}.  Separate scale factors ranging
from 0.6 to 2.0 are applied to light and heavy flavor jets to correct
the distribution of CSV values~\cite{CMS_ttH_8TeV}.

\section{Event selection}
\label{sec:selection}

Events for this analysis are divided into five mutually exclusive
channels, targeting different decay modes of the $\ttW$ and $\ttZ$
systems.  For all channels, at least one lepton is required to have
$\pt > 20\GeV$, and the remaining leptons must have $\pt > 10 \GeV $,
to satisfy the dilepton trigger requirements.
In addition, to reject leptons from $\PgU$, $\JPsi$, and off-shell photon decays,
no pair of leptons can have an invariant mass less than 12\GeV.
The selection requirements for each channel are described below
and summarized in Table~\ref{tab:selection}.

\begin{table}[htb]\small
\renewcommand{\arraystretch}{1.25}
\centering
\topcaption{Summary of selection requirements for each channel.}
\begin{tabular}{l|cc|cccccc|cc|cc|cc}
\hline
\multicolumn{1}{l|}{Channel} & \multicolumn{2}{c|}{OS $\ttZ$} & \multicolumn{6}{c|}{SS $\ttW$} &
\multicolumn{2}{c|}{3$\ell$ $\ttW$ } & \multicolumn{2}{c|}{ 3$\ell$ $\ttZ$ } &
\multicolumn{2}{c}{4$\ell$ $\ttZ$} \\
\hline
\multicolumn{1}{l|}{Lepton flavor} &
\multicolumn{1}{c|}{$\Pe\Pe/\PGm\PGm$} & \multicolumn{1}{c|}{$\textcolor{white}{\Pe}\Pe\PGm\textcolor{white}{\PGm}$} &
\multicolumn{2}{c|}{$\Pe\Pe$} & \multicolumn{2}{c|}{$\Pe\PGm$} & \multicolumn{2}{c|}{$\PGm\PGm$} &
\multicolumn{2}{c|}{Any} & \multicolumn{2}{c|}{Any} & \multicolumn{2}{c}{Any} \\
\hline
\multicolumn{1}{l|}{Lepton ID} & \multicolumn{2}{c|}{2 loose} & \multicolumn{6}{c|}{2 tight} &
\multicolumn{2}{c|}{SS tight} & \multicolumn{2}{c|}{SS tight} &
\multicolumn{2}{c}{4 loose} \\
\hline
\multicolumn{1}{l|}{Lepton charge ID} & \multicolumn{2}{c|}{${\geq}0$ pass} & \multicolumn{6}{c|}{2 pass} &
\multicolumn{2}{c|}{SS pass} & \multicolumn{2}{c|}{SS pass} &
\multicolumn{2}{c}{4 pass} \\
\hline
\multicolumn{1}{l|}{$\Z  \to \ell\ell$ candidates} & \multicolumn{2}{c|}{1} & \multicolumn{6}{c|}{0} &
\multicolumn{2}{c|}{0} & \multicolumn{2}{c|}{${\geq}1$} & \multicolumn{1}{c|}{ $ \ \ \ \ 2\ \ \ \ \ $ } & \multicolumn{1}{c}{1} \\
\hline
\multicolumn{1}{l|}{Number of jets} & \multicolumn{1}{c|}{5} & \multicolumn{1}{c|}{${\geq} 6$} &
\multicolumn{3}{c|}{ \hspace{0.01\textwidth} 3 \hspace{0.01\textwidth} } &
\multicolumn{3}{c|}{ ${\geq} 4$} &
\multicolumn{1}{c|}{ \hspace{0.01\textwidth}1 \hspace{0.005\textwidth} } &
\multicolumn{1}{c|}{ ${\geq} 2$ } &
\multicolumn{1}{c|}{ \hspace{0.01\textwidth} 3 \hspace{0.005\textwidth} } &
\multicolumn{1}{c|}{ ${\geq} 4$} &
\multicolumn{2}{c}{${\geq} 1$} \\
\hline
\multicolumn{1}{l|}{Number of \PQb tags} & \multicolumn{2}{c|}{${\geq} 1$ medium} & \multicolumn{10}{c|}{${\geq} 2$ loose or ${\geq} 1$ medium} &
\multicolumn{2}{c}{${\geq} 1$ loose} \\
\hline
\multicolumn{1}{l|}{Other} & \multicolumn{2}{c|}{} & \multicolumn{6}{c|}{$\Z  \to \Pe\Pe$ veto} &
\multicolumn{4}{c|}{} & \multicolumn{2}{c}{ \MHT$ > 30\GeV$} \\
\hline
\multicolumn{1}{l|}{Subchannels} & \multicolumn{2}{c|}{4} & \multicolumn{6}{c|}{6} &
\multicolumn{2}{c|}{2} & \multicolumn{2}{c|}{2} &
\multicolumn{2}{c}{2} \\
\hline
\end{tabular}
\label{tab:selection}
\end{table}

The opposite-sign (OS) dilepton channel targets $\ttZ$ events where the $\PZ$ boson
decays into an OS pair of electrons or muons, and the
$\ttbar$ system decays hadronically.  We select events with loose
OS leptons forming an invariant mass within 10\GeV of
the $\PZ$ boson mass and at least five jets, where one or more jets
pass the medium CSV working point. The channel is split into
categories with SF lepton pairs (targeting events with a $\PZ$ boson)
and different-flavor pairs (to calibrate the $\ttbar$ background).
It is further subdivided into events with exactly five jets and those
with six or more jets, which have a higher signal-to-background ratio.
This categorization provides an initial
separation of the $\ttZ$ signal from the dominant $\PZ$ boson and $\ttbar$
backgrounds, which are estimated from simulation.

The same-sign (SS) dilepton channel selects $\ttW$ events in which
the associated $\PW$ boson, and the $\PW$ boson of the same charge from
the $\ttbar$ system, each decay to a lepton and a neutrino, and the remaining $\PW$ boson decays
to quarks.  Events are selected with two SS tight leptons
which pass the charge ID criteria, plus three or more jets, of which at
least two pass the loose CSV threshold or at least one passes the medium CSV threshold.
In addition, in dielectron events, the $\Pe\Pe$ invariant mass
must be at least 10\GeV away from the $\PZ$ boson mass, to reject $\PZ$ boson decays in
which the charge of one electron is misidentified.
This channel is divided by lepton flavor ($\Pe\Pe$, $\Pe\PGm$,
and $\PGm\PGm$), and further into categories with exactly three jets
and four or more jets.  The dominant background is
$\ttbar$ with one non-prompt lepton, which is estimated
from data by computing a misidentification rate. Diboson $\PW \PZ$ events (modeled with simulation)
are selected if one lepton from the $\PZ$ boson decay does not pass the
preselection, or if the $\PZ$ boson decays to a pair of $\tau$ leptons, of
which only one produces a muon or electron.  For the $\Pe\Pe$ and $\Pe\PGm$
categories, dileptonic $\PZ$ boson and $\ttbar$ events with a charge-misidentified
electron also appear in the final selection.

The three-lepton (3$\ell$) $\ttW$ channel targets events in which both
the associated $\PW$ and the pair of $\PW$ bosons from the $\ttbar$
pair decay leptonically.  Events are selected in which
the lepton charges add up to ${\pm} 1$, and the
two leptons of the same charge pass the tight identification and the
charge ID criteria.  Furthermore, no SF OS
pair of leptons can have a mass within 10\GeV of the $\PZ$ boson mass.
Events must have at least one medium $\PQb$-tagged jet, or at least two
loose $\PQb$-tagged jets, and are divided into categories with
exactly one jet, or with two or more jets.  The main backgrounds
are $\ttbar$ decays with a non-prompt lepton, estimated
from data, and $\PW \PZ$ events, estimated using simulation.

The 3$\ell$ $\ttZ$ channel selects events in which
the $\PZ$ boson decays to a pair of electrons or muons, and
one $\PW$ boson from the $\ttbar$ system decays to a charged lepton
and a neutrino, with the remaining $\PW$ boson decaying to quarks.
The selection is identical to the one used for the 3$\ell$ $\ttW$ channel,
except that at least one SF OS pair
of leptons must have an invariant mass within 10\GeV of
the $\PZ$ boson mass, and the categories have exactly three jets, or
four or more jets.  The dominant backgrounds are $\PZ$ boson and
$\ttbar$ events with a non-prompt lepton, and $\PW \PZ$ events
with prompt leptons, estimated in the same manner as in the
3$\ell$ $\ttW$ channel.

Events with four leptons (4$\ell$) come from $\ttZ$ decays in which the $\PZ$ boson
and both $\PW$ bosons decay leptonically.
This channel requires four leptons that pass the loose identification
and the charge ID, and whose charges add up to zero.  At
least one SF OS dilepton pair must
have a mass within 10\GeV of the $\PZ$ boson mass, and at least one
loose $\PQb$-tagged jet must be present.  In addition, $H_\mathrm{T}^\text{miss}$
must exceed 30\GeV. These criteria, and the categorization of events into
those with exactly one lepton pair consistent with a $\PZ$ boson decay,
and those with two or more, help separate $\ttZ$ events
from the dominant $\PZ \PZ$ background, which is estimated
using simulation.  Small backgrounds from
$\ttbar$, $\PW \PZ$, and $\PZ$ boson events with one or two non-prompt leptons
are also estimated using simulation.

\section{Signal and background modeling}
\label{sec:backgrounds}

Events in the signal channels fall into three broad categories.
Signal and ``prompt'' background events have enough
leptons from $\PW$ or $\PZ$ boson decays, with the correct charges,
to satisfy the lepton selection of the channel.  ``Non-prompt''
backgrounds have at least one lepton which is a jet misidentified
as an electron, or which comes from the in-flight decay of a hadron,
or from photon conversion. The ``charge misidentified'' background
has an electron whose charge was misidentified. The expected yields for
these processes after the final selection are shown in
Tables~\ref{tab:yields_OS}--\ref{tab:yields_3l_4l} in Section~\ref{sec:bkg_yields}.

\subsection{Signal and prompt backgrounds}
\label{sec:bkg_prompt}

The signal and prompt backgrounds are estimated using simulation,
normalized to their predicted inclusive cross sections. We use next-to-next-to-leading-order (NNLO)
cross sections for $\ttbar$~\cite{PhysRevLett.110.252004} and single $\PZ$ boson~\cite{Gavin20112388} production;
next-to-leading-order (NLO) cross sections for $\ttW$ and $\ttZ$~\cite{Garzelli:2012bn},
$\ttH$~\cite{YellowReport3}, $\ttbar\gamma$~\cite{MadGraph5, Melnikov:2011ta},
$\PW \PZ$ and $\PZ \PZ$~\cite{Campbell201010}, and $\PW \PW \PW$, $\PW \PW \PZ$, and $\PQt \PQb \PZ$~\cite{MadGraph5} production;
and leading-order cross sections for $\PW^{\pm} \PW^{\pm}$, $\ttbar\gamma^{*}$, and $\ttbar\PW\PW$~\cite{MadGraph5} production.
Additional corrections are derived from data for $\PZ$ boson, $\PW \PZ$, and $\PZ \PZ$ processes with multiple extra jets.

Rare processes such as SS diboson (W$^{\pm}$W$^{\pm}$) and triboson
production ($\PW \PW \PW$, $\PW \PW \PZ$), associated production of a $\PZ$ boson with a single top quark (tbZ),
and $\ttbar$ with an on-shell or off-shell photon ($\ttbar\gamma$/$\ttbar\gamma^{*}$)
or two $\PW$ bosons ($\ttbar\PW\PW$) are subdominant backgrounds.
The associated production of a Higgs boson with a top quark pair
is included as a background, with uncertainties derived from
theoretical predictions. All of these are minor backgrounds,
with fewer expected events than the signal in each channel.

The main prompt backgrounds are $\ttbar$ and $\PZ$ boson production
(in the OS dilepton channel), $\PW \PZ$ events
(in the SS and 3$\ell$ channels), and $\PZ \PZ$ events (in
the 3$\ell$ and 4$\ell$ channels).  Because the $\PZ$ boson, $\PW \PZ$, and
$\PZ \PZ$ simulation samples are produced with fewer extra partons from QCD radiation than
there are jets in the final selection, their estimated contributions
to the signal channels are approximations with large
uncertainties.  To get a more accurate estimate of
these yields, scale factors are derived from events with SF OS
leptons consistent with a $\PZ$ boson decay and no medium $\PQb$-tagged jets.
Using about 5000 data events, of which 97\% are expected to come from
$\Z  \to \ell\ell$ events, we correct the predicted yield from the $\PZ$ boson
simulation as a function of the number of jets for events with five or more jets.
To validate this technique, we derive a scale factor from four jet events with
no medium \PQb tags and apply it to events with at least one medium \PQb tag, and find
that it yields good agreement between data and the $\PZ$ boson simulation.
These scale factors range from 1.35 to 1.7, and each has an uncertainty of 30\%,
based on the level of data-to-simulation agreement in $\PZ$ boson events with four jets. Additional
uncertainties in the $\eta$ distribution of jets in $\PZ$ boson and $\ttbar$ events,
and on the \ptmiss distribution in $\PZ$ boson events with extra jets, are assessed
due to possible data-to-simulation discrepancies in OS dilepton
events with four or more jets (excluding the OS $\ttZ$ signal region).
Scale factors for simulated $\PW \PZ$ and $\PZ \PZ$ events with three or more jets
are derived from 80 three-lepton data events (70\% from $\PW \PZ$) with no medium $\PQb$-tagged
jets and at most one loose $\PQb$-tagged jet. The scale factors of 1.4 for three-jet
events, and 1.6 for events with four or more jets, have 40\% and 60\% uncertainties,
respectively, based on the limited number of 3$\ell$ data events used to derive
the scale factors.

In addition, there is significant uncertainty associated with
the simulation of events with extra heavy flavor partons. Simulated
$\ttbar$, $\PW \PZ$, and $\PZ \PZ$ events with one or two extra \PQc jets,
an extra \PQb jet, or two extra \PQb jets are separated from
their inclusive samples and assigned extra rate uncertainties of 50\%.
The single $\PZ$ boson simulation is divided similarly.  However, by comparing the expected
and observed numbers of $\PQb$-tagged jets in SF OS events
with low \ptmiss and exactly four jets, we are able to constrain
the uncertainty in each of the $\PZ$ boson plus heavy flavor jet processes to 30\%.

The top quark $\pt$ spectrum in $\ttbar$ simulation (from \MADGRAPH) is corrected
to agree with the distribution predicted by higher-order
calculations~\cite{Kidonakis:2012rm} and
observed in $\ttbar$ differential cross section measurements in $\sqrt{s} = 8 \TeV$
data, using the techniques described in Ref.~\cite{Chatrchyan:2012saa}.

\subsection{Non-prompt backgrounds}
\label{sec:bkg_NP}

Backgrounds with at least one non-prompt lepton are expected to have
larger yields than the signal in the SS and 3$\ell$ $\ttW$
channels, about the same yields in the 3$\ell$ $\ttZ$ channel, and
very low yields in the 4$\ell$ channel. Non-prompt
backgrounds in the SS and 3$\ell$ channels are estimated
from data.  A sideband region dominated by non-prompt processes
is defined by events which pass the same selection as the signal channels,
but in which one or both of the preselected SS leptons fail the tight
lepton criteria.  Extrapolation to the signal region is performed by weighting
the sideband events by the probability for non-prompt leptons to pass the tight
lepton selection (the misidentification rate, $\epsilon$).
Events in which one of the SS leptons fails the tight
lepton requirement enter the signal region estimate with
weight $\epsilon/(1-\epsilon)$. Events where both SS leptons
fail the tight lepton selection get a negative weight
$-\epsilon_1\epsilon_2/[(1-\epsilon_1)(1-\epsilon_2)]$; this
accounts for events with two non-prompt leptons contaminating the
sideband sample of events with a single non-prompt lepton.

The misidentification rate is measured with SS and 3$\ell$ data events,
separately for electrons and muons, and as a function of the lepton $\pt$.
Same-sign dilepton events with two or more jets (excluding the $\ttW$ signal region)
are dominated by $\ttbar$ decays with a non-prompt lepton. Three-lepton
events with two or fewer jets, a lepton pair consistent with a $\PZ$ boson decay, and low
\ptmiss come mostly from $\PZ$ boson production with an extra non-prompt lepton.
These events usually have exactly one prompt and one non-prompt SS
lepton, so we use a modified tag-and-probe approach in which the prompt lepton is
tagged with the tight lepton selection, and the fraction of preselected probe leptons
passing the tight selection measures $\epsilon$. Because both leptons
in the numerator of this ratio are tight, there is a $\approx$50\% chance
that the tag lepton was actually non-prompt, and the probe lepton was prompt.  We estimate
the size of this contamination by weighting events where the tag lepton
fails the tight selection by $\epsilon/(1-\epsilon)$, and subtract those with a
tight probe lepton from the numerator, and those with a preselected
probe from the denominator.

Since this correction term depends on $\epsilon$ itself, we
cannot solve for $\epsilon$ explicitly. Instead, we find the
set of $\pt$- and flavor-dependent $\epsilon$ values that
minimizes the difference between the data and predicted
yields in the SS and 3$\ell$ derivation regions,
binned by lepton $\pt$ and flavor.
Events in which both SS leptons are non-prompt naturally
cancel to zero with the correction term, while those with two prompt
SS leptons are estimated from simulation and subtracted explicitly.
The misidentification rate in all the $\pt$ bins is computed to be $\approx$20\%
for muons and $\approx$15\% for electrons, except for the muon bin with
$\pt > 30 \GeV $, whose rate is 36\%.
This rate is uncorrelated with variables that do not depend on the lepton flavor or
$\pt$, including most of those used to separate signal from background events (Section~\ref{sec:signal}).
The relative uncertainty in $\epsilon$ is assessed at 40\% for electrons
and 60\% for muons, equal to the maximum observed discrepancy between predicted and
observed yields in any of 20 background-dominated selection regions with two SS leptons
and two or more jets, or three leptons and two or fewer jets. There is an additional
statistical uncertainty of 50\% for leptons with medium $\pt$ and 100\% for leptons with
high $\pt$, due to low event yields in the $\epsilon$ derivation regions.

In the 4$\ell$ channel, there are too few events passing the kinematic
requirements to use a data sideband to model the non-prompt background.
Instead we use simulated $\PW \PZ$, $\PZ$ boson, and $\ttbar$ samples to estimate
non-prompt yields after the final selection, which are expected to be
much smaller than the signal yields. We derive a scale
factor for the simulation estimate of non-prompt leptons passing the
loose selection using simulated $\PZ$ boson and $\ttbar$ events with exactly three
loose leptons and one or two jets, where at least one passes a
medium \PQb tag.  Events with a SF OS lepton pair close to the $\PZ$ boson mass are
dominated by $\PZ$ boson plus non-prompt lepton events; those without such a pair
are dominated by $\ttbar$ plus non-prompt lepton events.  The derived scale factor
of 2.0 per non-prompt lepton is then applied to the simulation in the 4$\ell$ category,
with 100\% rate uncertainties.

\subsection{Charge-misidentified backgrounds}
\label{sec:bkg_QF}

The misidentified charge background in SS dilepton events
is estimated from OS dilepton events in data that pass
all the other signal channel selections, weighted
by the probability for an electron passing the charge ID requirement
to have misidentified charge.  This probability is derived from data
as a function of electron $\eta$ from the ratio of SS
dielectron events with an invariant mass within 10\GeV of the
$\PZ$ boson mass and zero or more jets, to OS events with the same selection. The
probability ranges from 0.003\% for central electrons to 0.1\%
for endcap electrons.  The absence of a $\PZ$ boson mass peak in SS
dimuon events indicates that the probability is negligible
for muons.  Opposite-sign $\Pe\PGm$ events enter the SS
prediction region with a weight equal to the probability for
the electron to have its charge misidentified; $\Pe\Pe$ events
enter with the sum of the probabilities for each electron.
The charge misidentification probability has a 30\% rate
uncertainty, based on the agreement between predicted and observed
SS dielectron events with multiple jets and with the $\Pe\Pe$
invariant mass close to the $\PZ$ boson mass. We expect to see fewer
events with charge misidentified electrons than $\ttW$ signal
events in all the SS dilepton channels.

\subsection{Expected yields}
\label{sec:bkg_yields}

Expected yields for the signal and background processes after the final
fit described in Section~\ref{sec:results_sm}, along with the observed data
yields, are shown in Tables~\ref{tab:yields_OS}--\ref{tab:yields_3l_4l}.

\begin{table}[!htb]\small
\renewcommand{\arraystretch}{1.12}
\centering
\topcaption{Expected yields after the final fit described in Section~\ref{sec:results_sm}, compared to the observed data for OS $\ttZ$ final states.
         Here ``hf'' and ``lf'' stand for heavy and light flavors, respectively.}
\newcolumntype{x}{D{,}{\,\pm\,}{3}}
\begin{tabular}{l|xx|xx}
\hline
\multicolumn{1}{c|}{OS $\ttZ$} &
\multicolumn{2}{c|}{$\Pe^{\pm}\Pe^{\mp} / \PGm^{\pm}\PGm^{\mp}$} &
\multicolumn{2}{c}{$\Pe^{\pm}\PGm^{\mp}$} \\
\hline

 Process                  &     \multicolumn{1}{c}{5 jets}     &    \multicolumn{1}{c|}{${\geq}6$ jets} &
 \multicolumn{1}{c}{5 jets}     &   \multicolumn{1}{c}{${\geq}6$ jets} \\
 \hline
 $\PZ$+lf jets                &      265 , 57         &       93 , 20       &
 \multicolumn{1}{c}{${<}0.1$}          &    \multicolumn{1}{c}{${<} 0.1$}     \\
 $\PZ$+\ccbar jets            &      341 , 74         &       106 , 23      &
 \multicolumn{1}{c}{${<} 0.1$}          &    \multicolumn{1}{c}{${<} 0.1$}     \\
 $\PZ$+b jet                  &      236 , 59         &       68 , 18       &
 \multicolumn{1}{c}{${<} 0.1$}          &    \multicolumn{1}{c}{${<} 0.1$}     \\
 $\PZ$+\bbbar jets            &      378 , 72         &       136 , 25      &
 \multicolumn{1}{c}{${<} 0.1$}          &    \multicolumn{1}{c}{${<} 0.1$}     \\
 $\ttbar$+lf jets         &      188 , 19         &        58.4 , 7.3   &  180 , 16      &    57.8 , 6.4    \\
 $\ttbar$+hf jets         &      57 , 16          &        30.6 , 8.3   &  52 , 15       &    27.3 , 7.3    \\
 $\PQt \PQb \PZ$/$\ttbar\PW\PW$  &       4.2 , 1.8       &        1.8 , 0.7    &
 \multicolumn{1}{c}{${<} 0.1$}          &    \multicolumn{1}{c}{${<} 0.1$}     \\
 $\ttH$                   &       1.4 , 0.1       &        1.0 , 0.2    &   1.0 , 0.1    &    0.6 , 0.1     \\
 \hline
 Background total               &     1470 , 135        &       494 , 45      &  233 , 21      &    85.8 , 9.7    \\
 \hline
 $\ttZ$                   &       24.0 , 5.5      &        28.2 , 6.8   &   1.3 , 0.3    &    0.8 , 0.2     \\
 $\ttW$                   &       1.1 , 0.2       &        0.5 , 0.1    &   1.2 , 0.2    &    0.8 , 0.2     \\
 \hline
 Expected total                 &     1495 , 135        &       523 , 45      &  236 , 21      &    87.4 , 9.7    \\
 \hline
 Data                     &            \multicolumn{1}{c}{1493}           &             \multicolumn{1}{c|}{526}         &        \multicolumn{1}{c}{251}         &          \multicolumn{1}{c}{78}          \\
\hline
\end{tabular}
\label{tab:yields_OS}
\end{table}

\begin{table}[!htb]\small
\renewcommand{\arraystretch}{1.12}
\centering
\topcaption{Expected yields after the final fit described in Section~\ref{sec:results_sm}, compared to the observed data for SS $\ttW$ final states.
         The multiboson process includes $\PW \PW \PW$, $\PW \PW \PZ$, and $\PW^{\pm}\PW^{\pm}$; $\ttbar$+X includes $\ttbar\gamma$,
		 $\ttbar\gamma^{*}$, and $\ttbar\PW\PW$.}
\newcolumntype{x}{D{,}{\,\pm\,}{3}}
\begin{tabular}{l|xx|xx|xx}
\hline
\multicolumn{1}{c|}{SS $\ttW$} &
\multicolumn{2}{c|}{$\Pe^{\pm}\Pe^{\pm}$} &
\multicolumn{2}{c|}{$\Pe^{\pm}\PGm^{\pm}$} &
\multicolumn{2}{c}{$\PGm^{\pm}\PGm^{\pm}$} \\
\hline

  Process             &       \multicolumn{1}{c}{3 jets}     &     \multicolumn{1}{c|}{${\geq}4$ jets}   &         \multicolumn{1}{c}{3 jets}     &         \multicolumn{1}{c|}{${\geq}4$ jets} &           \multicolumn{1}{c}{3 jets}     &           \multicolumn{1}{c}{${\geq}4$ jets} \\
  \hline
  Non-prompt           &  16.0 , 3.7  &   12.9 , 3.1   &   57.0 , 5.4   &    40.5 , 4.2    &    29.0 , 4.7    &     26.0 , 4.4     \\
  Charge-misidentified        &  3.3 , 1.6   &   1.7 , 0.8    &   2.9 , 0.7    &    1.6 , 0.4     &          \multicolumn{1}{c}{\NA}           &           \multicolumn{1}{c}{\NA}           \\
  $\PW \PZ$                  &  1.6 , 0.5   &   0.9 , 0.3    &   4.5 , 1.4    &    2.2 , 0.8     &    3.1 , 1.0     &     1.3 , 0.5      \\
  $\PZ \PZ$                  &  0.2 , 0.1   &   0.1 , 0.1    &   0.3 , 0.1    &    0.2 , 0.1     &    0.2 , 0.1     &     0.1 , 0.1      \\
  Multiboson          &  0.8 , 0.3   &   0.5 , 0.2    &   1.5 , 0.5    &    1.2 , 0.4     &    1.2 , 0.5     &     1.1 , 0.4      \\
  $ \PQt \PQb\PZ$/$\ttbar$+X &  1.4 , 0.4   &   2.5 , 1.3    &   4.1 , 1.4    &    5.8 , 2.2     &    0.9 , 0.3     &     1.2 , 0.4      \\
  $\ttH$              &  0.3 , 0.1   &   1.4 , 0.2    &   1.1 , 0.1    &    4.0 , 0.5     &    0.7 , 0.1     &     3.0 , 0.5      \\
  \hline
  Background total          &  23.7 , 4.1  &   20.1 , 3.5   &   71.4 , 5.8   &    55.4 , 4.9    &    35.1 , 4.8    &     32.8 , 4.5     \\
  \hline
  $\ttW$              &  5.5 , 1.4   &   8.1 , 1.9    &   13.9 , 3.7   &    25.2 , 5.5    &    10.4 , 2.8    &     17.7 , 4.0     \\
  $\ttZ$              &  0.4 , 0.1   &   1.3 , 0.3    &   1.1 , 0.2    &    3.0 , 0.6     &    0.7 , 0.1     &     2.1 , 0.4      \\
  \hline
   Expected total           &  29.6 , 4.4  &   29.4 , 4.0   &   86.4 , 6.9   &    83.6 , 7.3    &    46.2 , 5.6    &     52.6 , 6.0     \\
   \hline
   Data               &        \multicolumn{1}{c}{31}        &         \multicolumn{1}{c|}{32 }        &         \multicolumn{1}{c}{89}         &          \multicolumn{1}{c|}{69}          &          \multicolumn{1}{c}{47}          &           \multicolumn{1}{c}{61}           \\
\hline
\end{tabular}
\label{tab:yields_SS}
\end{table}

\begin{table}[!htb]\small
\renewcommand{\arraystretch}{1.12}
\centering
\topcaption{Expected yields after the final fit described in Section~\ref{sec:results_sm}, compared to the observed data for 3$\ell$ $\ttW$
         and three and 4$\ell$ $\ttZ$ final states. The 4$\ell$ ``Z-veto'' channel has exactly one lepton pair consistent with a $\PZ$ boson decay; the ``Z''
         channel has two. The multiboson process includes $\PW \PW \PW$ and $\PW \PW \PZ$; $\ttbar$+X includes $\ttbar\gamma$, $\ttbar\gamma^{*}$, and $\ttbar\PW\PW$.}
\newcolumntype{y}[1]{D{,}{\,\pm\,}{#1}}
\begin{tabular}{l|y{3}y{3}|y{3}y{3}|y{3}y{5}}
\hline
\multicolumn{1}{c|}{} &
\multicolumn{2}{c|}{3$\ell$ $\ttW$} &
\multicolumn{2}{c|}{3$\ell$ $\ttZ$} &
\multicolumn{2}{c}{4$\ell$ $\ttZ$} \\
\hline
   Process   &                \multicolumn{1}{c}{1 jet} &                \multicolumn{1}{c|}{${\geq}2$ jets} &                \multicolumn{1}{c}{3 jets} &                \multicolumn{1}{c|}{${\geq}4$ jets} &           \multicolumn{1}{c}{${\geq}1$ jet+$\PZ$} &           \multicolumn{1}{c}{${\geq}1$ jet+$\PZ$-veto} \\
   \hline
      Non-prompt  &      44.6 , 5.3     &       54.8 , 6.4       &      8.2 , 2.8       &       5.4 , 2.1        &          \multicolumn{1}{c}{\NA}           &             \multicolumn{1}{c}{\NA}             \\
    Non-prompt $\PW \PZ$/Z   &          \multicolumn{1}{c}{\NA}          &           \multicolumn{1}{c|}{\NA}            &          \multicolumn{1}{c}{\NA}           &           \multicolumn{1}{c|}{\NA}            &      \multicolumn{1}{c}{${<}0.1$}       &         \multicolumn{1}{c}{${<}0.1$}         \\
 Non-prompt $\ttbar$ &          \multicolumn{1}{c}{\NA}          &           \multicolumn{1}{c|}{\NA}            &          \multicolumn{1}{c}{\NA}           &           \multicolumn{1}{c|}{\NA}            &      \multicolumn{1}{c}{${<}0.1$}       &         0.2 , 0.2         \\
       $\PW \PZ$     &      3.2 , 0.8      &       8.0 , 1.7        &      11.7 , 2.9      &       5.4 , 1.6        &          \multicolumn{1}{c}{\NA}           &             \multicolumn{1}{c}{\NA}             \\
         $\PZ \PZ$     &      1.0 , 0.2      &       1.5 , 0.3        &      1.6 , 0.4       &       0.9 , 0.3        &      3.3 , 0.5       &         1.8 , 0.3         \\
   Multiboson &      0.1 , 0.1      &       0.4 , 0.2        &      0.5 , 0.2       &       0.5 , 0.2        &      \multicolumn{1}{c}{${<}0.1$}       &         0.3 , 0.1         \\
   $ \PQt \PQb\PZ$/$\ttbar$+X &      0.4 , 0.1      &       3.8 , 1.1        &      1.6 , 0.6       &       0.7 , 0.3        &      \multicolumn{1}{c}{${<}0.1$}       &         \multicolumn{1}{c}{${<}0.1$}         \\
     $\ttH$   &      0.2 , 0.1      &       4.7 , 0.4        &      0.3 , 0.1       &       0.4 , 0.1        &      \multicolumn{1}{c}{${<}0.1$}       &         0.2 , 0.1         \\
 \hline
   Background total &      49.5 , 5.4     &       73.1 , 6.7       &      23.9 , 4.1      &       13.3 , 2.7       &      3.3 , 0.5       &         2.4 , 0.4         \\
   \hline
       $\ttW$   &      2.5 , 0.8      &       18.8 , 4.7       &      0.5 , 0.1       &       0.2 , 0.1        &          \multicolumn{1}{c}{\NA}           &             \multicolumn{1}{c}{\NA}             \\
       $\ttZ$   &      0.3 , 0.1      &       7.5 , 1.2        &      8.8 , 1.9       &       16.9 , 3.6       &      0.4 , 0.1       &         4.3 , 1.0         \\
   \hline
     Expected total  &      52.3 , 5.4     &       99.4 , 8.3       &      33.2 , 4.5      &       30.4 , 4.5       &      3.7 , 0.5       &         6.7 , 1.1         \\
     \hline
      Data    &            \multicolumn{1}{c}{51}           &             \multicolumn{1}{c|}{97}             &            \multicolumn{1}{c}{32}            &             \multicolumn{1}{c|}{30}             &            \multicolumn{1}{c}{3}             &               \multicolumn{1}{c}{6}               \\
\hline
\end{tabular}
\label{tab:yields_3l_4l}
\end{table}

\section{Full event reconstruction}
\label{sec:reconstruction}

Even after the selection requirements have been applied, the final signal
categories are dominated by background events.  To help identify the $\ttW$
and $\ttZ$ signals, and the $\ttbar$ background, we attempt a full reconstruction
of the events, by matching leptons, jets, and \ptmiss to the decaying W
and $\PZ$ bosons, and to the top quark and antiquark.

In all channels targeting the $\ttZ$ signal, the SF OS
pair of leptons with an invariant mass closest to the $\PZ$ boson mass is assumed
to be from the $\PZ$ boson decay.
In selected $\ttW$ events, there are at least two leptons and two undetected
neutrinos, so the associated $\PW$ boson cannot be reconstructed. Thus, for both
$\ttW$ and $\ttZ$ events, as well as $\ttbar$ events, it is the $\ttbar$ system
which remains to be reconstructed.  In selected OS $\ttZ$ events, both
$\PW$ bosons from the $\ttbar$ pair decay into quarks; we refer to this as
a fully hadronic $\ttbar$ decay.  In SS $\ttW$ and 3$\ell$ $\ttZ$
events, the $\ttbar$ pair decays semileptonically. The 3$\ell$ $\ttW$ and
4$\ell$ $\ttZ$ channels target leptonic $\ttbar$ decays. While
background $\ttbar$ events have genuine top quarks to reconstruct, they decay in a
different mode than the signal does, \eg in OS $\ttbar$ events both W
bosons decay leptonically, and in SS and 3$\ell$ $\ttbar$ events one lepton usually
comes from a $\PQb$-hadron decay.

The leptons, jets, and \ptmiss from $\ttbar$ decays preserve information about
their parent particles. Pairs of jets from hadronic $\PW$ boson decays have an
invariant mass close to the $\PW$ boson mass; adding the \PQb jet from
the same top quark decay gives three jets with an invariant mass close to the top
quark mass.  In semileptonic $\ttbar$ decays,
the transverse component of the lepton momentum vector and \ptvecmiss give a
Jacobian mass distribution which peaks around 60\GeV and quickly drops as it
approaches the $\PW$ boson mass.  Additionally, the lepton and \PQb jet coming from
the same top quark decay will have an invariant mass smaller than the top quark mass.
Jets from \PQb quarks tend to have higher CSV values, while light flavor jets
have lower values, and \PQc jets have an intermediate distribution.  The jet
charges of \PQb jets from top quarks and quark jets from $\PW$ boson decays are also
used. Finally, the ratio of the invariant mass using only the transverse component
of momentum vectors (\MT) to the full invariant mass tends to be higher for the
set of jets coming from top quark and $\PW$ boson decays than for sets with jets from
extra radiated partons. These variables are all used in the event reconstruction
described below, and are listed in Appendix~\ref{sec:appendix_match_vars},
Table~\ref{tab:match_vars_ttW_ttZ}.

\begin{figure}[!htbp]
  \centering
  \includegraphics[width=0.32\textwidth]{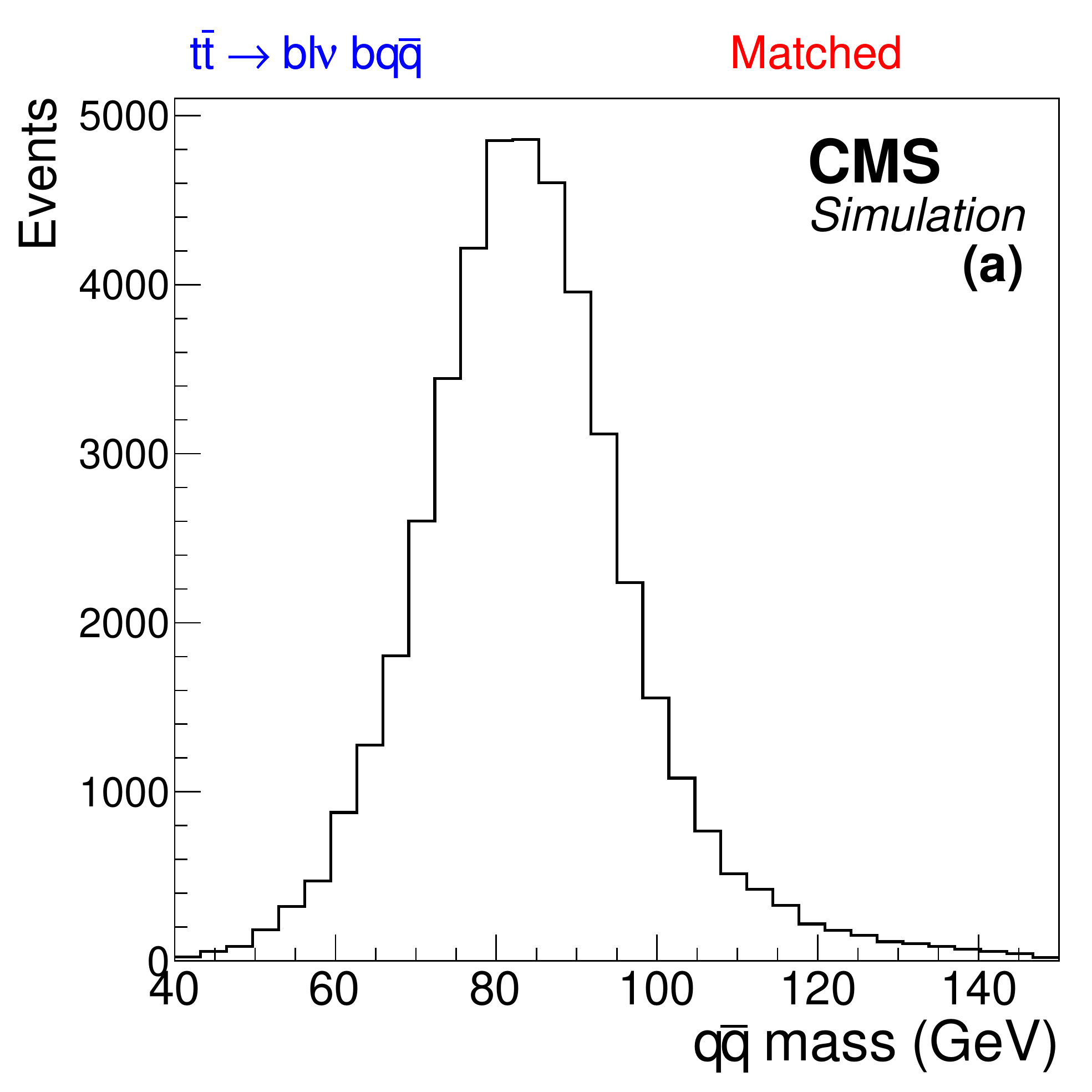}
  \includegraphics[width=0.32\textwidth]{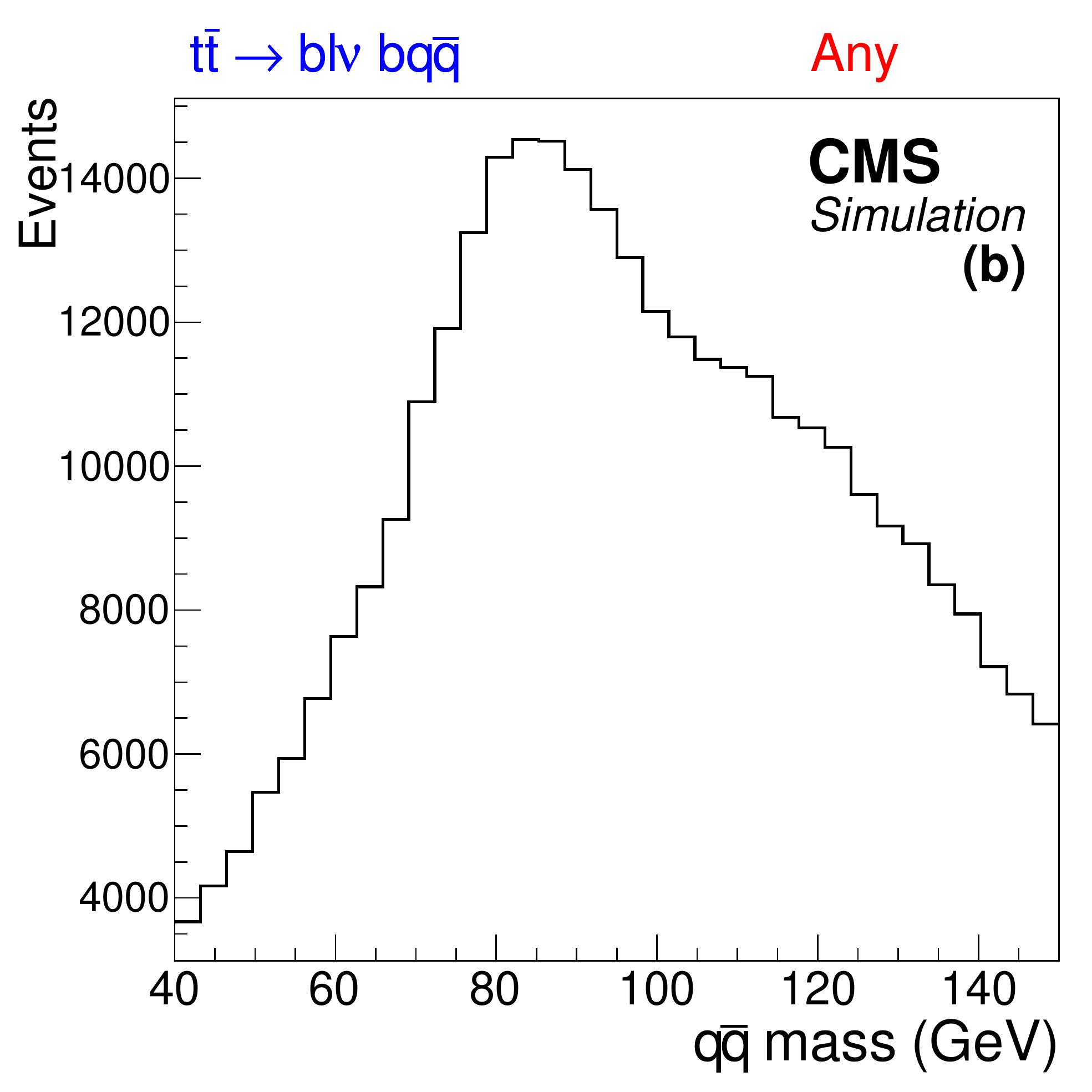}
  \includegraphics[width=0.32\textwidth]{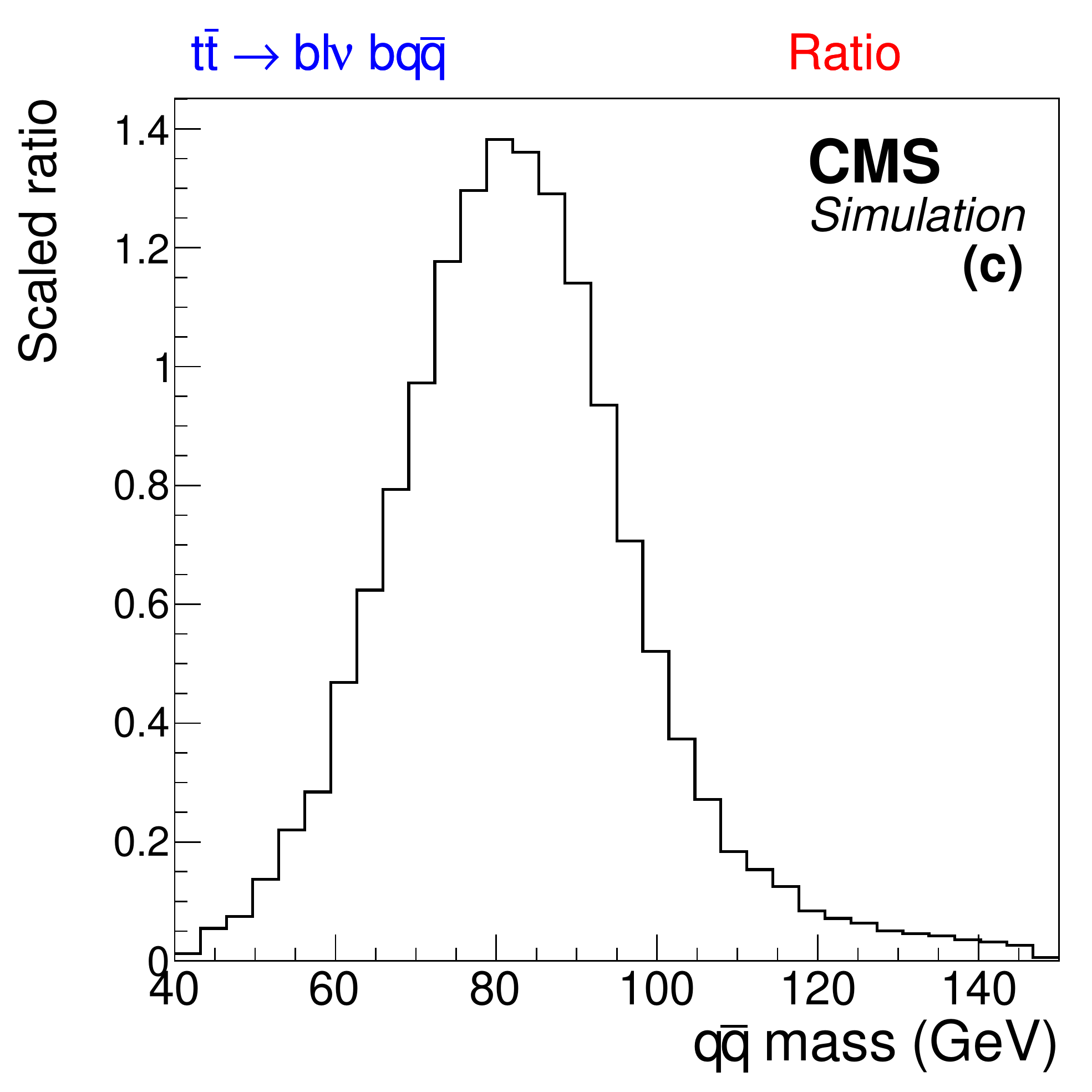}
  \caption{Distributions from simulated $\ttbar  \to \PQb\ell\nu$ $\PAQb\PQq\PAQq$ events with exactly
           four jets. Shown are the invariant mass of two jets matched to a hadronic $\PW$ boson decay (a), the
           invariant mass of any two jets (b), and the rescaled ratio of the two (c).}
  \label{fig:match_input_histos}
\end{figure}

To optimally match jets and leptons to their top quark and $\PW$ boson parents
in $\ttbar$ decays, we construct a linear discriminant, similar to a likelihood
ratio, which evaluates different permutations of object-parent pairings. The
discriminant is created using millions of simulated $\ttbar$ events, so
the true parentage of each object is known, and the variable distribution
shapes have high precision. For each input variable to the discriminant,
we take the ratio of the distribution using correctly matched objects (\eg the
invariant mass of two jets coming from the same $\PW$ boson decay) to the
distribution using any set of objects (\eg the invariant mass of any two jets
in the event), and rescaled the ratio histogram to have a mean value of one
for correctly matched objects, as shown in Fig.~\ref{fig:match_input_histos}.
Variables with more discriminating power, such as the reconstructed $\PW$ boson
mass, have ratio histograms with some bin values very close to zero, and others
above one; less discriminating variables such as \PQb jet charge have values
well above zero in all bins.

\begin{figure}[htb]
  \centering
  \includegraphics[width=0.32\textwidth]{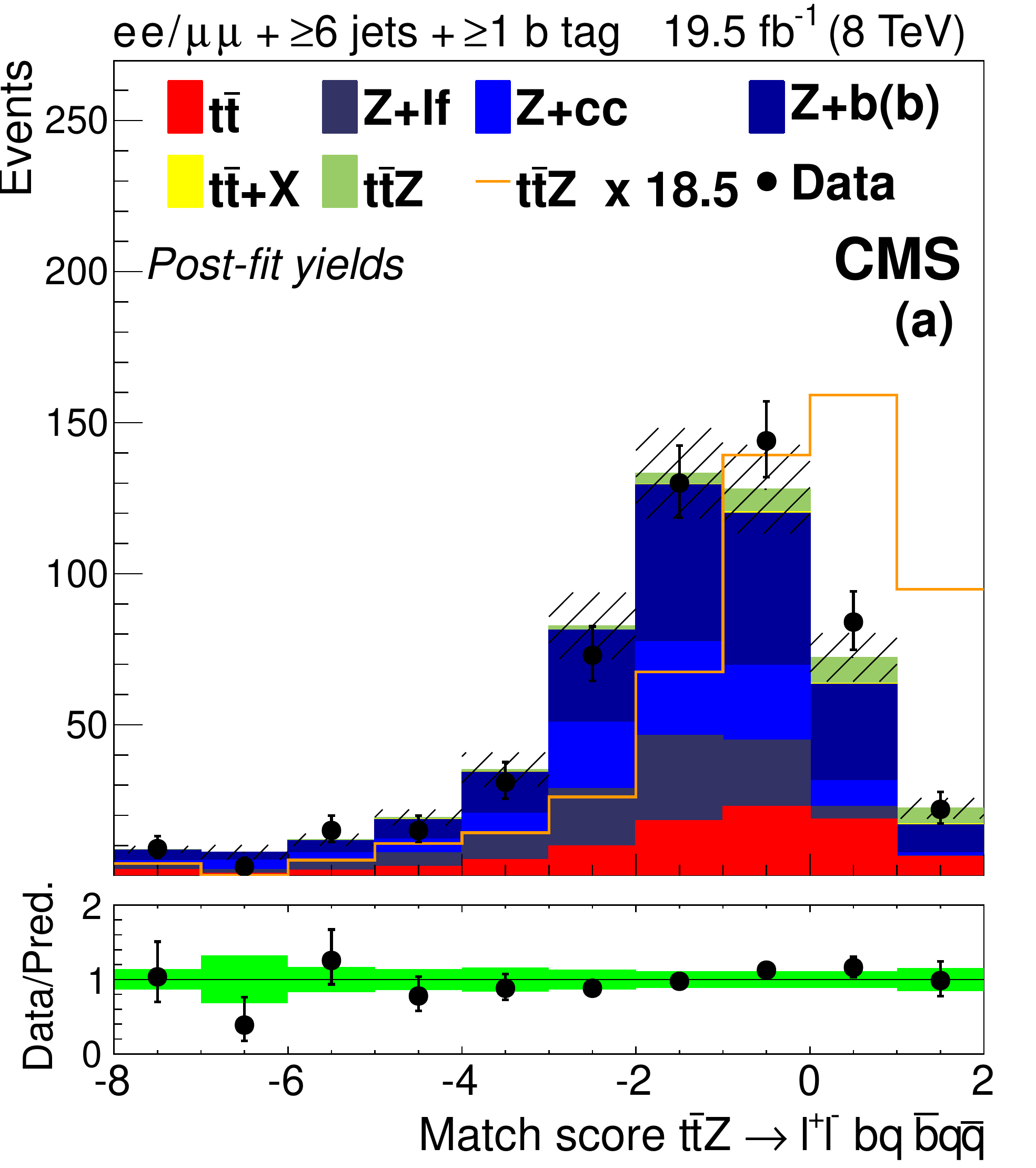}
  \includegraphics[width=0.32\textwidth]{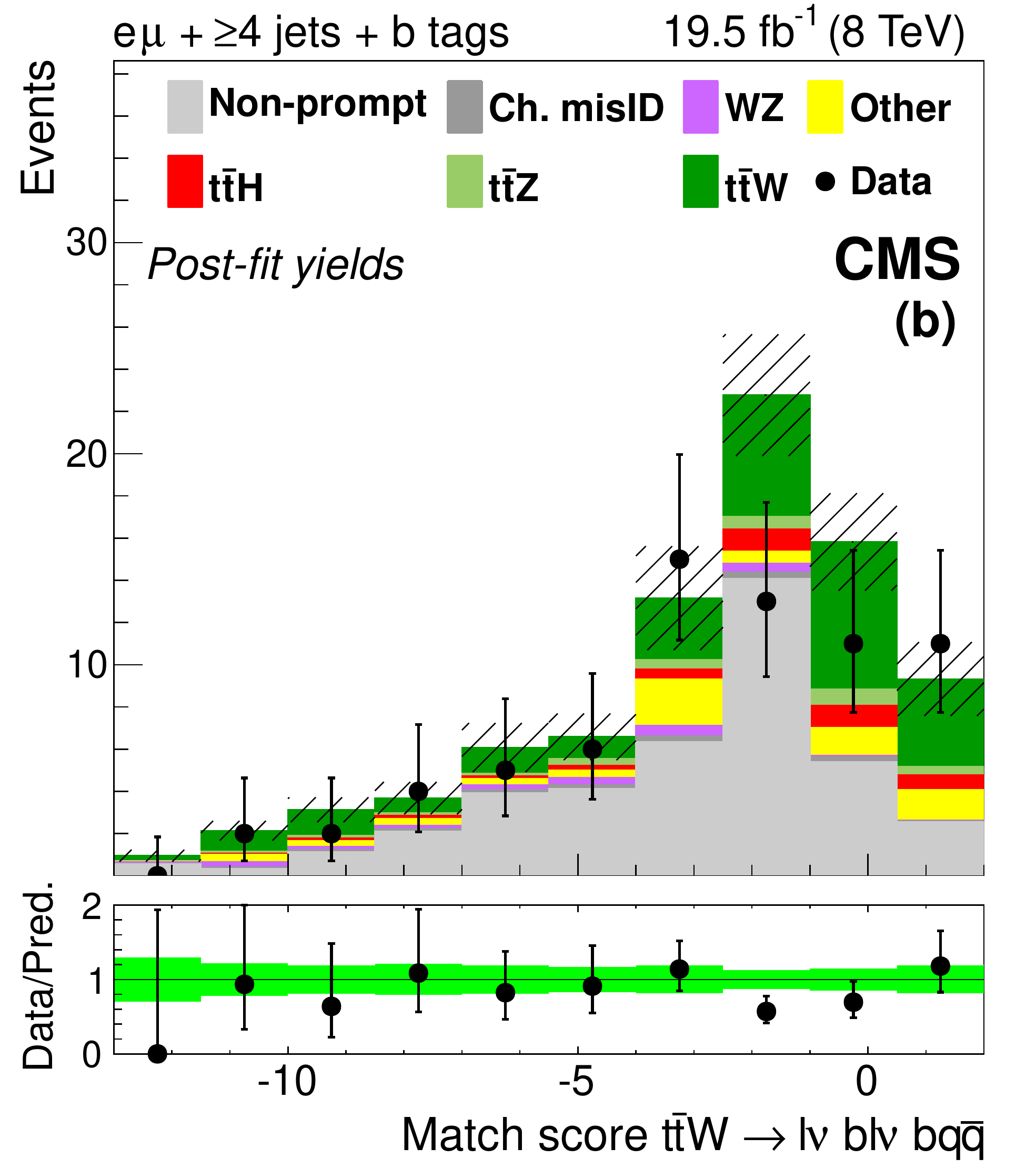}
  \includegraphics[width=0.32\textwidth]{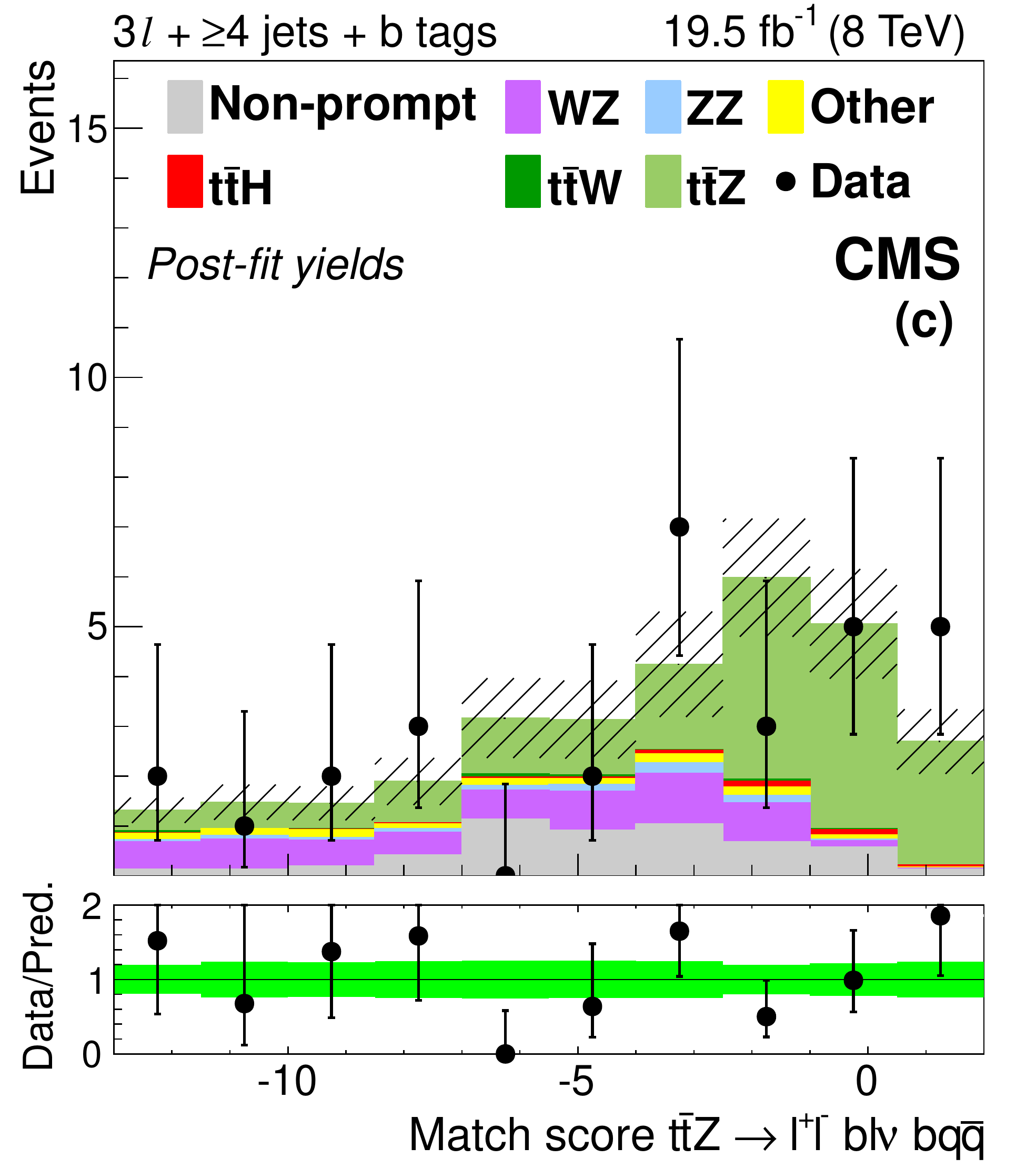}
  \caption{Distributions for match scores with signal and background yields from the final
           fit described in Section~\ref{sec:results_sm}. Plot (a) shows the match
           score for partially reconstructed hadronic $\ttbar$ systems in SF
           OS dilepton events with six or more jets.  Plots (b) and (c)
           show scores for fully reconstructed semileptonic $\ttbar$ systems in
           events with at least four jets, and a SS $\Pe\PGm$ pair, or three
           leptons, respectively. The 68\% CL uncertainty in the signal plus background
           prediction is represented by hash marks in the stack histogram, and a green
	       shaded region in the data-to-prediction ratio plot. The orange line in plot (a)
           shows the shape of the $\ttZ$ signal, suitably normalized. ``Ch.~misID'' indicates the charge-misidentified
           background. ``Other''  backgrounds include  $\ttbar\gamma$, $\ttbar\gamma^{*}$,
           $\ttbar\PW\PW$, $ \PQt \PQb\PZ$, $\PW \PW \PW$, $\PW \PW \PZ$, and $\PW^{\pm} \PW^{\pm}$.}
  \label{fig:match_output_histos}
\end{figure}

We use these ratio histograms to match objects to $\ttbar$ decays in selected
events in the $\ttW$ and $\ttZ$ signal regions, where the object parentage is
not known. In $\ttZ$ channels the leptons matched to the $\PZ$ boson decay are
excluded from the $\ttbar$ reconstruction, and in $\ttW$ channels the lepton with
the worst match to a $\ttbar$ decay is assumed to come from the associated $\PW$ boson.
For each permutation of leptons and jets matched to parent particles, we
find the value of every variable (mass, CSV, charge, \etc) associated with
an object-parent pairing. The matching discriminant is then computed
as the product of the corresponding bin values from all the ratio histograms.
The permutation with the highest discriminant value is considered to be the
best reconstruction of the $\ttbar$ system. To more easily display the full
range of values, we take the log of the discriminant value of the best
reconstruction to calculate the match score. Events that contain all of the jets
and leptons from the $\ttbar$ decay have match scores around zero, while events
without all the decay products typically get negative scores. For semileptonic
$\ttbar$ decays in events with exactly four jets, all from the $\ttbar$ system, the
highest scored permutation is the correct assignment 75\% of the time.  For
events with five or more jets, of which four are from the $\ttbar$ system, the
exact correct match is achieved in 40\% of cases, as there are five times
as many permutations to choose from.  Since one or two jets from the $\ttbar$
decay often fail to be reconstructed, we also attempt to match partial $\ttW$
and $\ttZ$ systems, with one or two jets missing.  Output match scores in the
OS $\ttZ$, SS $\ttW$, and 3$\ell$ $\ttZ$ channels
are shown in Fig.~\ref{fig:match_output_histos},
along with the 68\% confidence level (CL) uncertainty in the signal plus background prediction.

Since the background processes do not have the same parent particles as
the signal in each channel, their best reconstructed match scores are typically
lower. Thus, the match scores for full or partial reconstructions of the
$\ttbar$ system in $\ttW$, $\ttZ$, and $\ttbar$ decays, along with the values of input variables to the
chosen match (\eg dijet mass of the hadronically decaying $\PW$ boson in a semileptonic $\ttbar$ decay),
provide good discrimination between signal and background processes,
especially those without any genuine top quarks.

\section{Signal extraction}
\label{sec:signal}

The match scores and other event reconstruction variables are
combined with kinematic quantities (\eg lepton $\pt$ and jet CSV
values) in boosted decision trees (BDTs)~\cite{bdt_nim} to
distinguish signal events from background processes.
The linear discriminant for event reconstruction combines a large number
of variables into maximally distinctive observables, achieving better
separation than a BDT alone would, since fewer variables can be used in a BDT
when the number of simulated events for training is limited.
A separate BDT is trained for each jet category in each analysis
channel, for a total of 10 BDTs. The input variables to
these BDTs are described below, and listed in Appendix~\ref{sec:appendix_BDT_vars},
Tables~\ref{tab:BDT_inputs_SS_ttW}--\ref{tab:BDT_inputs_OS_ttZ}.

An initial BDT in the OS channel is trained with $\ttZ$ events against the
$\ttbar$ background, using the $\PZ$ boson mass and $\pt$, and $\Delta R$ separation between
leptons as inputs, as well as \MHT, the number of jets with $\pt > 40 \GeV $,
and the ratio of the \MT to the mass of a four-momentum vector composed of all
the jets in the event. Event reconstruction variables include match scores to leptonic and
fully hadronic $\ttbar$ decays, and the CSV values of jets matched to \PQb quarks from the leptonic
$\ttbar$ decay. The final BDT is then trained against $\PZ$ boson and $\ttbar$ events,
using the $\ttZ$ vs.~$\ttbar$ BDT as an input, along with the two highest
jet CSV values, the fifth-highest jet $\pt$, the number of
jets with $\pt > 40 \GeV $, and the ratio of the \MT to the mass of all the selected
leptons and jets. Match scores to the partial five-jet and full six-jet
hadronic $\ttbar$ system are also included, along with the minimum
$\chi^{2}$ value of a fit to the full hadronic $\ttbar$ system that
uses only the $\PW$ boson and top quark masses as inputs.

The SS channel BDT is trained with $\ttW$ events against $\ttbar$ simulation,
using the lepton $\pt$ values, \ptmiss, the second-highest jet CSV value, and the
\MT of the system formed by the leptons, jets, and \ptvecmiss.  Event reconstruction
variables include match scores to three- and four-jet $\ttW$ decays and three-jet
$\ttbar$ decays, the matched top quark candidate mass from two jets from the $\PW$ boson
and the non-prompt lepton from the $\PQb$-hadron decay, and the other top quark
candidate \MT from the prompt lepton, \ptvecmiss, and the \PQb jet in $\ttbar$ decays.

The BDT for the 3$\ell$ $\ttW$ channel is trained against
$\ttbar$ simulation, using the $\pt$ of the SS leptons,
the highest jet $\pt$, the second-highest jet CSV value, \ptmiss,
and the \MT of the leptons, jets, and \ptvecmiss. Match
scores for the two-jet $\ttW$ system and one-jet $\ttbar$ systems,
along with the invariant mass of the prompt and non-prompt leptons
matched to the same top quark in a $\ttbar$ decay, are also used.

The 3$\ell$ $\ttZ$ BDT is trained against simulated $\PW \PZ$ and
$\ttbar$ events, which contribute equally to the background in
this channel. The input kinematic variables are the reconstructed
$\PZ$ boson mass (which discriminates against $\ttbar$), the \MT
of the \ptvecmiss, leptons and jets, and the number of
medium $\PQb$-tagged jets.  In the three-jet category, match scores
for $\ttZ$ reconstructions with one or two jets missing from
the semileptonic $\ttbar$ decay are used as inputs; in the four jet
category, match scores for three-jet systems and the full four-jet
system are used.

The 4$\ell$ channel has too few signal and background
events to train a BDT; here the number of medium $\PQb$-tagged jets
is used as a discriminant instead.  This variable effectively
separates $\ttZ$ events from the dominant $\PZ \PZ$ background, and
from subdominant non-prompt $\PW \PZ$, $\PZ$ boson, and $\ttbar$ backgrounds.

\begin{figure}[htb]
\centering
  \includegraphics[width=0.32\textwidth]{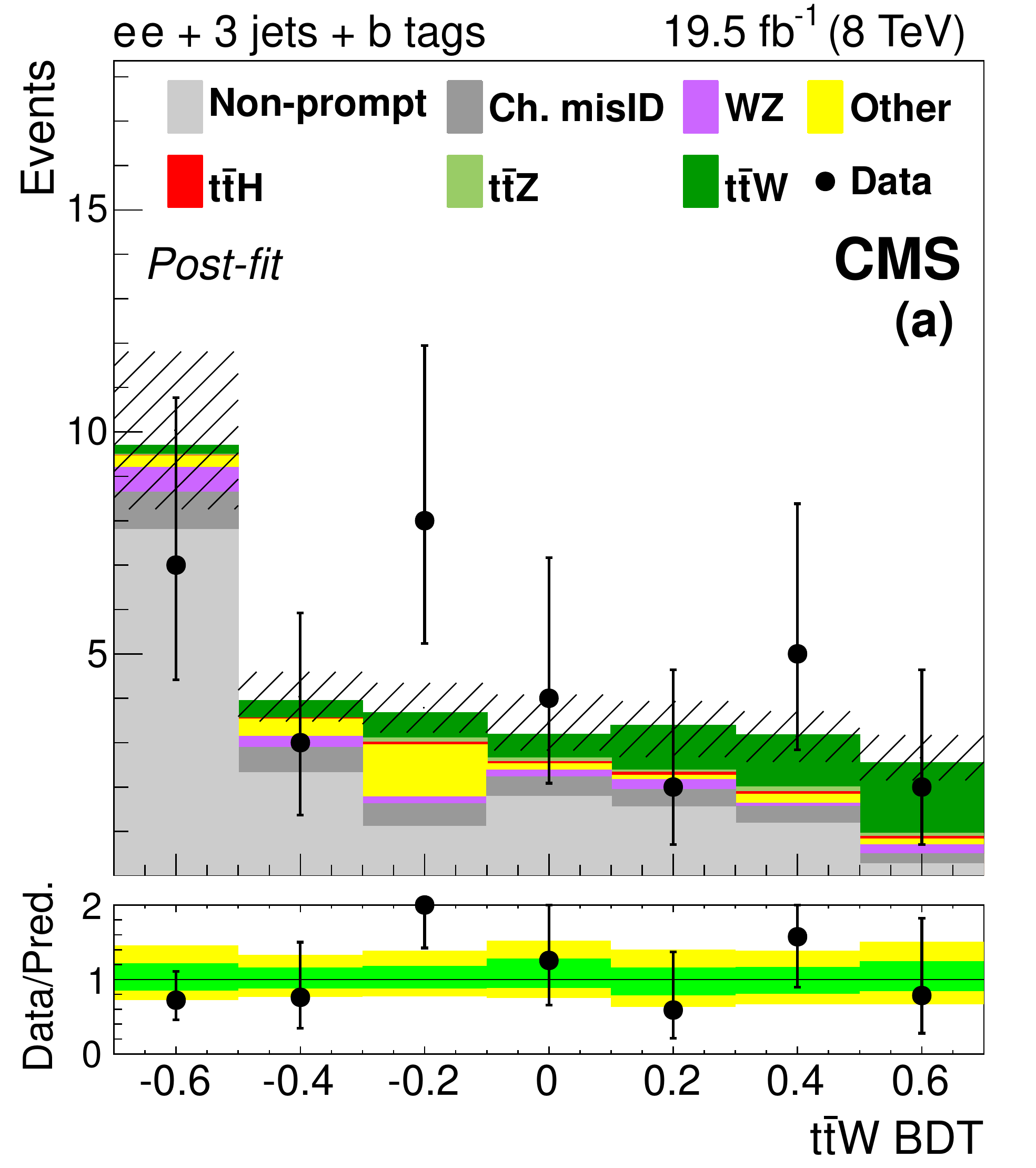}
  \includegraphics[width=0.32\textwidth]{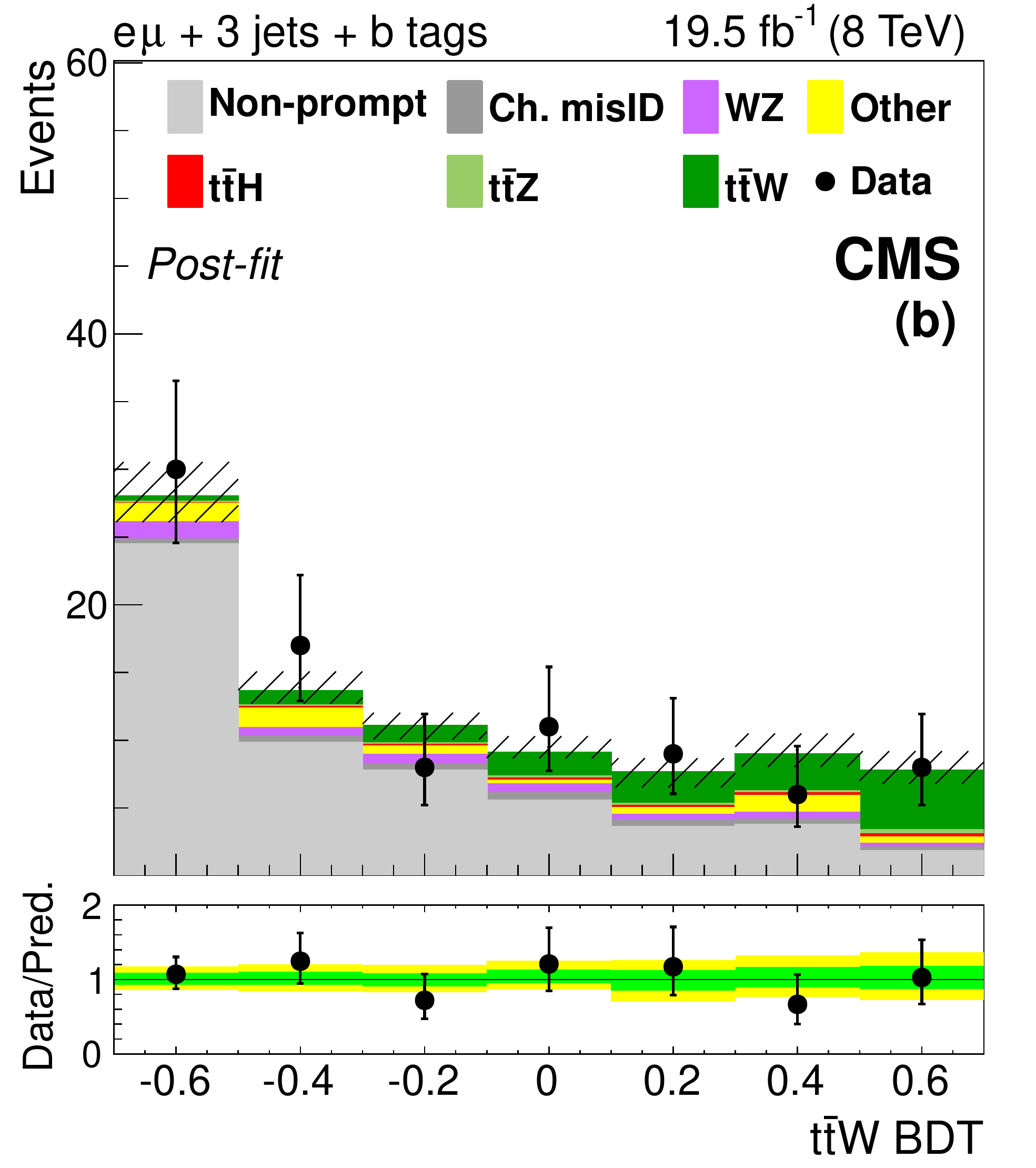}
  \includegraphics[width=0.32\textwidth]{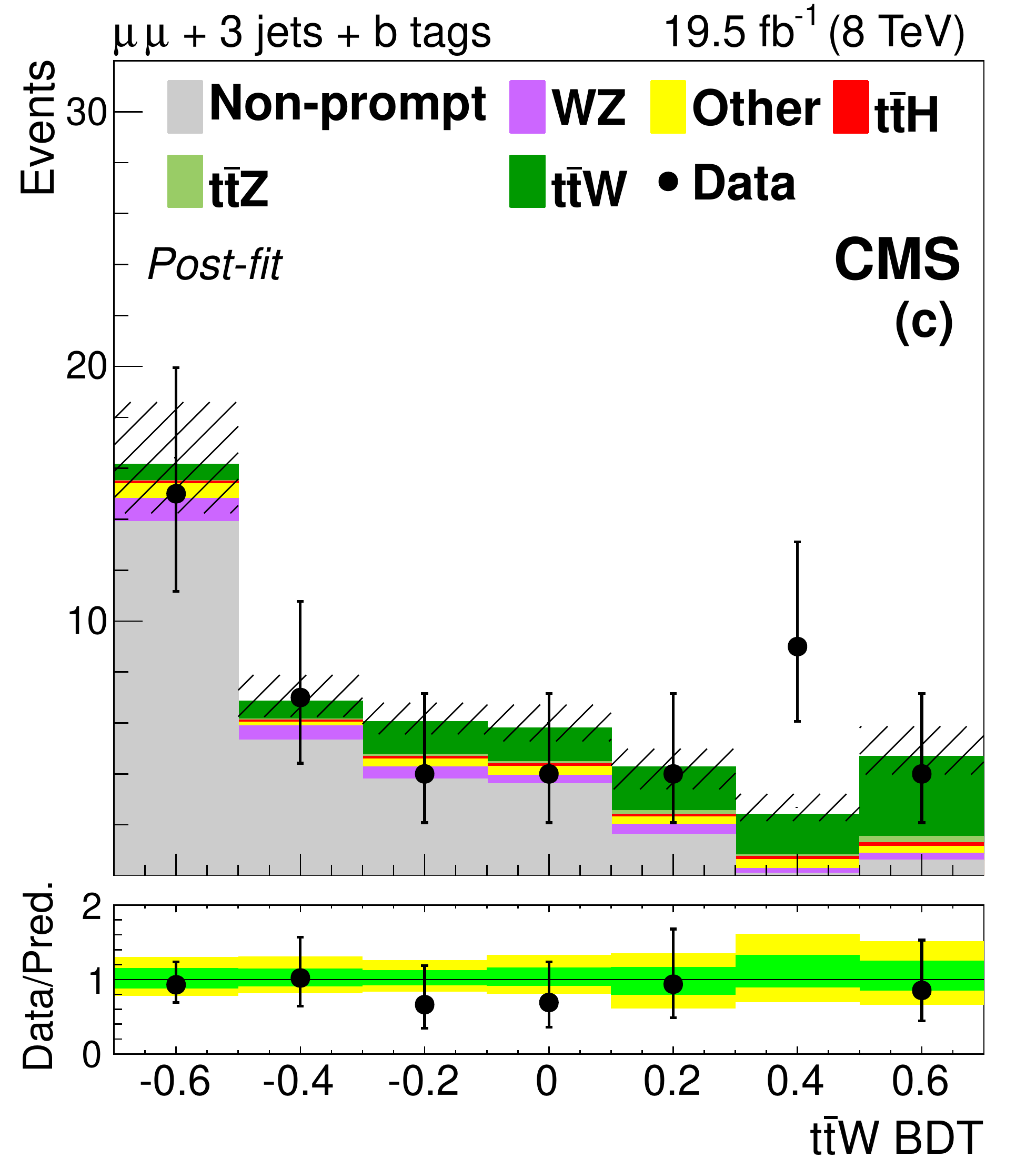}
  \includegraphics[width=0.32\textwidth]{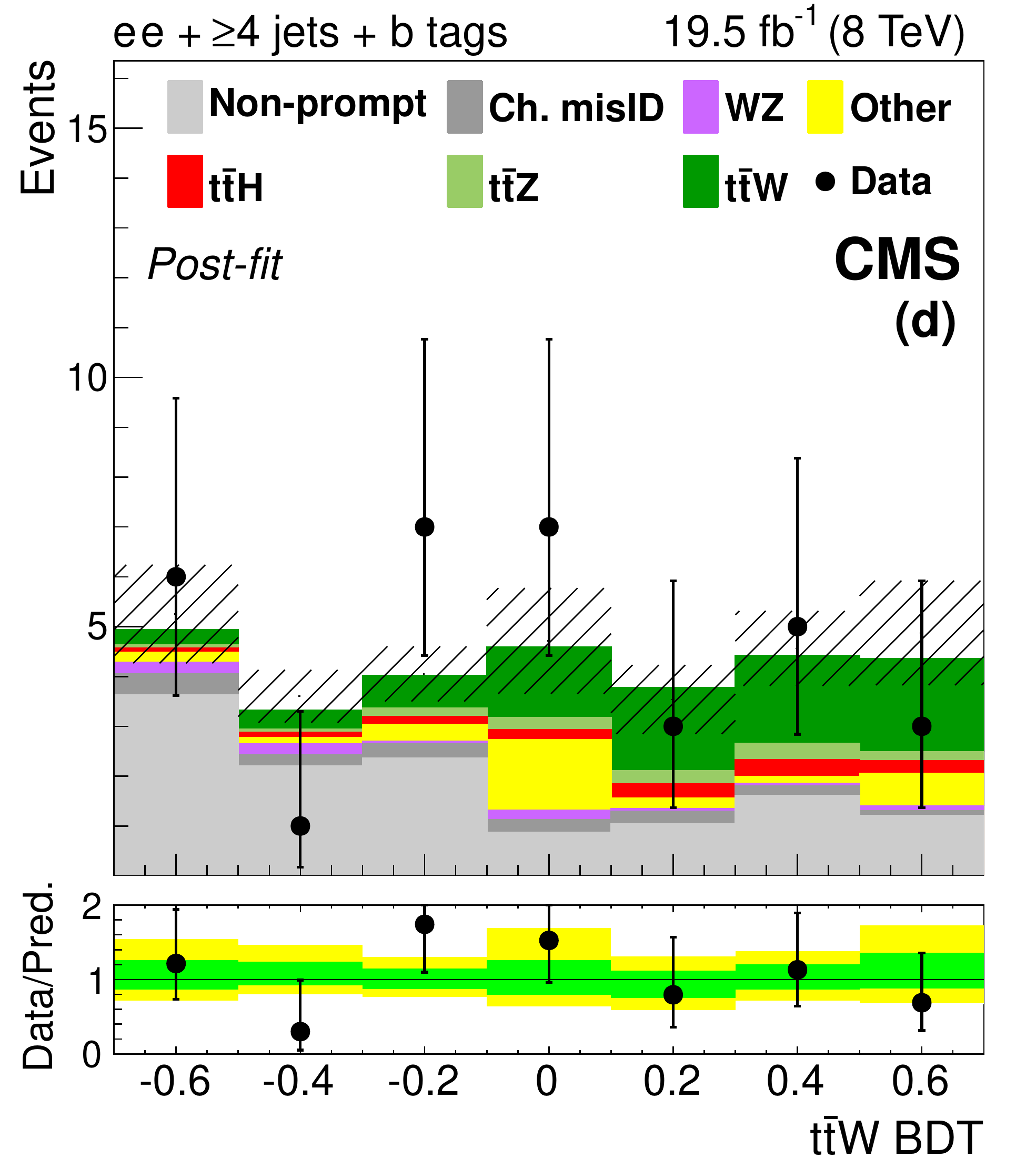}
  \includegraphics[width=0.32\textwidth]{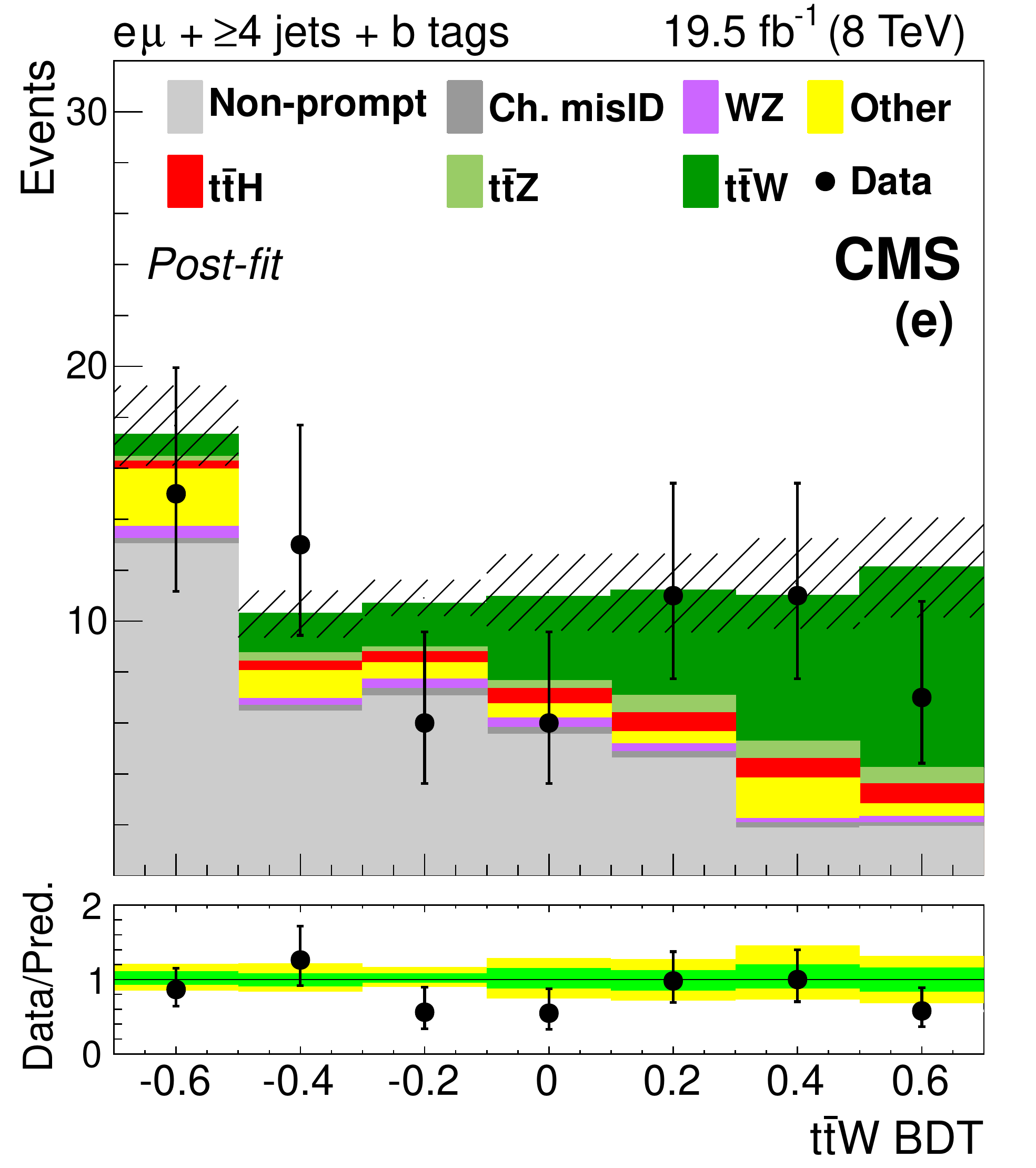}
  \includegraphics[width=0.32\textwidth]{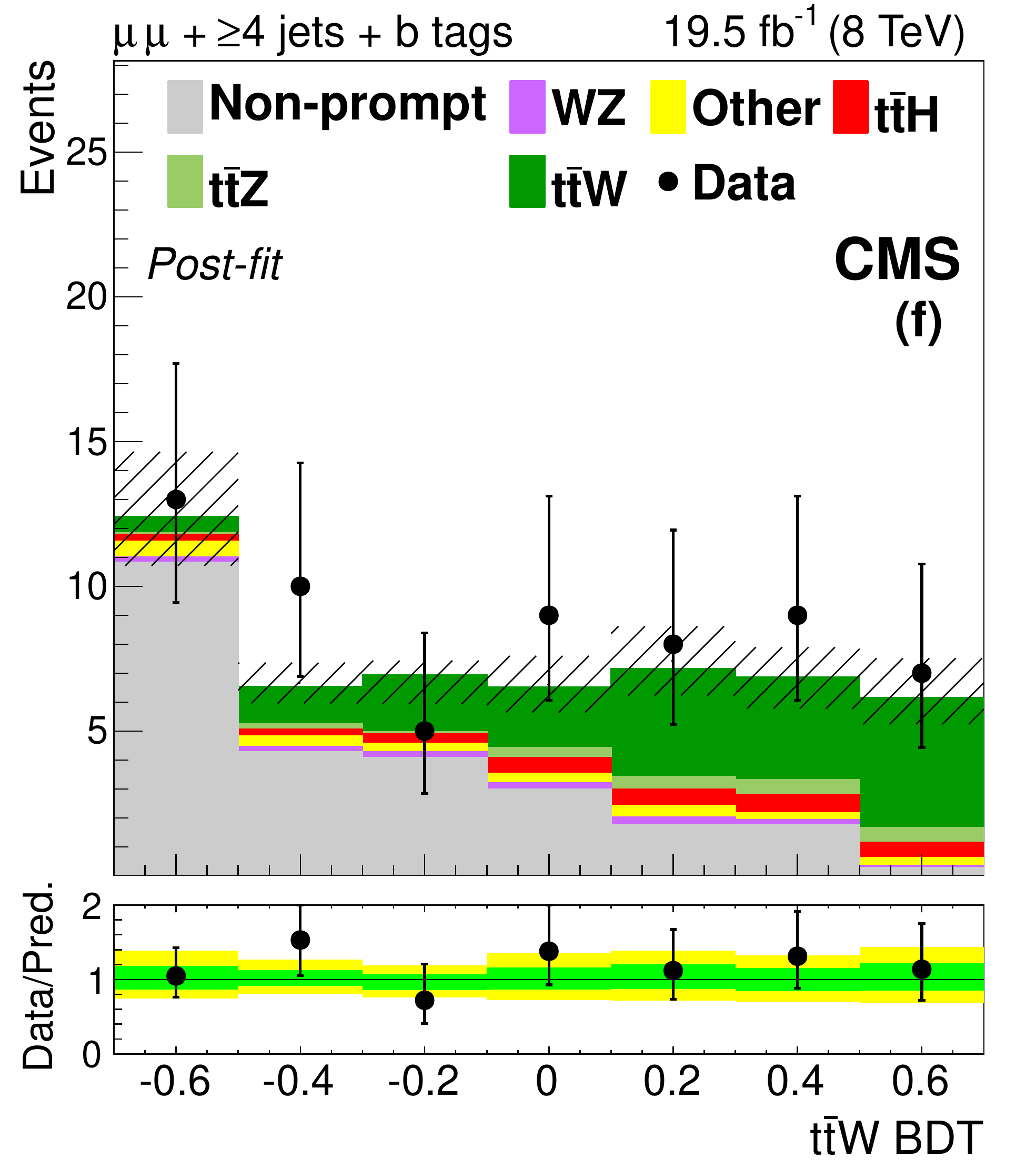}
  \caption{The final discriminant for SS $\ttW$ channel events with 3 jets (top) and
           ${\geq}4$ jets (bottom), after the final fit described in Section~\ref{sec:results_sm}.
           The lepton flavors are $\Pe\Pe$ (a, d), $\Pe\PGm$ (b, e), and $\PGm\PGm$ (c, f).
           The 68\% CL uncertainty in the fitted signal plus
           background is represented by hash marks in the stack histogram,
           and a green shaded region in the data-to-prediction ratio plot.
           The 95\% CL band from the fit is shown in yellow. ``Ch.~misID'' indicates the
           charge-misidentified background. ``Other'' backgrounds include $\ttbar\gamma$,
           $\ttbar\gamma^{*}$, $\ttbar\PW\PW$, $ \PQt \PQb\PZ$, $\PW \PW \PW$, $\PW \PW \PZ$, and $\PW^{\pm}\PW^{\pm}$.}
  \label{fig:BDT_output_ttW_SS}
\end{figure}

\begin{figure}[htb]
  \centering
  \includegraphics[width=0.32\textwidth]{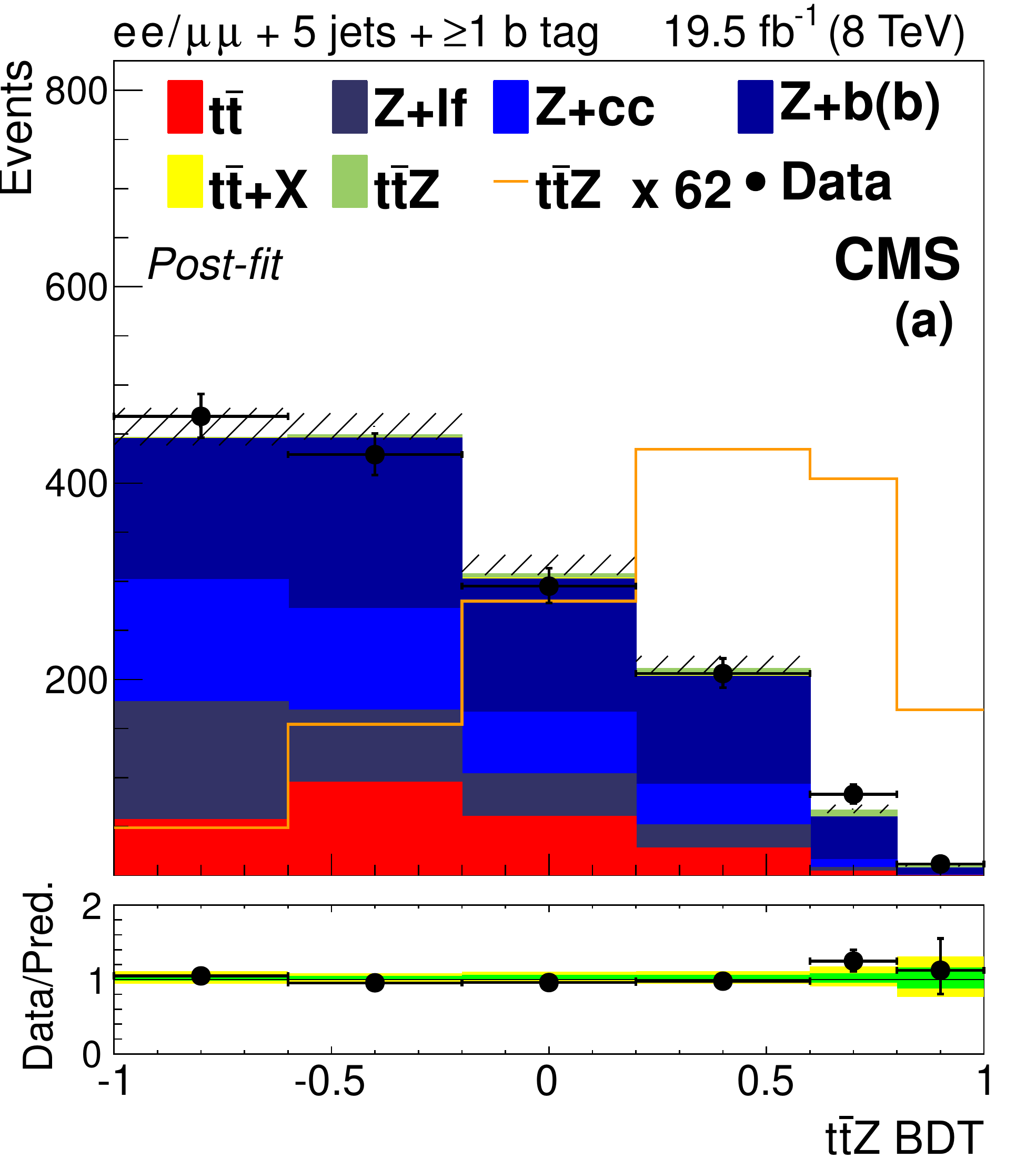}
  \includegraphics[width=0.32\textwidth]{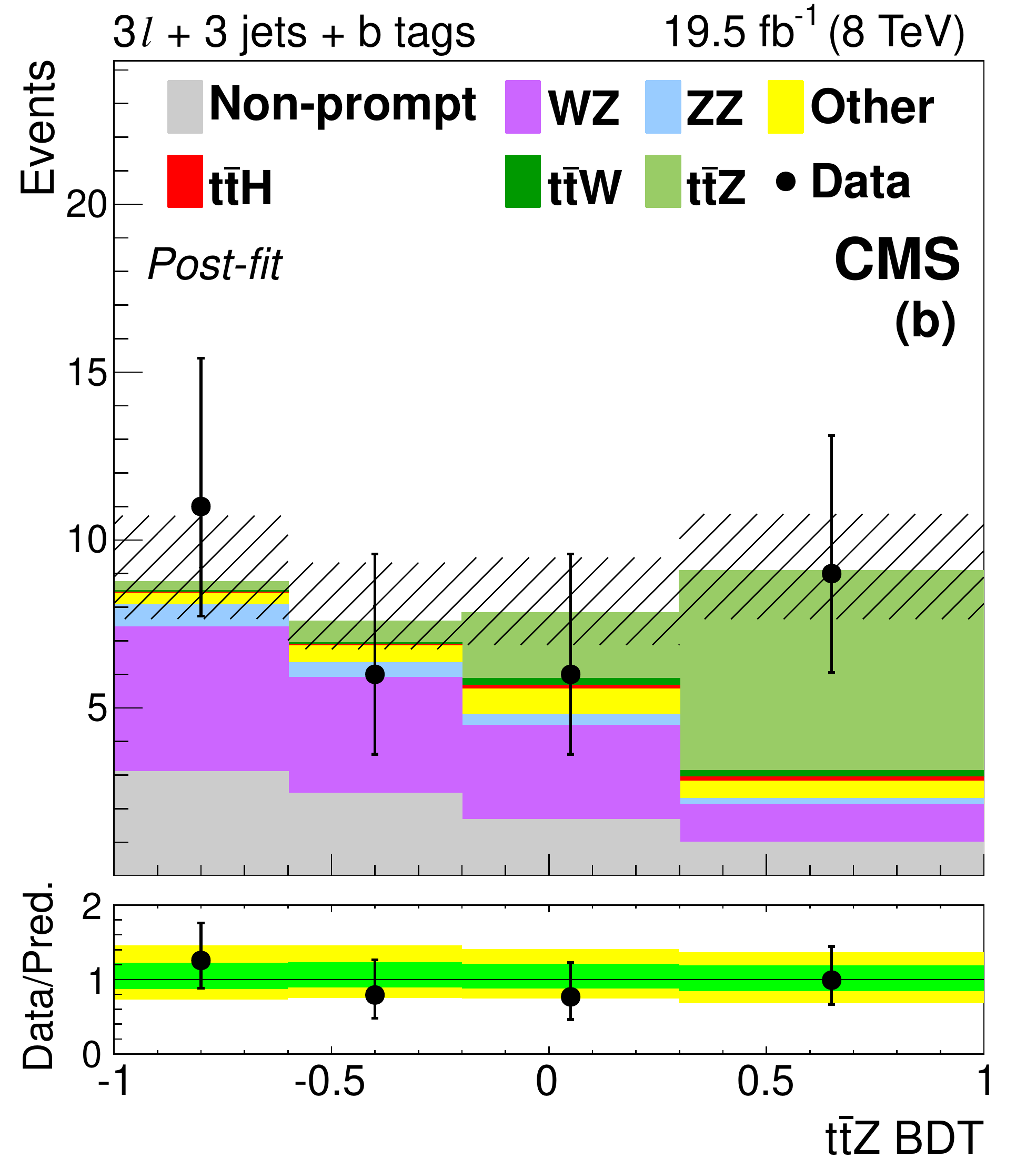}
  \includegraphics[width=0.32\textwidth]{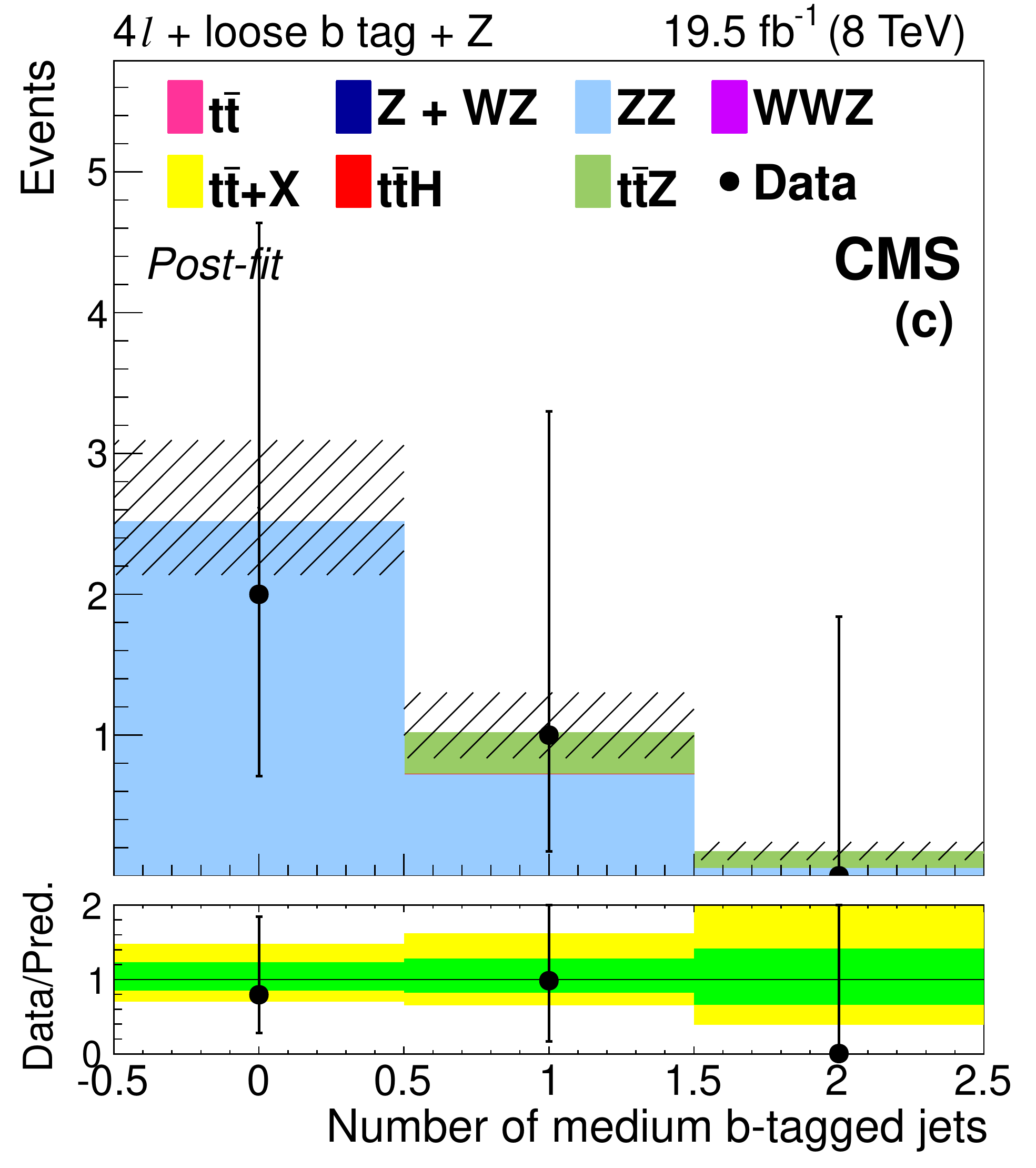}
  \includegraphics[width=0.32\textwidth]{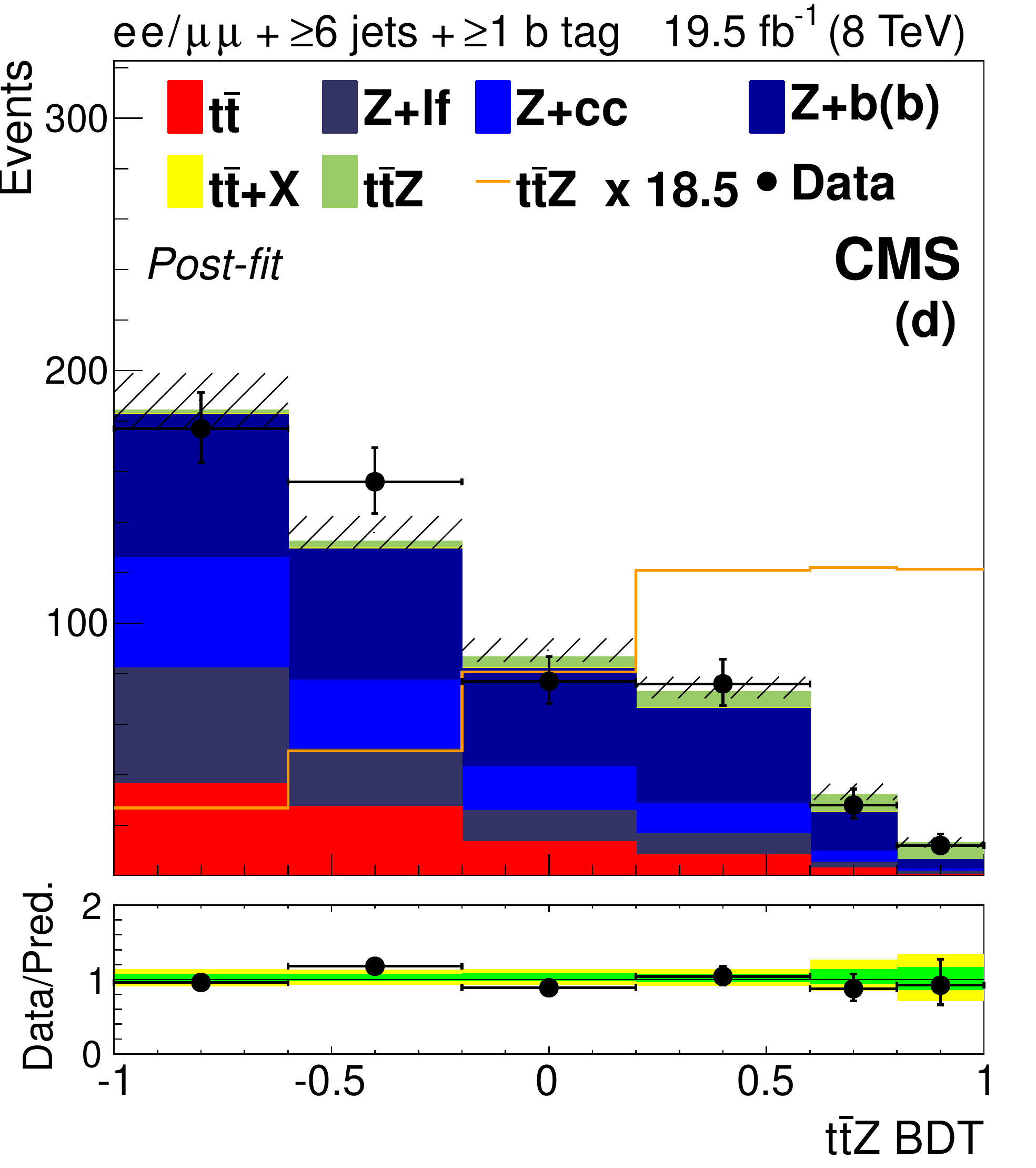}
  \includegraphics[width=0.32\textwidth]{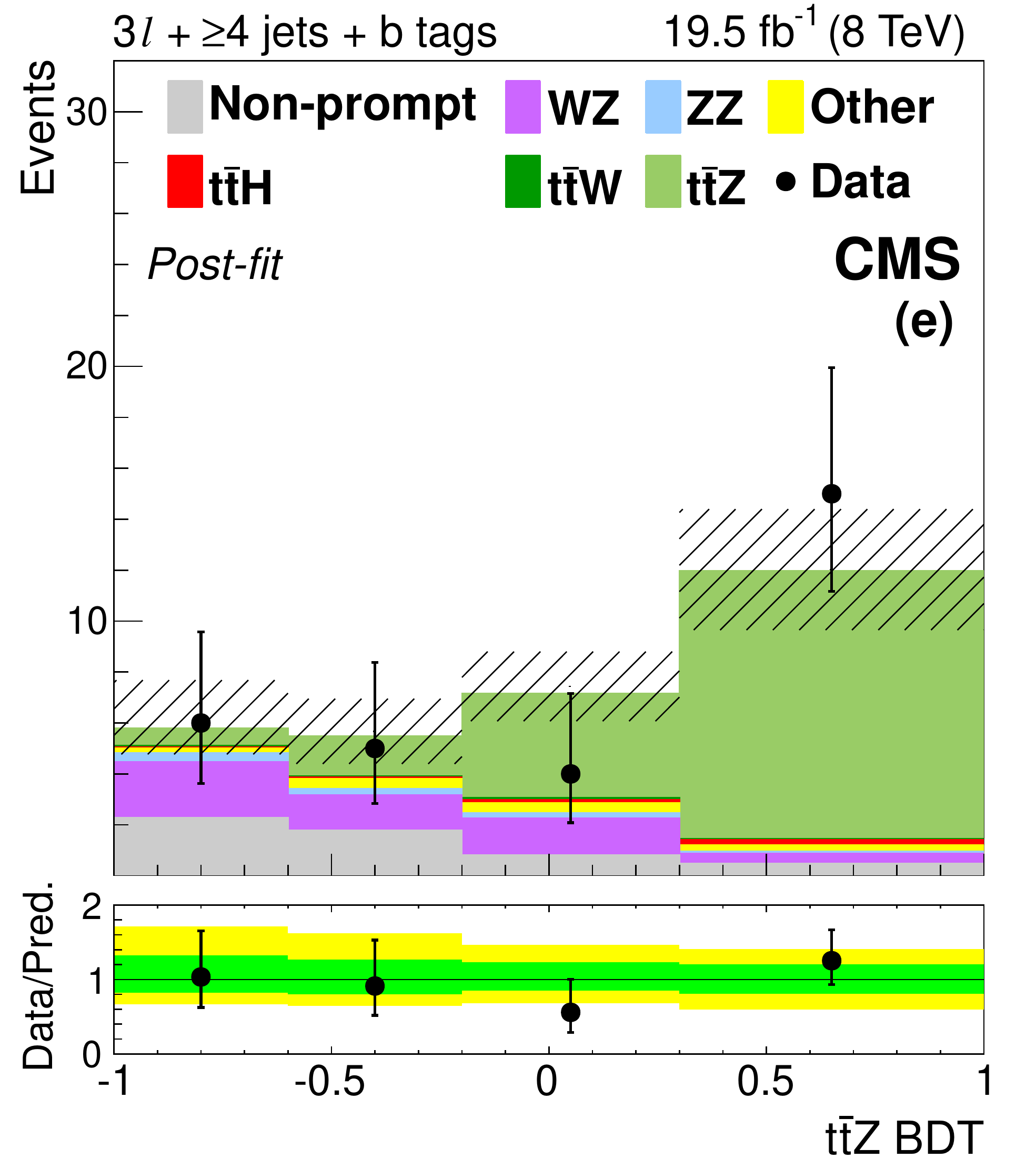}
  \includegraphics[width=0.32\textwidth]{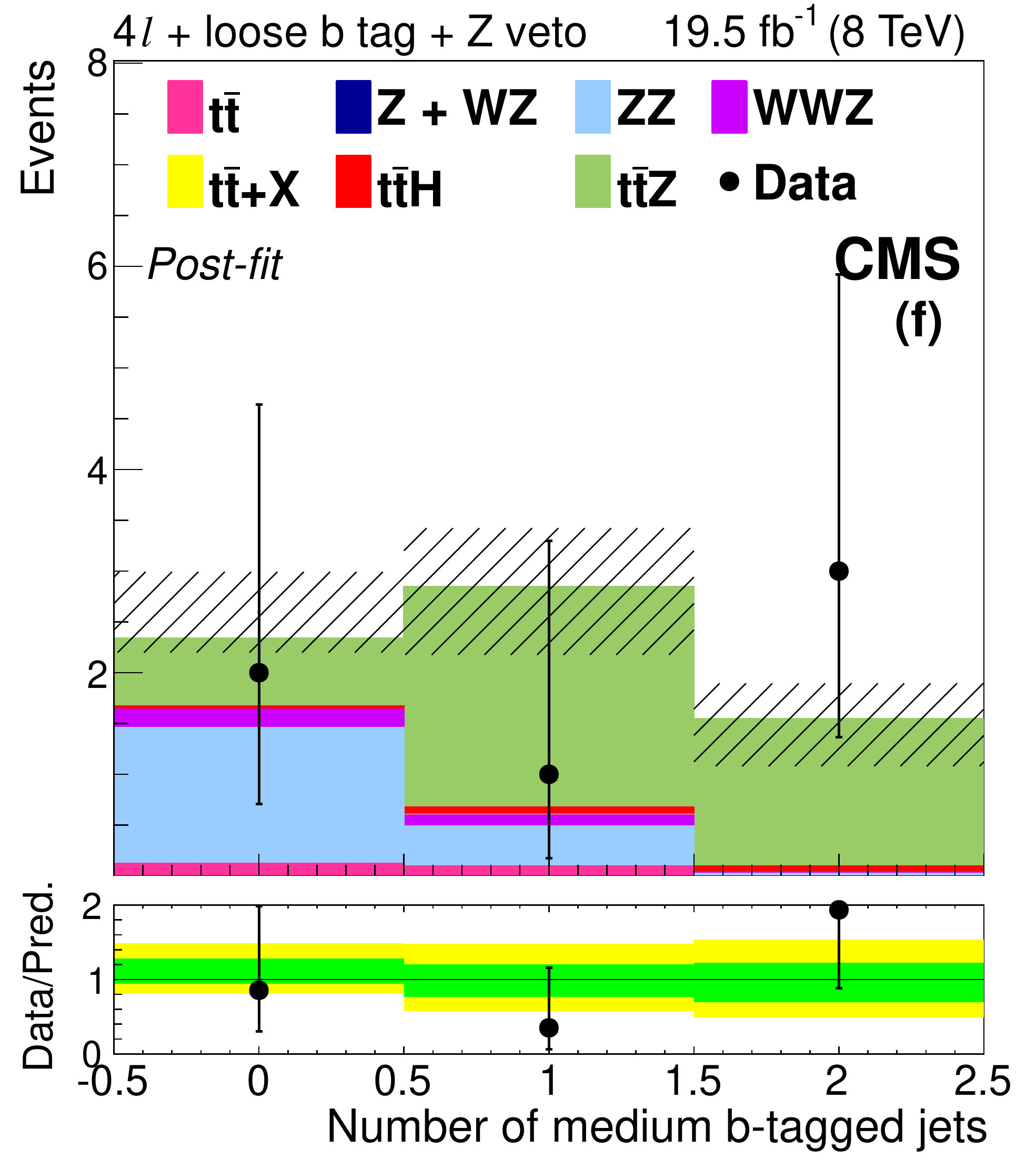}
  \caption{The final discriminant for $\ttZ$ channel events with two OS leptons
           and 5 jets (a) or ${\geq}6$ jets (d), three leptons and 3 jets (b) or
           $\geq$4 jets (e), or four leptons and two lepton pairs (c) or exactly one
           lepton pair (f) consistent with a $\Z  \to \ell\ell$ decay, after the final
           fit described in Section~\ref{sec:results_sm}.
           The 68\% CL uncertainty in the fitted signal plus
           background is represented by hash marks in the stack histogram,
           and a green shaded region in the data-to-prediction ratio plot.
           The 95\% CL band from the fit is shown in yellow.
           The orange line shows the shape of the $\ttZ$ signal, suitably normalized.
		   The $\ttbar$+X background includes $\ttW$, $\ttH$, and $\ttbar\PW\PW$;
		   ``Other'' backgrounds include $\ttbar\gamma$, $\ttbar\gamma^{*}$, $\ttbar\PW\PW$,
		   $ \PQt \PQb\PZ$, $\PW \PW \PW$, and $\PW \PW \PZ$.}
\label{fig:BDT_output_ttZ}
\end{figure}

The expected and observed distributions of the BDT output
for each channel and category are shown in
Figs.~\ref{fig:BDT_output_ttW_SS}--\ref{fig:BDT_output_ttZ}.  The expected
signal and background distributions represent the best fit to the data of the
SM predicted backgrounds and signal, where the signal cross section
is allowed to float freely.  The 68\% CL
uncertainty in the fitted signal plus background is represented
by hash marks in the stack histogram, and a green shaded region in the
data-to-prediction ratio plot.  The 95\% CL band from the fit is shown in yellow.

\begin{figure}[htb]
\centering
  \includegraphics[width=0.32\textwidth]{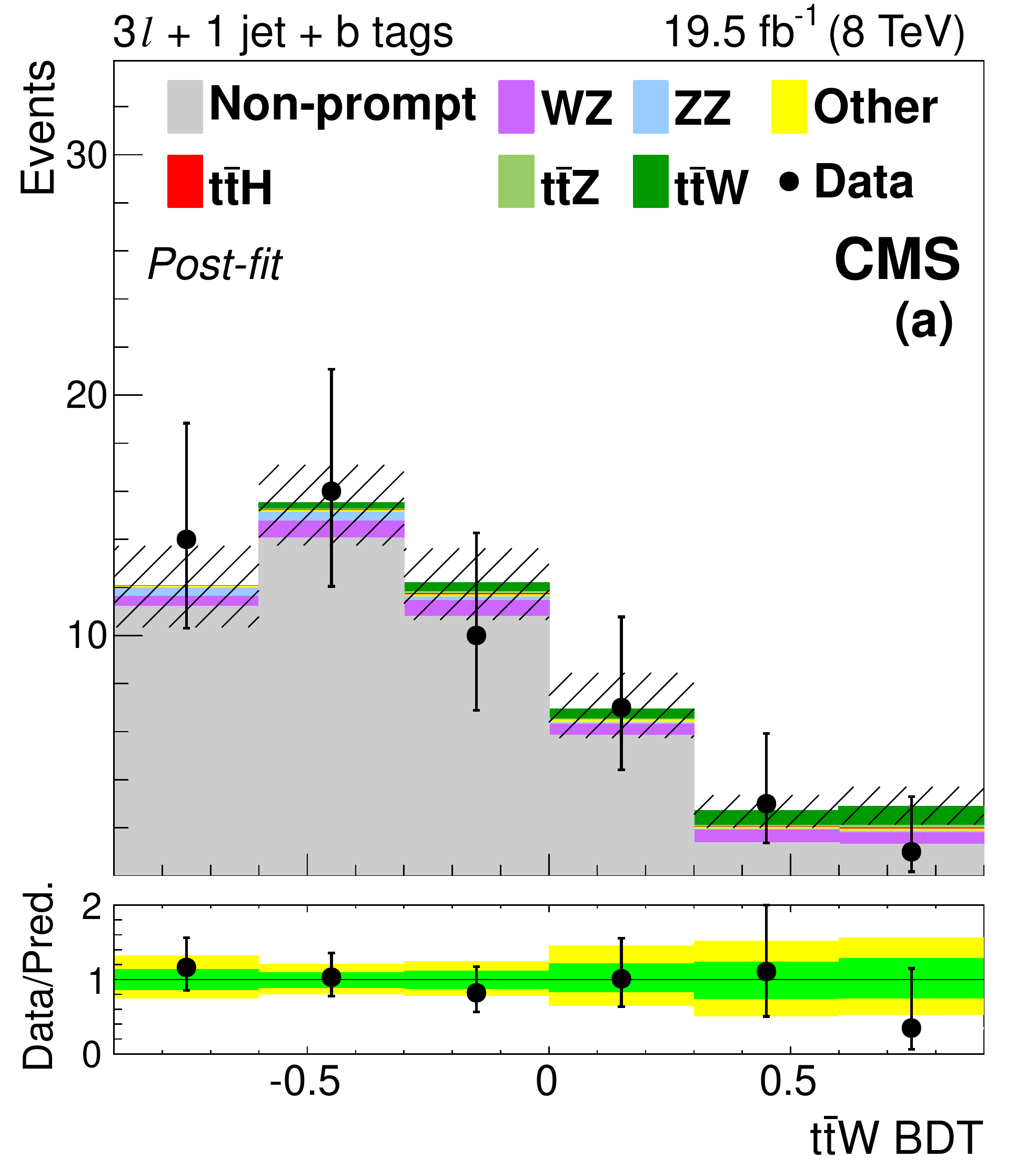}
  \includegraphics[width=0.32\textwidth]{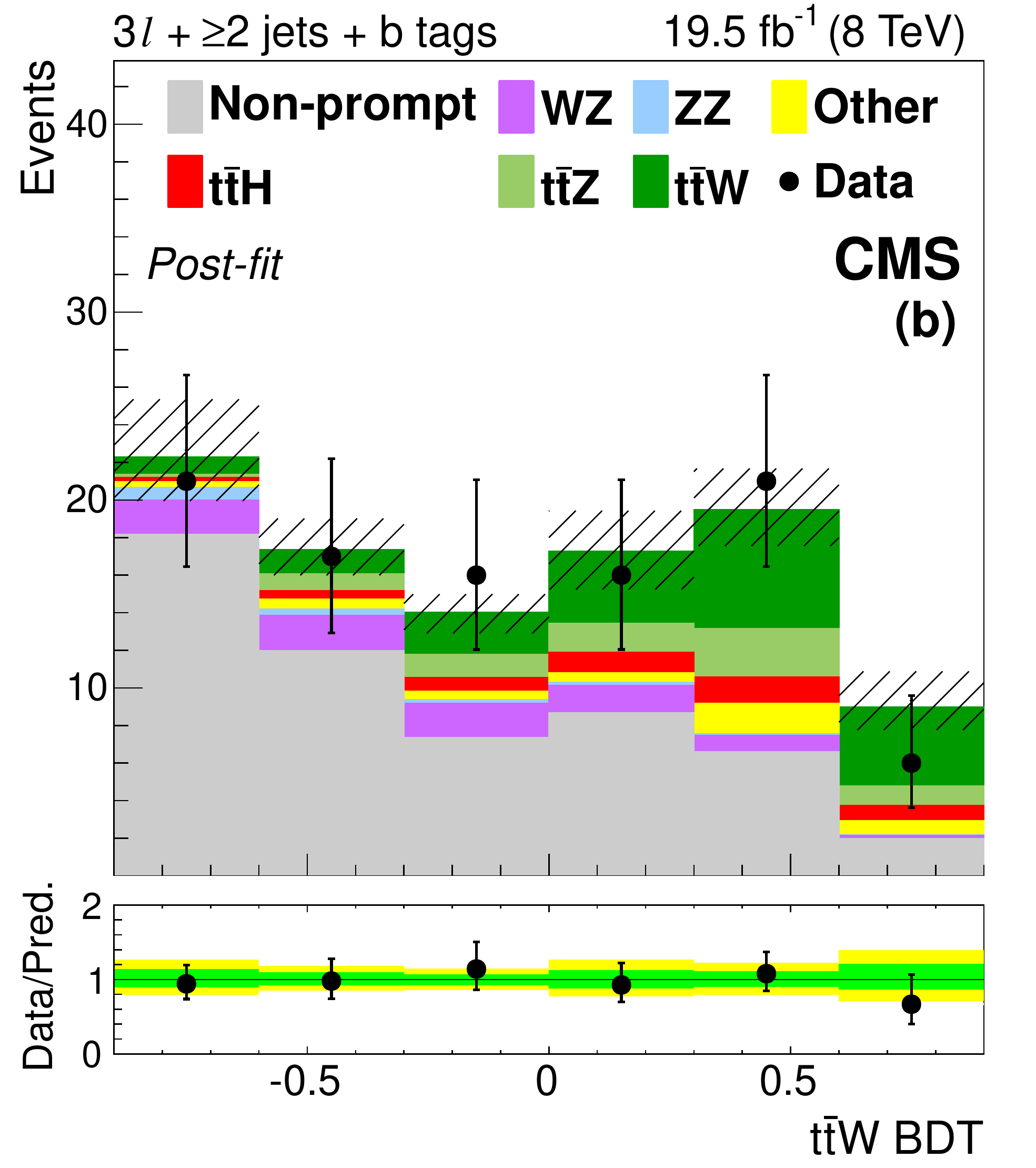}
  \caption{The final discriminant for 3$\ell$ $\ttW$ channel events with 1
           jet (a) and ${\geq}2$ jets (b) after the final fit
           described in Section~\ref{sec:results_sm}.
           The 68\% CL uncertainty in the fitted signal plus
           background is represented by hash marks in the stack histogram,
           and a green shaded region in the data-to-prediction ratio plot.
           The 95\% CL band from the fit is shown in yellow.
		   ``Other'' backgrounds include $\ttbar\gamma$, $\ttbar\gamma^{*}$, $\ttbar\PW\PW$,
		   $\PQb \PQt \PZ$, $\PW \PW \PW$, and $\PW \PW \PZ$.}
  \label{fig:BDT_output_ttW_3l}
\end{figure}

\begin{figure}[htb]
\centering
  \includegraphics[width=0.32\textwidth]{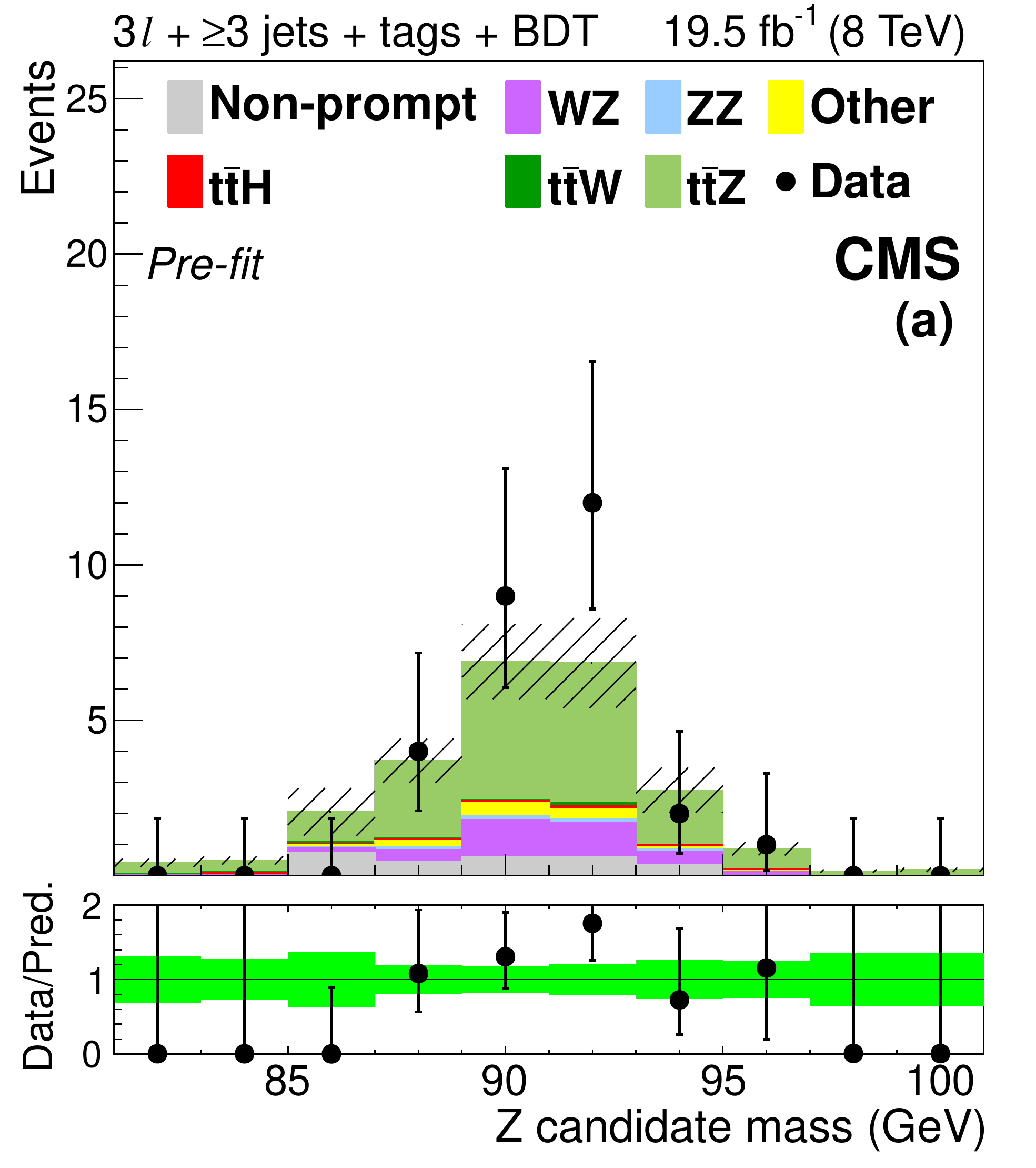}
  \includegraphics[width=0.32\textwidth]{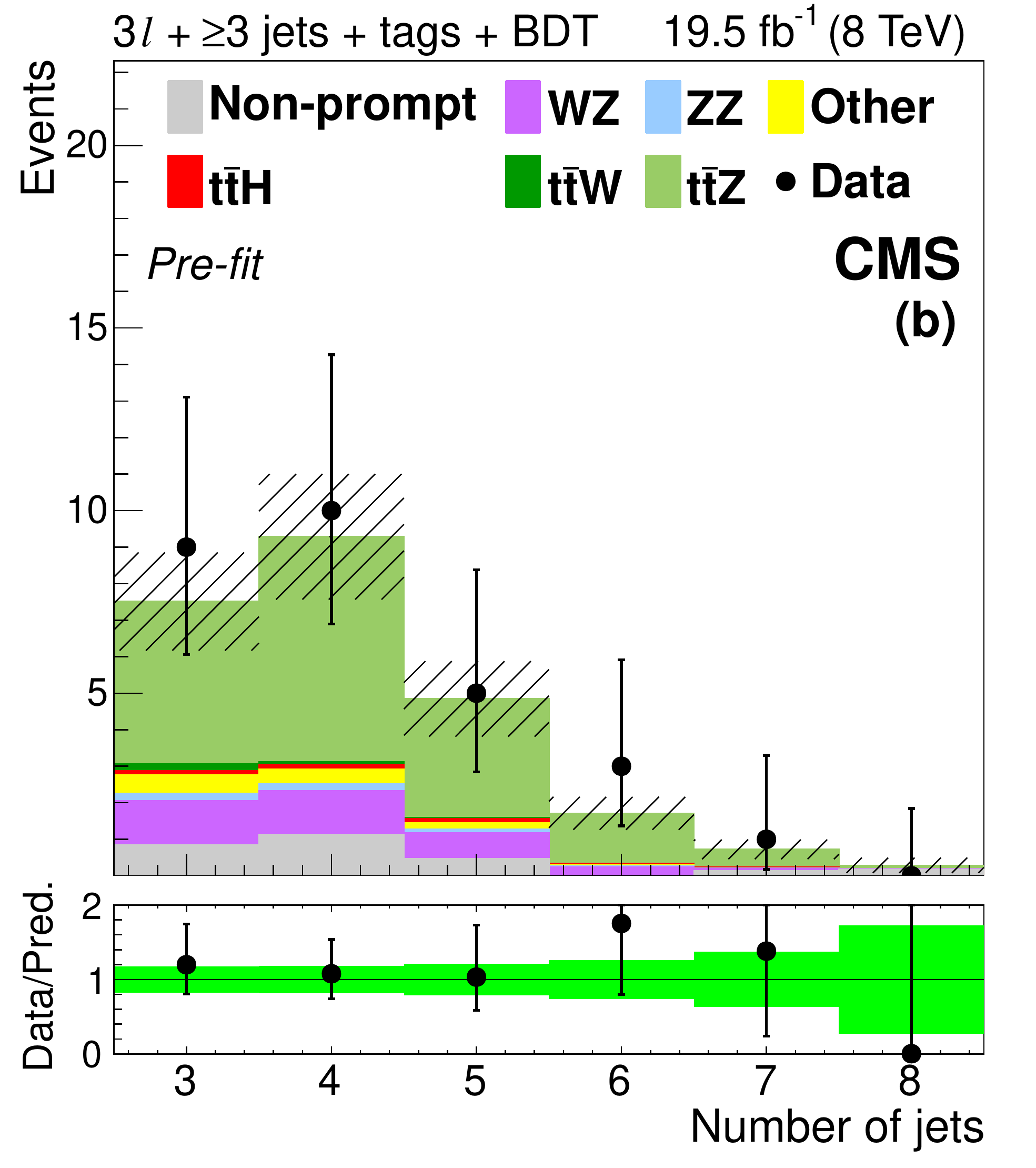}
  \includegraphics[width=0.32\textwidth]{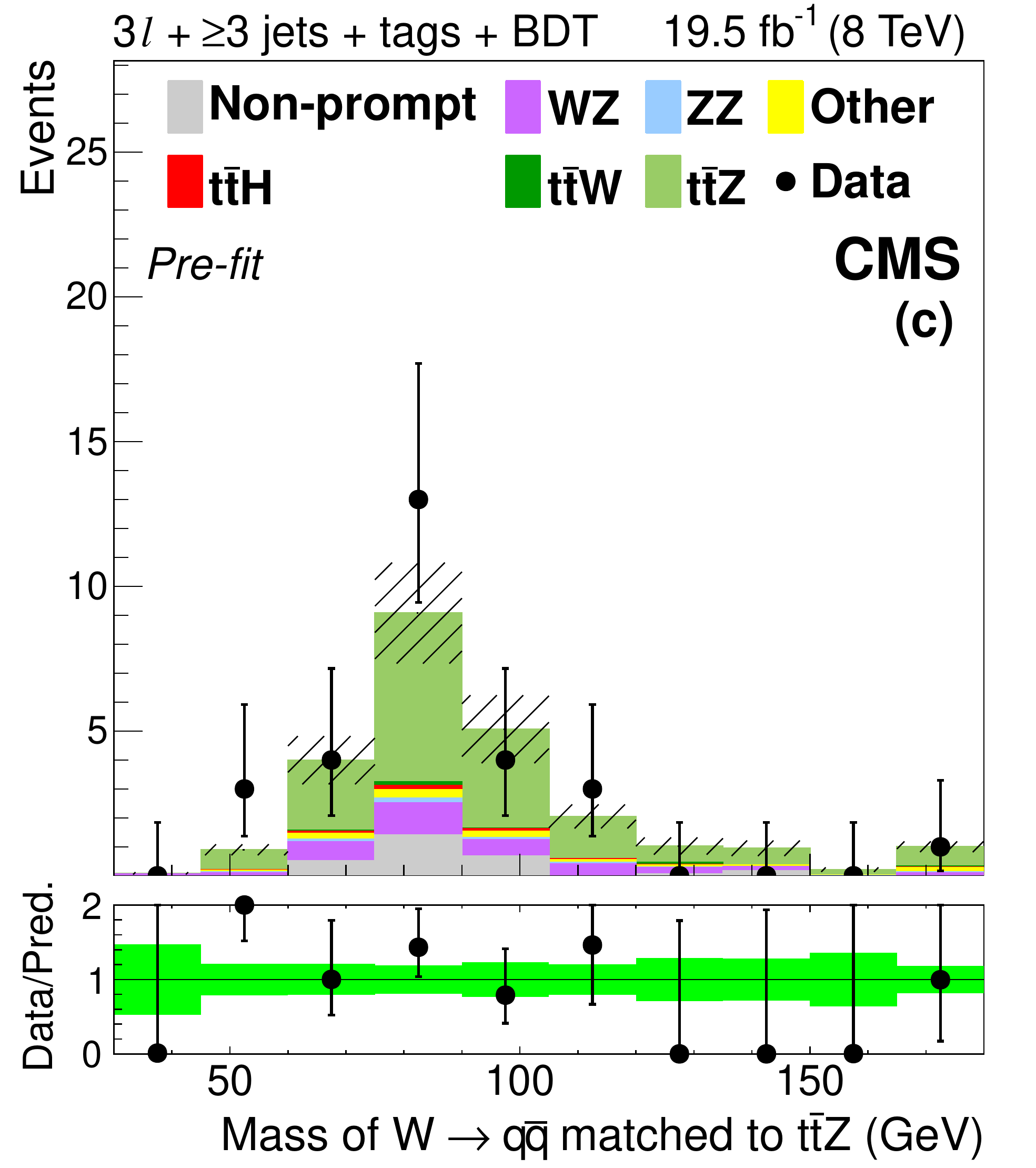}
  \includegraphics[width=0.32\textwidth]{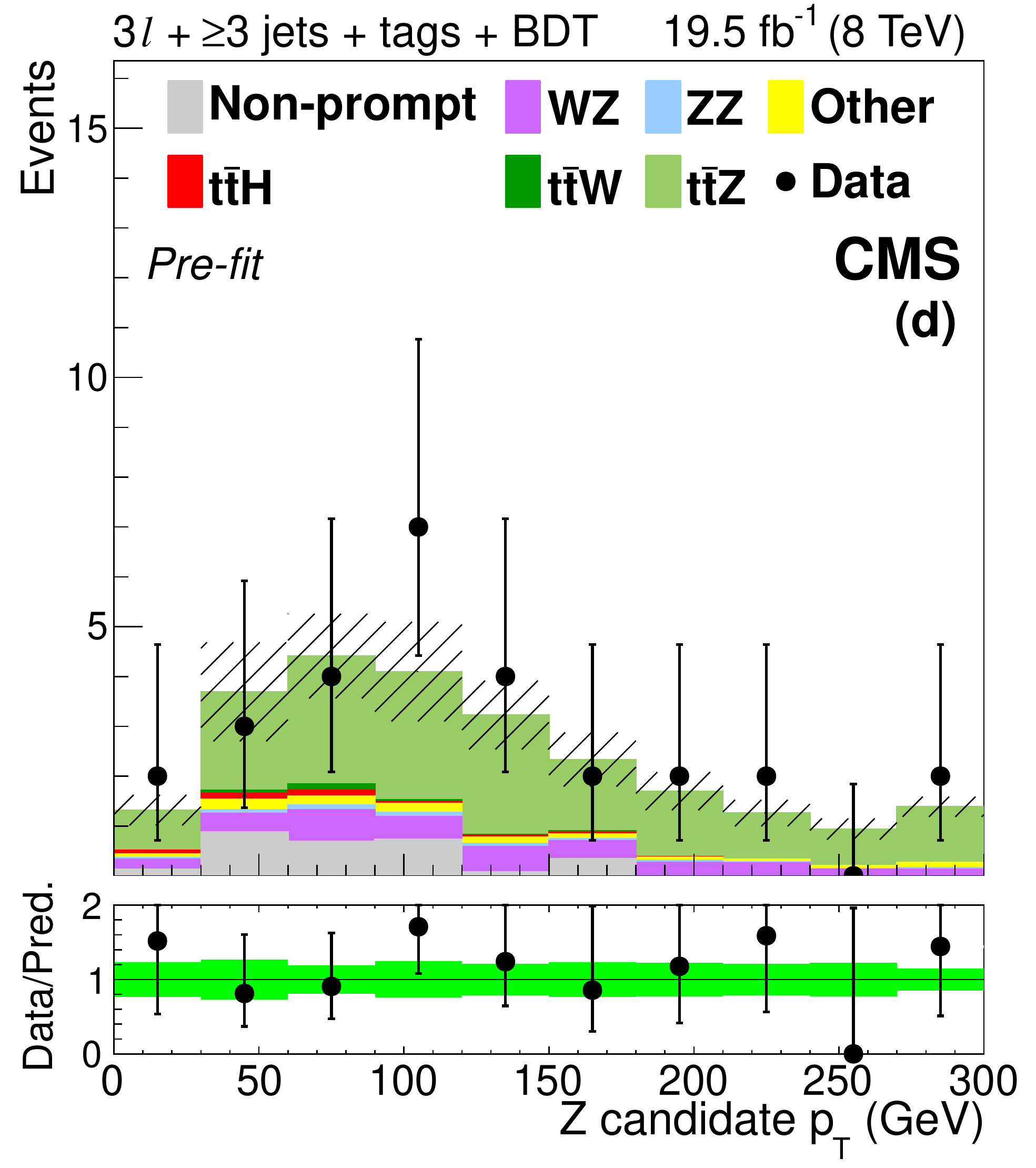}
  \includegraphics[width=0.32\textwidth]{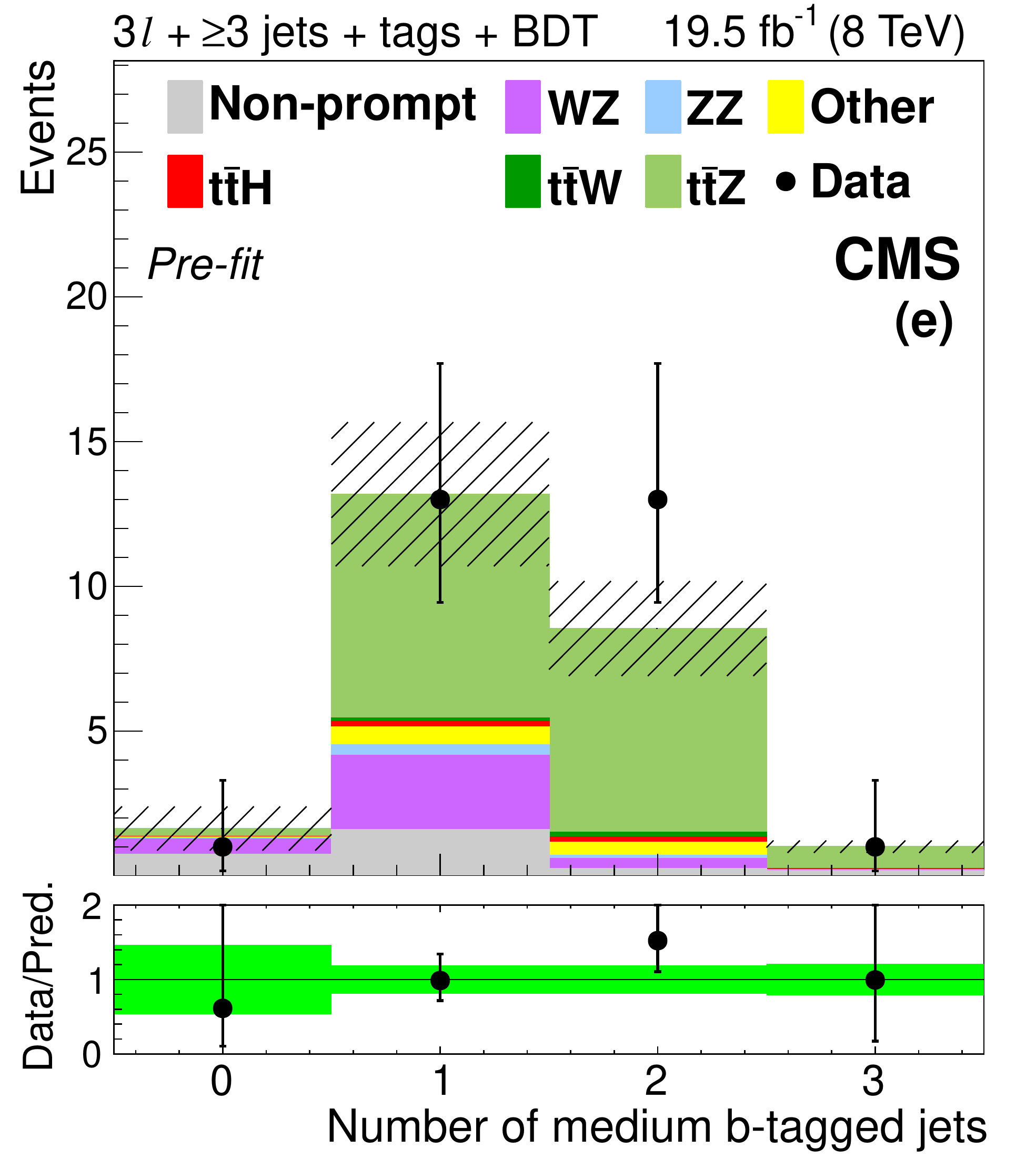}
  \includegraphics[width=0.32\textwidth]{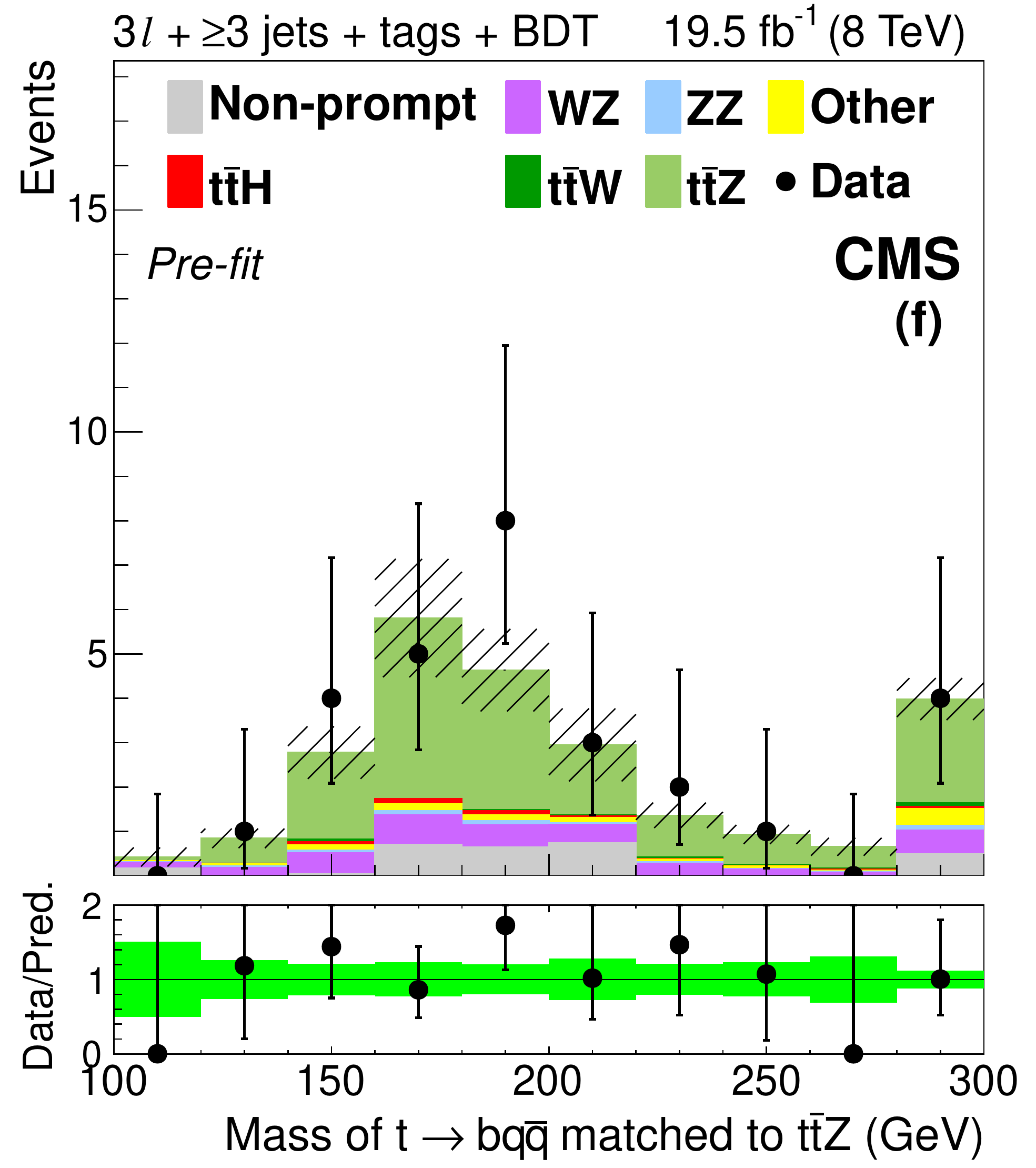}
  \caption{Distributions of the mass (a) and $\pt$ (d) of the lepton pair identified with the $\PZ$ boson decay,
           the number of jets (b) and medium $\PQb$-tagged jets (e), and the mass of the best fit dijet pair from
		   a $\PW$ boson decay (c) and trijet system from a top quark decay (f). The plots show signal-like events from the
		   3$\ell$ $\ttZ$ channel (3 jets with BDT $ > 0.3$ and ${\geq}4$ jets with $\mathrm{BDT} > -0.2$) before the final
		   fit described in Section~\ref{sec:results_sm} is performed. The green band in the data-to-prediction
		   ratio plot denotes the 68\% CL rate and shape uncertainties in the signal plus background
		   prediction. ``Other'' backgrounds include $\ttbar\gamma$, $\ttbar\gamma^{*}$, $\ttbar\PW\PW$, $ \PQt \PQb\PZ$, $\PW \PW \PW$, and $\PW \PW \PZ$.}
  \label{fig:distributions_ttZ_3l}
\end{figure}

Events in the 3$\ell$ $\ttZ$ channel with high BDT values (${>}0.3$
for three jet events, ${>}-0.2$ for events with four or more jets) should
provide a high-purity sample of $\ttZ$ events.  Data distributions of
the reconstructed $\PZ$ boson and top quark properties are consistent
with the SM $\ttZ$ signal, as shown in Fig.~\ref{fig:distributions_ttZ_3l}.

\section{Systematic uncertainties}
\label{sec:systematics}

There are several systematic uncertainties that affect the expected
rates for signal and background processes, the shape of input
variables to the BDTs, or both. The most important uncertainties are on
the \PQb tagging efficiency, signal modeling, and the rates of non-prompt
backgrounds and prompt processes with extra jets.
		
Some uncertainties affect the simulation in
all of the channels, and are correlated across the entire
analysis.  The integrated luminosity has an uncertainty of
2.6\%~\cite{CMS-PAS-LUM-13-001}.
The total inelastic proton-proton cross section is varied up and
down by 5\%, which affects the number of pileup vertices, and
is propagated to the output distributions~\cite{tagkey20135}.

The properties and reconstruction efficiencies of different objects
have their own uncertainties.  The uncertainty in the jet energy
scale~\cite{cmsJEC} is accounted for by shifting the energy
scale up and down by one standard deviation for all simulated
processes, and evaluating the output distributions with the
shifted energy scale.  The shape of the CSV distribution for
light flavor or gluon jets, \PQc jets, and \PQb jets has
uncertainties associated with the method used to match the CSV
shapes in data and simulation, as detailed in Ref.~\cite{CMS_ttH_8TeV}.
Calibration regions for light flavor jets have some contamination
from heavy flavor jets, and vice versa.
The associated uncertainty in the final light or heavy flavor
CSV shape is accounted for by varying the expected yields of
contaminating jets up and down by one standard deviation, and
propagating the result to the final CSV distribution.  The weights
for these alternate shapes are applied to produce alternate
final discriminant histograms in each channel. Likewise there
are uncertainties from the limited number of events in the calibration
regions; these are assessed using the maximum linear and quadratic
deformations of the CSV shape within an envelope whose size
is determined by the magnitude of the statistical uncertainty.
Because there is no calibration region to determine the CSV shape
of \PQc jets in data, they receive no correction factors, but
have all the \PQb jet uncertainties applied to them, multiplied
by a factor of two so that they include the scale factor values for
b jets.

Prompt electron and muon efficiency uncertainties are computed using
high-purity dilepton samples in data from $\PZ$ boson decays. These include
rate uncertainties associated with the trigger efficiency,
reconstruction efficiency, and the fraction of prompt leptons
passing the tight, loose, and charge ID selection criteria.

The rate of non-prompt leptons passing the tight lepton selection
receives a 40\% uncertainty for electrons and a 60\% uncertainty
for muons, based on the agreement between expected and observed
yields in control regions in data, as described in Section~\ref{sec:backgrounds}.
Additional uncertainties of 50\% and 100\% are assessed on the rates of non-prompt
leptons with medium and high $\pt$, respectively, because of the limited number of
events in the sample used to find the misidentification rates. These uncertainties
are applied separately for electrons and muons, and are uncorrelated
between the SS and 3$\ell$ channels, to account for possible
differences in the sources of non-prompt leptons. While the uncertainties
on event yields with non-prompt electrons and muons are initially large, the
final fit constrains them to 10--15\% using bins in the final discriminants
which contain mostly non-prompt backgrounds.

The rates of charge misidentified electrons in the SS
channels receive a 30\% rate uncertainty, based on the agreement
between predicted and observed SS dielectron events consistent
with a $\PZ$ boson decay.

Theoretical uncertainties from the PDFs of different simulated
processes, as well as the choice of renormalization and factorization
scales, are accounted for with rate uncertainties in all signal
and backgrounds processes.  The rate uncertainties for $\ttW$ and
$\ttZ$ are 10\% and 11\%, respectively, from the choice of
scales~\cite{Garzelli:2012bn}, and 7.2\% and 8.2\% from the PDFs~\cite{Alekhin:2011sk, Botje:2011sn}.
In addition, shape uncertainties derived from simulation generated with
different PDF sets and \PYTHIA tunes are applied to the $\ttW$, $\ttZ$,
and $\ttH$ processes using linear and quadratic deformations of
10--11\% on the final discriminant shape.  The $\ttW$ and $\ttZ$
rate uncertainties are not included in the $\ttW$ and $\ttZ$ cross
section measurements, respectively, and neither is used in the
simultaneous measurement of the $\ttW$ and $\ttZ$ cross sections.

The systematic uncertainty in top quark $\pt$ reweighting in simulated
$\ttbar$ events is assessed by applying no top quark $\pt$ weight
for the lower systematic uncertainty, and twice the weight for the upper
systematic uncertainty. Since neither higher-order theoretical
calculations~\cite{Bredenstein:2010rs} nor independent control region
studies currently constrain the normalization of the
$\mathrm{t}\bar{\mathrm{t}}$+\ccbar, $\mathrm{t}\bar{\mathrm{t}}$+b, or
$\mathrm{t}\bar{\mathrm{t}}$+\bbbar processes to better than 50\% accuracy,
an extra 50\% uncorrelated rate uncertainty is assigned to each process.
An additional shape uncertainty is applied to the ratio of the \MT to
the invariant mass of the system of jets in $\ttbar$ events with five,
or six or more jets.

Because the $\PZ$ boson, $\PW \PZ$, and $\PZ \PZ$ simulations are used to model events
with more jets than there are extra partons in the generated event,
rate uncertainties are assigned to these processes.
Events with a $\PZ$ boson plus five jets and six or more jets receive uncorrelated 30\% rate
uncertainties, based on the extrapolation from $\PZ$ boson events with four jets
and no medium $\PQb$-tagged jets to those with at least one medium \PQb tag.
Diboson $\PW \PZ$ and $\PZ \PZ$ events with three jets and four or more jets have uncorrelated
40\% and 60\% uncertainties, respectively, due to the limited number of events in the
light flavor sideband used to calibrate jet multiplicity.  Diboson events with
extra heavy flavor jets receive uncertainties identical to the $\ttbar$ plus
heavy flavor simulation.  The good data-to-simulation agreement in dileptonic $\PZ$ boson events with four
jets and one or two medium \PQb tags constrains the $\PZ+\ccbar$, $\PZ+\PQb$, and $\PZ+\bbbar$
uncertainties to 30\% each. Simulated $\PZ$ boson events have extra shape uncertainties
in \MHT~and the \MT-to-mass ratio of jets, uncorrelated between events with
five and six or more jets, and between the different jet flavor subsamples.
These account for possible data-to-simulation differences seen in $\PZ$ boson events with four
or more jets (excluding the $\ttZ$ signal region). Although these uncertainties
are large, the $\PZ$ boson and diboson backgrounds are well separated from the
$\ttZ$ signal using the final discriminants, so they have a small effect on
the final measurement.

Rare processes with low expected yields such as triboson production ($\PW \PW \PW$, $\PW \PW \PZ$),
associated production of a $\PZ$ boson with a single top quark (tbZ),
and $\ttbar$ with an on-shell or off-shell photon ($\ttbar\gamma$/$\ttbar\gamma^{*}$)
or two $\PW$ bosons ($\ttbar\PW\PW$) get 50\% rate uncertainties, because
they are either calculated at leading order or require extra jets or
b jets to enter the signal region.

The expected impact of different sources of systematic uncertainty is estimated by
removing groups of uncertainties one at a time and gauging the improvement in the
signal strength precision, as measured using pseudo-data from simulation. (The
measurement technique is described in the next section.)
If we expect to measure a signal strength of $1 \pm \delta_{i}$ with all the systematic
uncertainties included, and expect to measure $1 \pm \delta_{i \neq j}$ with fewer uncertainties,
a large reduction in uncertainty $\epsilon_{j} = \delta_{i} - \delta_{i \neq j}$ indicates that the
removed uncertainties have a significant impact on the measurement.
Uncertainties in \PQb tagging efficiency, signal modeling, and rates of prompt processes
with extra jets are found to have the greatest effect on the $\ttZ$ signal precision,
while the $\ttW$ measurement is most impacted by uncertainties in the non-prompt
backgrounds, \PQb tagging efficiency, and signal modeling.  The full set of systematic
uncertainties and their expected effects are shown in Table~\ref{tab:syst_effect}.
Because we are measuring $\epsilon_{j}$ and not $\delta_{j}$, we do not expect the
quantities in Table~\ref{tab:syst_effect} to add in quadrature.

\begin{table}[htb]\small
\renewcommand{\arraystretch}{1.2}
\centering
\topcaption{Reduction in the expected signal strength uncertainties produced by removing sets of systematic
         uncertainties. The quantities in each column are not expected to add in quadrature.}
\begin{tabular}{lrr}
\hline
 Systematic uncertainties removed  & \multicolumn{1}{c}{$\ttW$} & \multicolumn{1}{c}{$\ttZ$} \\
\hline
 Signal modeling & 5.2\% & 7.1\% \\
 Non-prompt backgrounds & 12.5\% & 0.5\% \\
 Inclusive prompt backgrounds & 0.7\% & 2.6\% \\
 Prompt backgrounds with extra jets & 0.2\% & 3.4\% \\
 Prompt backgrounds with extra heavy flavor jets & ${<}0.1$\% & 1.1\% \\
 \PQb tagging efficiency & 6.1\% & 7.3\% \\
 Jet energy scale & 1.4\% & ${<}0.1$\% \\
 Lepton ID and trigger efficiency & 0.3\% & 0.5\% \\
 Integrated luminosity and pileup & 0.7\% & 0.5\% \\
 Bin-by-bin statistical uncertainty in the prediction & 4.4\% & 1.2\% \\
\hline
 All systematic uncertainties removed & 31\% & 29\% \\
\hline
\end{tabular}
\label{tab:syst_effect}
\end{table}
		
\section{Cross section measurement}
\label{sec:results_sm}

The statistical procedure used to compute the $\ttW$ and $\ttZ$ cross sections and their
corresponding significances is the same as the one used for the LHC Higgs boson analyses,
and is described in detail in Refs.~\cite{CMS-NOTE-2011-005,Collaboration2012}.
A binned likelihood function $L(\mu, \theta)$
is constructed, which is the product of Poisson probabilities for all bins in the final discriminants
of every channel. The signal strength parameter $\mu$ characterizes the amount of signal, with $\mu=1$
corresponding to the SM signal hypothesis, and $\mu=0$ corresponding to the background-only hypothesis.
Systematic uncertainties in the signal and background predictions are represented by a set of nuisance
parameters, denoted $\theta$.  Each nuisance parameter represents a different source of uncertainty.
When multiple channels have the same source of uncertainty, the nuisance parameter is correlated
across the channels, allowing certain initially large systematic uncertainties (such as the rate of
non-prompt leptons passing the tight selection) to be constrained in bins with a large number of
data events but few expected signal events.

To test how consistent the data are with a hypothesized value of $\mu$, we consider the
profile likelihood ratio test statistic
$q(\mu) = -2 \ln L(\mu, \hat{\theta}_{\mu})/L(\hat{\mu}, \hat{\theta})$,
where $\hat{\theta}_{\mu}$ denotes the set of values of the nuisance parameters $\theta$ that
maximizes the likelihood $L$ for the given $\mu$. The denominator
is the likelihood maximized over all $\mu$ and $\theta$. This test statistic is
integrated using asymptotic formulae~\cite{Asymptotic} to obtain the $p$-value, \ie the probability
under the signal-plus-background hypothesis of finding data of equal or greater incompatibility
with the background-only hypothesis.
Results are reported both in terms of the best fit cross section and $\mu$ values and their
associated uncertainties, and in terms of the significance of observation of the
two signal processes.

We perform separate one-dimensional fits for the $\ttW$ and $\ttZ$ cross sections using the relevant
channels for each process. The fit for each cross section is performed with the other cross section set
to the SM value with the uncertainty coming from theory calculations. The resulting measurements
and significances are reported in Tables~\ref{tab:results_ttW} and~\ref{tab:results_ttZ}.  The $\ttZ$ cross section
is measured with a precision of 25\%, and agrees well with the SM prediction. The observed $\ttW$ cross section
is higher than expected, driven by an excess of signal-like SS dimuon events in the data.
Most of the signal-like dimuon events with four or more jets also contributed to a similar
excess seen in the CMS $\ttH$ search~\cite{CMS_ttH_8TeV}. In both analyses, a close examination
yielded no evidence of mismodeling or underestimated backgrounds, and the excess events appear
consistent with a $\ttW$ or $\ttH$ signal.
The best fit values for the $\ttW$ and $\ttZ$ cross sections are compatible with the SM
expectation at the 13\% and 60\% CL, respectively. Taking into account significant
differences in event selection, this result is also consistent with the previous CMS
measurement~\cite{CMS-TOP-12-036}, which it supersedes.

\begin{figure}[!htbp]
\centering
  \includegraphics[width=0.80\textwidth]{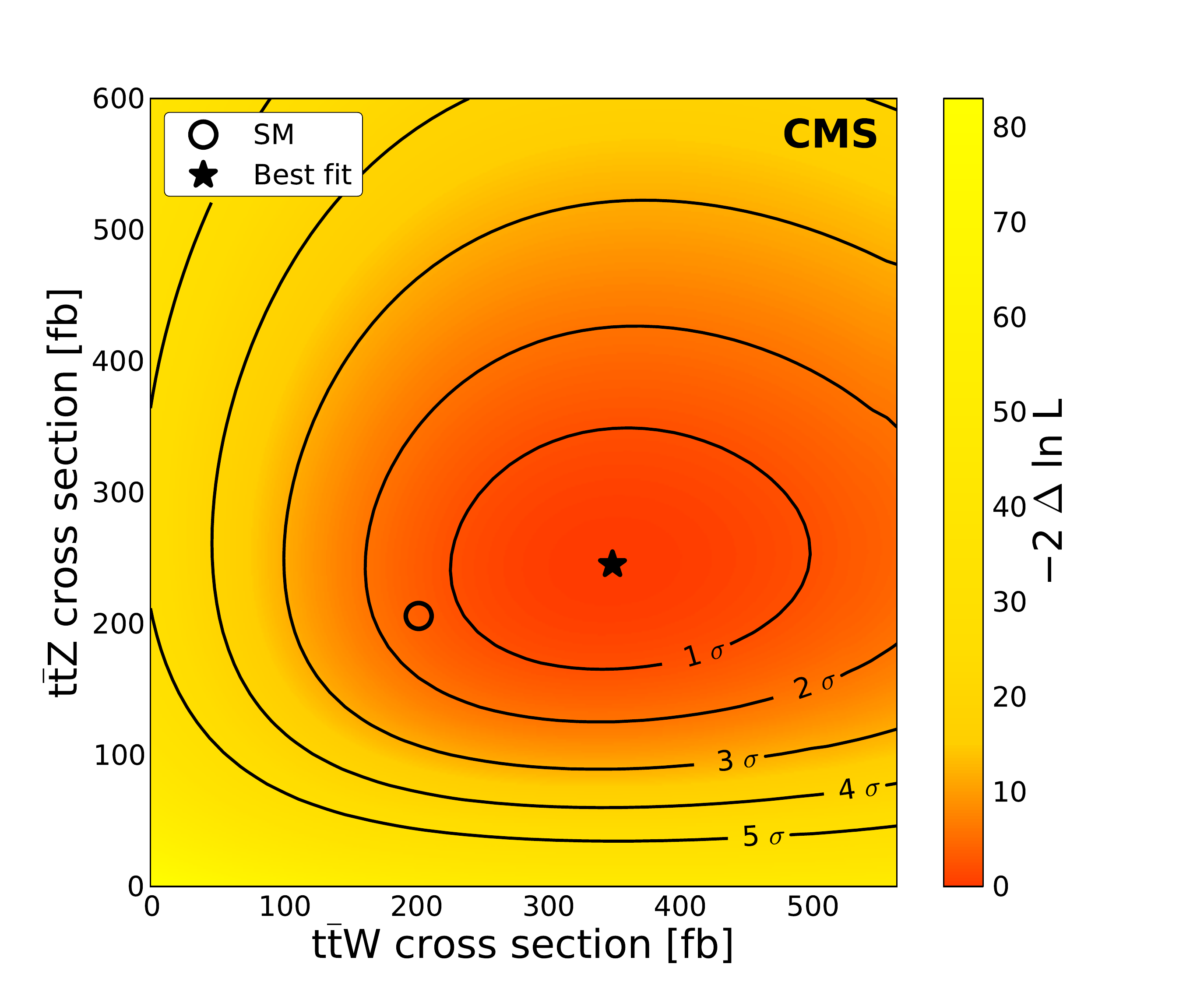}
\caption{Profile likelihood as a function of $\sigma(\ttW)$ and $\sigma(\ttZ)$.
		 Lines denote the 1 to 5 standard deviation ($\sigma$) CL contours.}
\label{fig:2dttZttW}
\end{figure}

We also perform a simultaneous fit of both processes using all the channels.
Figure~\ref{fig:2dttZttW} shows the two-dimensional likelihood scan over
$\sigma(\ttW)$ and $\sigma(\ttZ)$. The respective best fit values are found to be
$350^{+150}_{-123}\unit{fb}$ and $245^{+104}_{-80}\unit{fb}$,
compatible at the 15\% CL with the SM expectation~\cite{Garzelli:2012bn,Campbell:2012dh}.

\begin{table}[htb]\small
\renewcommand{\arraystretch}{1.5}
\centering
\topcaption{Expected and observed measurements of the cross section and signal strength with 68\% CL ranges and significances for $\ttW$, in
         SS dilepton and 3$\ell$ channels.}
\begin{tabular}{c|cc|cc|cc}
\hline
\multicolumn{1}{c|}{$\ttW$} &
\multicolumn{2}{c|}{Cross section (fb)} &
\multicolumn{2}{c|}{Signal strength ($\mu$)} &
\multicolumn{2}{c}{Significance ($\sigma$)} \\
\hline
Channels & Expected & Observed & Expected & Observed & Expected & Observed\\
\hline
SS           & $203^{+88}_{-73}$  & $414^{+135}_{-112}$ & $1.00^{+0.45}_{-0.36}$ & $2.04^{+0.74}_{-0.61}$ & 3.4 & 4.9 \\
3$\ell$      & $203^{+215}_{-194}$ & $210^{+225}_{-203}$ & $1.00^{+1.09}_{-0.96}$ & $1.03^{+1.07}_{-0.99}$ & 1.0 & 1.0 \\
SS + 3$\ell$ & $203^{+84}_{-71}$   & $382^{+117}_{-102}$ & $1.00^{+0.43}_{-0.35}$ & $1.88^{+0.66}_{-0.56}$ & 3.5 & 4.8 \\
\hline
\end{tabular}
\label{tab:results_ttW}
\end{table}

\begin{table}[htb]\small
\renewcommand{\arraystretch}{1.5}
\centering
\topcaption{Expected and observed measurements of the cross section and signal strength with 68\% CL ranges and significances for $\ttZ$,
         in OS dilepton, 3$\ell$, and 4$\ell$ channels.}
\begin{tabular}{c|cc|cc|cc}
\hline
\multicolumn{1}{c|}{$\ttZ$} &
\multicolumn{2}{c|}{Cross section (fb)} &
\multicolumn{2}{c|}{Signal strength ($\mu$)} &
\multicolumn{2}{c}{Significance ($\sigma$)} \\
\hline
Channels & Expected & Observed & Expected & Observed & Expected & Observed\\
\hline
OS                     & $206^{+142}_{-118}$ & $257^{+158}_{-129}$ & $1.00^{+0.72}_{-0.57}$ & $1.25^{+0.76}_{-0.62}$ & 1.8 & 2.1 \\
3$\ell$                & $206^{+79}_{-63}$   & $257^{+85}_{-67}$ & $1.00^{+0.42}_{-0.32}$ & $1.25^{+0.45}_{-0.36}$ & 4.6 & 5.1 \\
4$\ell$                & $206^{+153}_{-109}$ & $228^{+150}_{-107}$ & $1.00^{+0.77}_{-0.53}$ & $1.11^{+0.76}_{-0.52}$ & 2.7 & 3.4 \\
OS + 3$\ell$ + 4$\ell$ & $206^{+62}_{-52}$   & $242^{+65}_{-55}$ & $1.00^{+0.34}_{-0.27}$ & $1.18^{+0.35}_{-0.29}$ & 5.7 & 6.4 \\
\hline
\end{tabular}
\label{tab:results_ttZ}
\end{table}

\section{Extended interpretation}
\label{sec:results_extended}

Direct measurement of the $\ttZ$ and $\ttW$ cross sections can
be applied to searches for new physics (NP) within the framework
of an effective field theory. The effects of new particles or interactions
can be captured in a model-independent way by supplementing the SM
Lagrangian with higher-dimensional operators involving SM fields.
The effective Lagrangian can be written~\cite{Buchmuller1986}
as an expansion in the inverse of the cutoff energy scale, 1/$\Lambda$:

\begin{equation}
  \begin{aligned}
  \mathcal{L}_{\text{eff}} &= \mathcal{L}_{\mathrm{SM}} + \frac{1}{\Lambda}\mathcal{L}_{1} + \frac{1}{\Lambda^{2}}\mathcal{L}_{2} + \cdots\\
                             &= \mathcal{L}_{\mathrm{SM}} + \frac{1}{\Lambda}\sum_{i}(c_i\mathcal{O}_i + \text{h.c.}) +
                                \frac{1}{\Lambda^{2}}\sum_{j}(c_j\mathcal{O}_j + \text{h.c.}) + \cdots,
  \end{aligned}
\end{equation}

where $\mathcal{L}_{\mathrm{SM}}$ is the SM Lagrangian density of
dimension four, $\mathcal{L}_{1}$ is of dimension five, etc.
The Wilson coefficients $c_i$ and $c_j$ are numerical constants that parameterize
the strength of the nonstandard interactions, and $\mathcal{O}_i$ and
$\mathcal{O}_j$ are operators corresponding to combinations of SM fields. Hermitian
conjugate terms are denoted h.c. Good agreement between
data and SM expectations suggests that deviations due to NP are small and it
is reasonable to work in the first order of $c_i$ and $c_j$~\cite{Berger2009};
we limit ourselves to this domain.

It is not possible to construct a dimension-five
operator that conserves lepton number~\cite{Buchmuller1986}, so only dimension-six operators
are considered in this work.  Assuming baryon number conservation, there
are 59 independent dimension-six operators~\cite{Grzadkowski2010}. We
follow the notation and operator naming scheme introduced in Ref.~\cite{Alloul2014}.
We study the effect of these operators on the $\PQt \PZ$ coupling constants and the
$\ttW$ and $\ttZ$ cross sections, and compare them to the measured values.

\subsection{Constraints on the axial and vector components of the \texorpdfstring{$\PQt \PZ$}{tZ} coupling}
\label{sec:results_vector_axial}

Indirect measurements of the top quark to $\PZ$ boson coupling include, for example,
precision studies of the $\Z  \to \PQb\PAQb$ branching fraction at
LEP and the SLC~\cite{Abdallah2009, Barate:1997kr, Abreu:1998xf, Acciarri:1999ue, Abbiendi:1998eh, Abe:2005nqa}.
The $\ttZ$ process provides the first experimentally accessible direct probe of the tZ coupling.

The SM $\ttZ$ interaction Lagrangian can be written in terms of the vector and
axial couplings $C_V^{\mathrm{SM}}$ and $C_A^{\mathrm{SM}}$, which can be precisely calculated. In
the effective field theory approach, the modified couplings $C_{1,V}$ and
$C_{1,A}$ are considered, which can be written in terms of the SM contribution
plus deviations due to the Wilson coefficients $c_j$ of dimension-six operators~\cite{Rontsch2014},
scaled by $\Lambda$, the Higgs field vacuum expectation value $v$, and the weak mixing angle $\theta_{w}$:

\begin{equation}
    \begin{aligned}
C_{1,V} &= C_V^{\mathrm{SM}} + \frac{1}{4 \sin \theta_{w} \cos \theta_{w}} \frac{v^2}{\Lambda^2}\mathrm{Re} [\overline{c}^{\prime}_{HQ} - \overline{c}_{HQ} - \overline{c}_{Hu}], \\
C_{1,A} &= C_A^{\mathrm{SM}} - \frac{1}{4 \sin \theta_{w} \cos \theta_{w}} \frac{v^2}{\Lambda^2}\mathrm{Re} [\overline{c}^{\prime}_{HQ} - \overline{c}_{HQ} + \overline{c}_{Hu}].
    \end{aligned}
\end{equation}

A method for calculating $\sigma(\ttZ)$ in terms of $C_{1,V}$
and $C_{1,A}$ has been presented in Ref.~\cite{Rontsch2014}. The cross section depends
on a constant term, linear and quadratic terms in $C_{1,V}$ and $C_{1,A}$,
and a mixed term.  Each of these six terms is scaled by a factor which was evaluated in Ref.~\cite{Rontsch2014}
by calculating the 7\TeV cross section at six points and solving the system of equations.
To extrapolate to 8\TeV, we scale $\sigma(\ttZ)(C_{1,V}, C_{1,A})$ linearly by the ratio
of the theoretical $\ttZ$ cross sections at 7 and 8\TeV.
From this we define the signal strength parameter $\mu_{\ttZ}$ in terms of $C_{1,V}$ and $C_{1,A}$, and
a profile likelihood ratio test statistic, as described in Section~\ref{sec:results_sm}.
We perform a two-dimensional scan of the ($C_{1,V}$, $C_{1,A}$) phase space to extract the
best fit values, which are found to satisfy the constraint:
\begin{equation}
    \begin{aligned}
	74.6 + 0.5\,C_{1,V} + 189.4\,C_{1,V}^{2} - 16.3\,C_{1,A} + 359.7\,C_{1,A}^2 &= 242. \\
	\end{aligned}
\end{equation}

\begin{figure}[htb]
  \centering
  \includegraphics[width=0.70\textwidth]{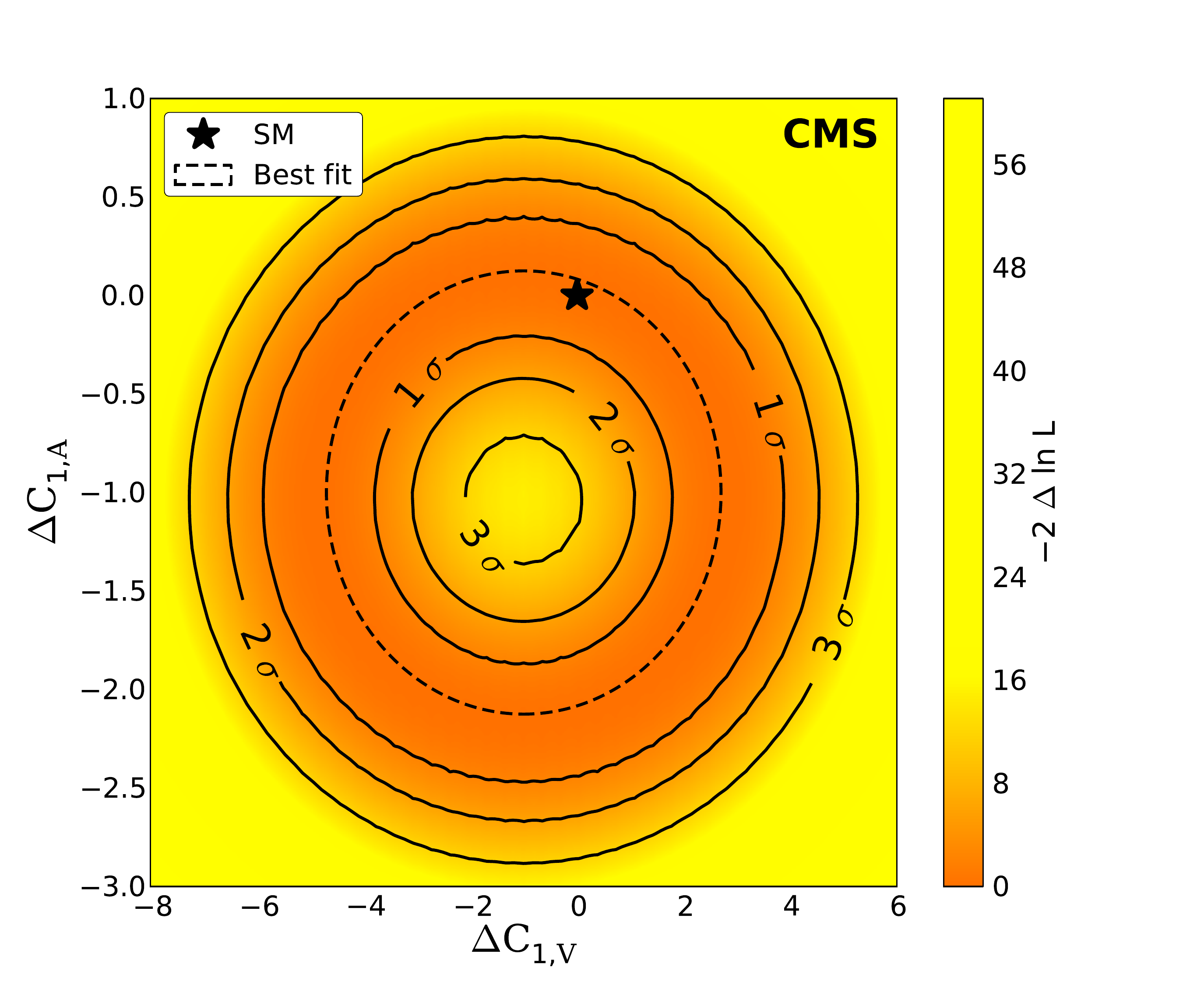}
  \caption{Difference between the profile likelihood and the best fit profile
           likelihood functions for the relative vector and axial components of
		   the $\PQt \PZ$ coupling. Contours corresponding to the best fit and the
		   1, 2, and 3 standard deviation ($\sigma$) CLs are shown in lines.}
  \label{fig:cVcA}
\end{figure}

The difference between the profile likelihood and the best fit profile likelihood
is plotted as a function of the relative vector and axial components
$\Delta C_{1,V} = C_{1,V} / C^{\mathrm{SM}}_{V} - 1$ and
$\Delta C_{1,A} = C_{1,A} / C^{\mathrm{SM}}_{A} - 1$ in Fig.~\ref{fig:cVcA}.

\subsection{Constraints on dimension-six operators}
\label{sec:results_dimension_six}

\begin{figure}[!htbp]
  \centering
  \includegraphics[width=0.49\textwidth]{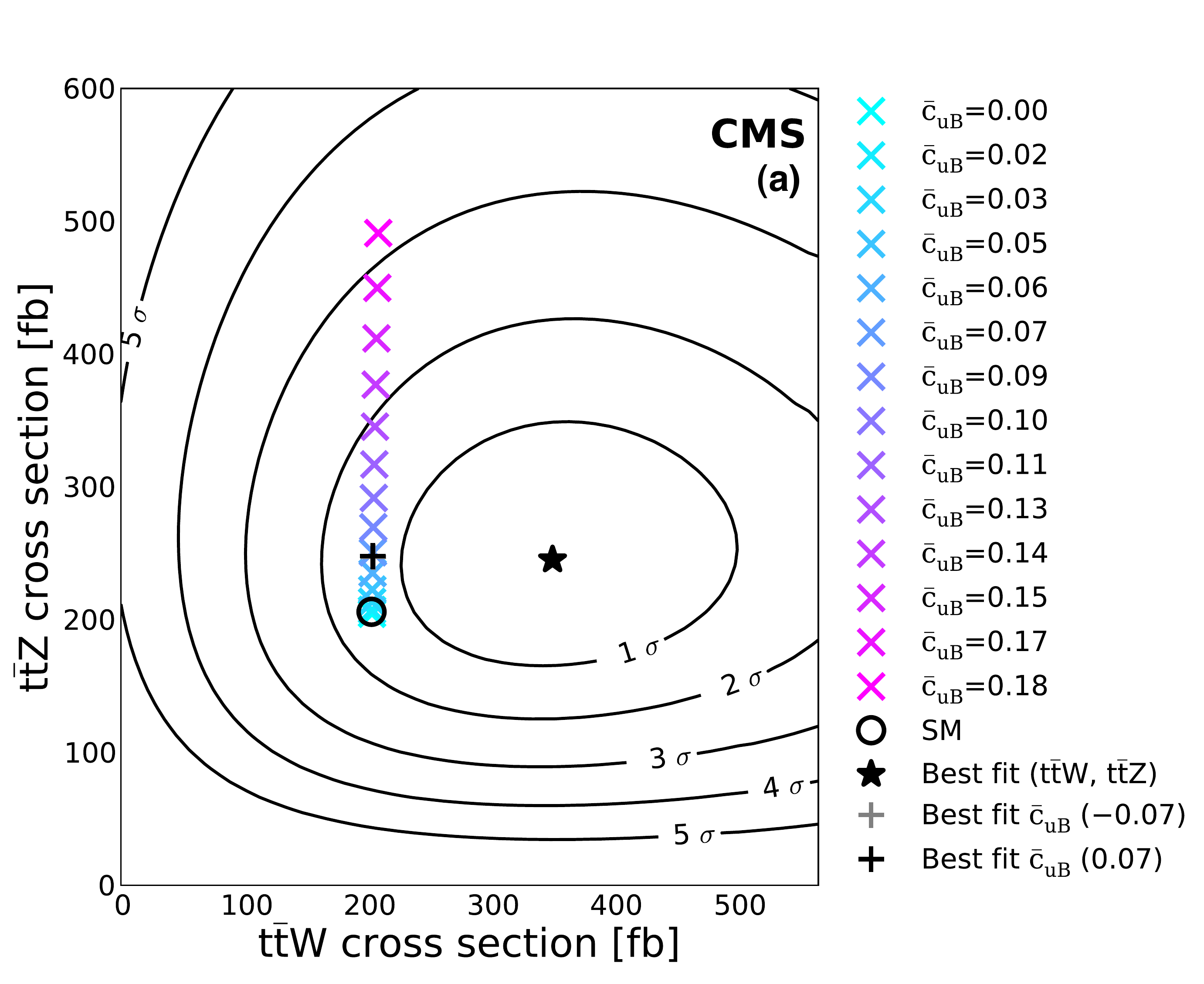}
  \includegraphics[width=0.49\textwidth]{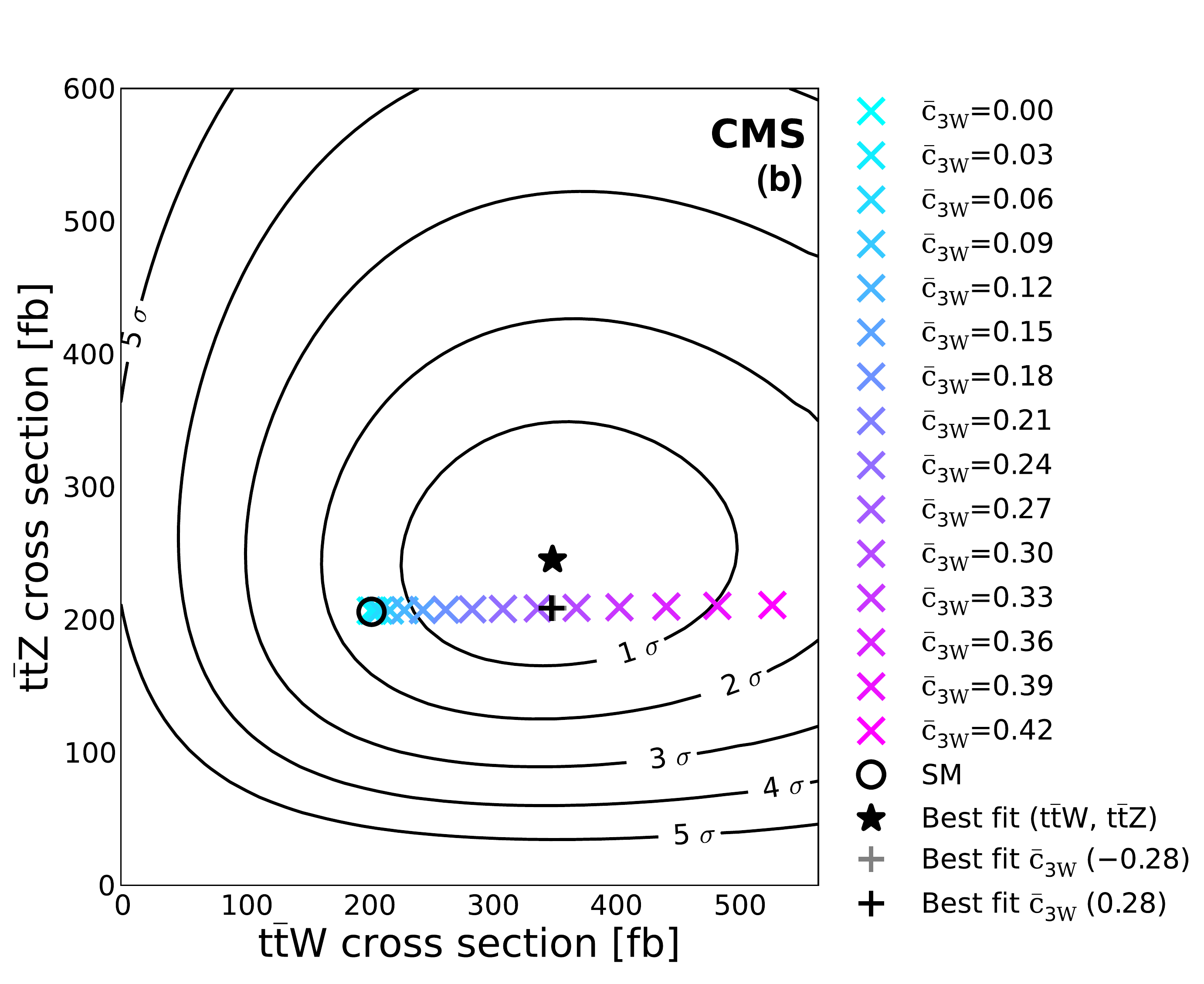}
  \includegraphics[width=0.49\textwidth]{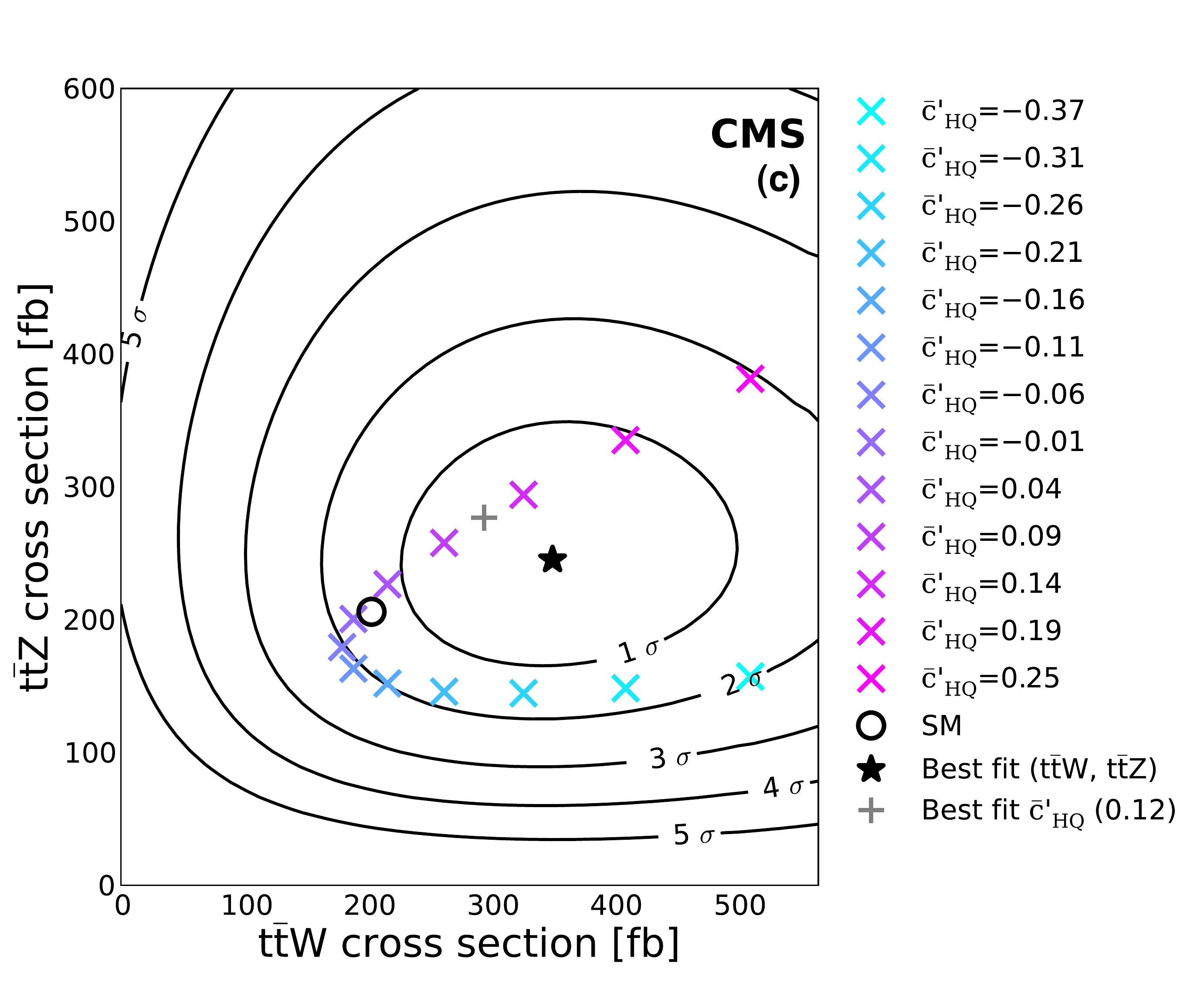}
  \includegraphics[width=0.49\textwidth]{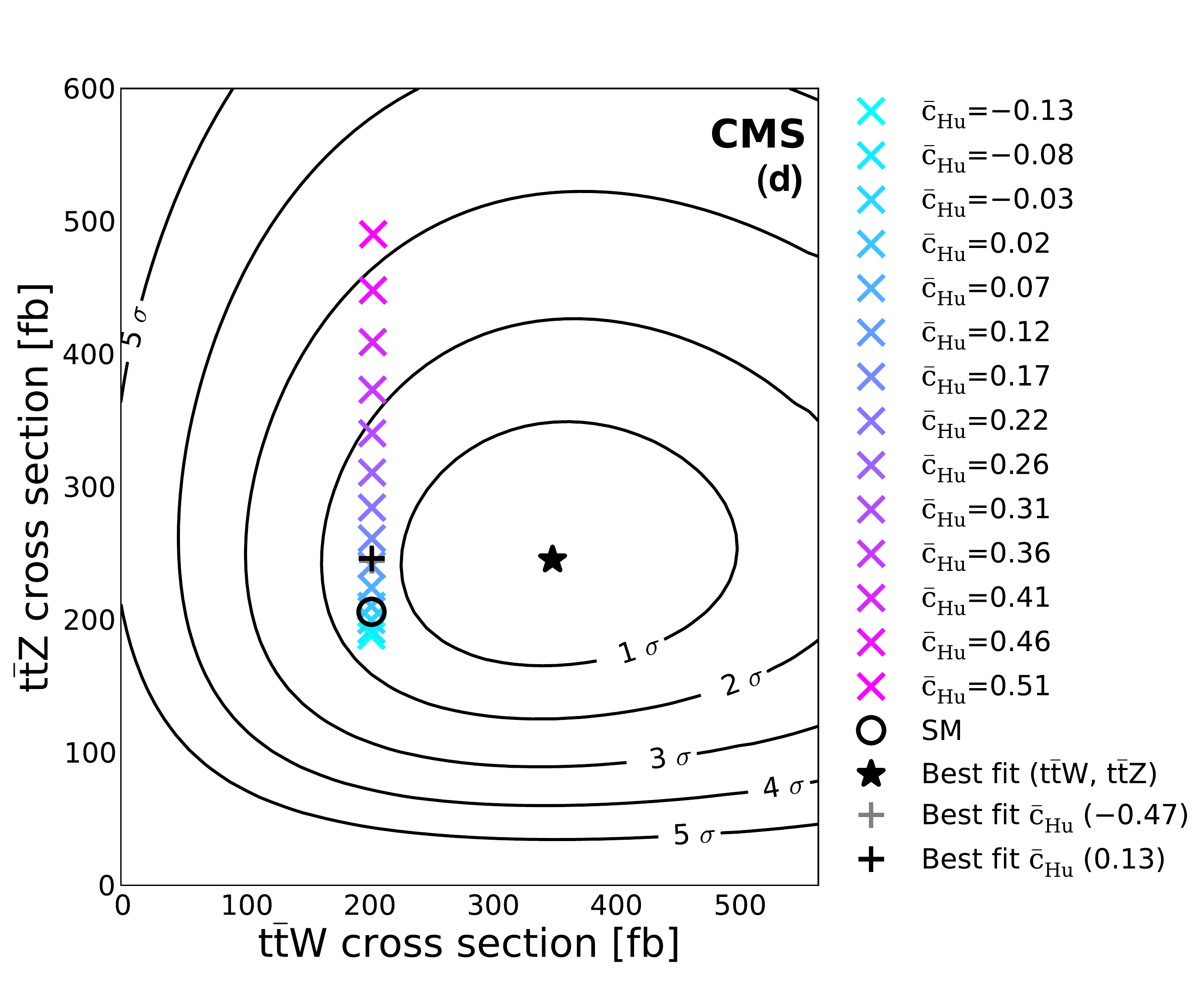}
  \includegraphics[width=0.49\textwidth]{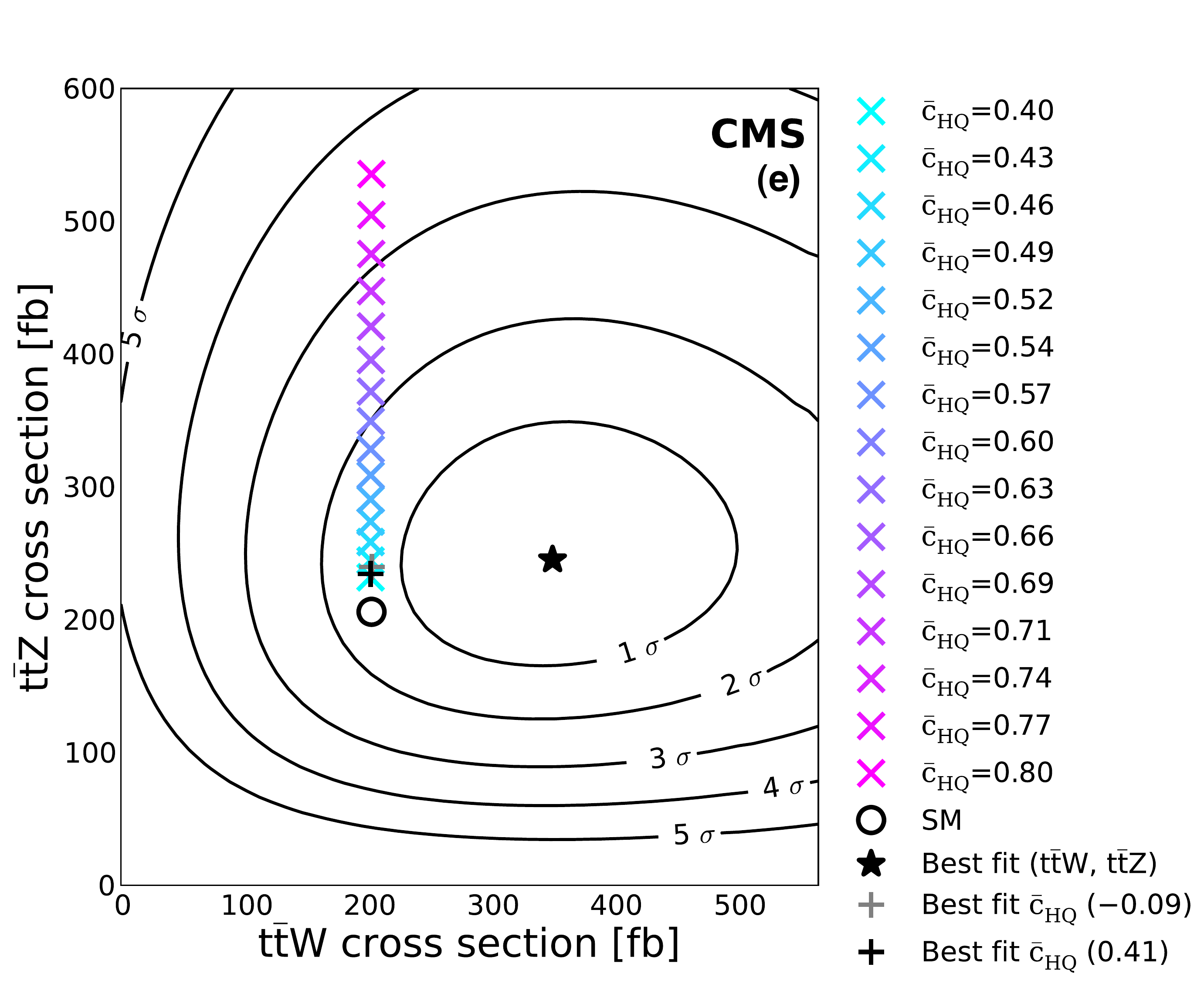}

  \caption{Sampled coefficient values for $\cuB$ (a), $\cThreeW$ (b), $\cpHQ$ (c), $\cHu$ (d),
           and $\cHQ$ (e), plotted in the ($\sigma(\ttW)$, $\sigma(\ttZ)$) plane. There are
           typically two best fit values, one greater and one less than zero, which lie on
           top of one another in the plane.}
  \label{fig:op_points}
\end{figure}

Both indirect and direct constraints on dimension-six operators are documented
in Refs.~\cite{Ellis2014, Whisnant1997, Berger2009, Rontsch2014, PhysRevD.50.4462, Zhang2012, Tonero2014}.
To study the effects of NP on the $\ttW$ and $\ttZ$ processes, we use the \textsc{FeynRules}~\cite{Alloul2013}
implementation from Ref.~\cite{Alloul2014}. This implementation is used with \MADGRAPH 5~\cite{MadGraph5}
to compute cross sections as a function of $(v^2/\Lambda^2)$\,$c_j$, henceforward simply
denoted by $c_j$. Cross sections were computed for the production of
$\ttbar$, a Higgs boson, $\ttZ$, and $\ttW$, sampling 20 points for each $c_j$.
For each sampled point, all $c_{k \neq j}$ were fixed at zero. From this survey,
we select five operators as of particular interest because they have a small effect on inclusive
Higgs boson and $\ttbar$ production, and a large effect on $\ttZ$, $\ttW$, or both: $\cuB$,
$\cpHQ$, $\cHQ$, $\cHu$, and $\cThreeW$. An alternative way to display the effect of each
$c_j$ is shown in Fig.~\ref{fig:op_points}, where sampled values are
plotted in the ($\sigma(\ttW)$, $\sigma(\ttZ)$) plane. From these it is clear that $\cuB$, $\cHu$,
and $\cHQ$ affect only $\ttZ$, whereas $\cThreeW$ only affects $\ttW$, and $\cpHQ$ affects both
processes. For each of the five operators, we perform a finer scan of 200 cross section points
and use a spline fit to obtain an expression for the cross section in terms of $c_j$,
$\sigma(\ttZ)_{\mathrm{SM+NP}}(c_j)$. We define the signal strength $\mu_{\ttZ}(c_j)$
to be the ratio of the $\ttZ$ production cross section to the combined expectations from SM and NP
$\sigma(\ttZ)_{\mathrm{SM+NP}}(c_j)$, and likewise for $\ttW$. From this we can define a profile
likelihood ratio in terms of $c_j$, similarly to what is described in Section~\ref{sec:results_sm}.

Best fit values, along with ${1}\sigma$ and ${2}\sigma$ CL ranges are summarized in
Table~\ref{tab:6dim_op_constraints}.  Operators that affect either the $\ttW$
or the $\ttZ$ cross section, but not both, have symmetric likelihood distributions
and thus have two best fit values.
Bounds on $\cpHQ$, $\cHQ$, and $\cHu$ are stricter than those
derived in Ref.~\cite{Rontsch2014} from CMS and ATLAS searches for $\ttZ$ using LHC
data at 7\TeV. Constraints on $\cuB$ are tighter than those
derived in Ref.~\cite{Tonero2014}.

\begin{table}[!htbp]\small
  \renewcommand{\arraystretch}{1.2}
  \topcaption{Constraints from this $\ttZ$ and $\ttW$ measurement on selected dimension-six operators.}
  \label{tab:6dim_op_constraints}
  \centering
  \begin{tabular}{c|c|c|c}
    \hline

   Operator & Best fit point(s) & 1 standard deviation CL & 2 standard deviation CL\\
   \hline
   $\cuB$     & $-$0.07 and 0.07 & [$-$0.11, 0.11]                    & [$-$0.14, 0.14]\\
   $\cThreeW$ & $-$0.28 and 0.28 & [$-$0.36, $-$0.18] and [0.18, 0.36]  & [$-$0.43, 0.43]\\
   $\cpHQ$    & 0.12           & [$-$0.07, 0.18]                    & [$-$0.33, $-$0.24] and [$-$0.02, 0.23]\\
   $\cHu$     & $-$0.47 and 0.13 & [$-$0.60, $-$0.23] and [$-$0.11, 0.26] & [$-$0.71, 0.37]\\
   $\cHQ$     & $-$0.09 and 0.41 & [$-$0.22, 0.08] and [0.24, 0.54]   & [$-$0.31, 0.63]\\
    \hline
    \end{tabular}
\end{table}

\section{Summary}
\label{sec:summary}

An observation of top quark pairs produced in association with a $\PZ$ boson and
measurements of the $\ttW$ and $\ttZ$ cross sections have been made,
using 19.5\fbinv of 8\TeV pp collision data collected by the CMS detector at the LHC. Signatures
from different decay modes of the top quark pair resulting in final states with two,
three, and four leptons have been analyzed. Signal events have been identified by
uniquely matching reconstructed leptons and jets to final state particles from
$\ttW$ and $\ttZ$ decays. Results from two independent $\ttW$
channels and three $\ttZ$ channels have been presented, along with combined
measurements.  The combined $\ttW$ cross section measurement in
same-sign dilepton and three-lepton events is $382^{+117}_{-102}\unit{fb}$,
4.8 standard deviations from the background-only hypothesis,
where a significance of 3.5 standard deviations is expected in the standard model.
Combining opposite-sign dilepton, three-lepton, and four-lepton channels, the $\ttZ$ cross
section is measured to be $242^{+65}_{-55}\unit{fb}$, an observation
with a significance of 6.4 standard deviations from the background-only
hypothesis, and in agreement with the standard model expectation.  Using the measured cross
sections, limits have been placed on the vector and axial couplings of the $\PZ$ boson to the
top quark, and on the Wilson coefficients of five dimension-six operators parameterizing new
physics: $\cuB$, $\cpHQ$, $\cHQ$, $\cHu$, and $\cThreeW$. These measurements are
compatible with the standard model predictions, and are the
most sensitive reported to date to these high mass scale processes.

\section*{Acknowledgements}
\label{sec:acknowledgements}

\hyphenation{Bundes-ministerium Forschungs-gemeinschaft Forschungs-zentren} We congratulate our colleagues in the CERN accelerator departments for the excellent performance of the LHC and thank the technical and administrative staffs at CERN and at other CMS institutes for their contributions to the success of the CMS effort. In addition, we gratefully acknowledge the computing centers and personnel of the Worldwide LHC Computing Grid for delivering so effectively the computing infrastructure essential to our analyses. Finally, we acknowledge the enduring support for the construction and operation of the LHC and the CMS detector provided by the following funding agencies: the Austrian Federal Ministry of Science, Research and Economy and the Austrian Science Fund; the Belgian Fonds de la Recherche Scientifique, and Fonds voor Wetenschappelijk Onderzoek; the Brazilian Funding Agencies (CNPq, CAPES, FAPERJ, and FAPESP); the Bulgarian Ministry of Education and Science; CERN; the Chinese Academy of Sciences, Ministry of Science and Technology, and National Natural Science Foundation of China; the Colombian Funding Agency (COLCIENCIAS); the Croatian Ministry of Science, Education and Sport, and the Croatian Science Foundation; the Research Promotion Foundation, Cyprus; the Ministry of Education and Research, Estonian Research Council via IUT23-4 and IUT23-6 and European Regional Development Fund, Estonia; the Academy of Finland, Finnish Ministry of Education and Culture, and Helsinki Institute of Physics; the Institut National de Physique Nucl\'eaire et de Physique des Particules~/~CNRS, and Commissariat \`a l'\'Energie Atomique et aux \'Energies Alternatives~/~CEA, France; the Bundesministerium f\"ur Bildung und Forschung, Deutsche Forschungsgemeinschaft, and Helmholtz-Gemeinschaft Deutscher Forschungszentren, Germany; the General Secretariat for Research and Technology, Greece; the National Scientific Research Foundation, and National Innovation Office, Hungary; the Department of Atomic Energy and the Department of Science and Technology, India; the Institute for Studies in Theoretical Physics and Mathematics, Iran; the Science Foundation, Ireland; the Istituto Nazionale di Fisica Nucleare, Italy; the Ministry of Science, ICT and Future Planning, and National Research Foundation (NRF), Republic of Korea; the Lithuanian Academy of Sciences; the Ministry of Education, and University of Malaya (Malaysia); the Mexican Funding Agencies (CINVESTAV, CONACYT, SEP, and UASLP-FAI); the Ministry of Business, Innovation and Employment, New Zealand; the Pakistan Atomic Energy Commission; the Ministry of Science and Higher Education and the National Science Center, Poland; the Funda\c{c}\~ao para a Ci\^encia e a Tecnologia, Portugal; JINR, Dubna; the Ministry of Education and Science of the Russian Federation, the Federal Agency of Atomic Energy of the Russian Federation, Russian Academy of Sciences, and the Russian Foundation for Basic Research; the Ministry of Education, Science and Technological Development of Serbia; the Secretar\'{\i}a de Estado de Investigaci\'on, Desarrollo e Innovaci\'on and Programa Consolider-Ingenio 2010, Spain; the Swiss Funding Agencies (ETH Board, ETH Zurich, PSI, SNF, UniZH, Canton Zurich, and SER); the Ministry of Science and Technology, Taipei; the Thailand Center of Excellence in Physics, the Institute for the Promotion of Teaching Science and Technology of Thailand, Special Task Force for Activating Research and the National Science and Technology Development Agency of Thailand; the Scientific and Technical Research Council of Turkey, and Turkish Atomic Energy Authority; the National Academy of Sciences of Ukraine, and State Fund for Fundamental Researches, Ukraine; the Science and Technology Facilities Council, UK; the US Department of Energy, and the US National Science Foundation.

Individuals have received support from the Marie-Curie program and the European Research Council and EPLANET (European Union); the Leventis Foundation; the A. P. Sloan Foundation; the Alexander von Humboldt Foundation; the Belgian Federal Science Policy Office; the Fonds pour la Formation \`a la Recherche dans l'Industrie et dans l'Agriculture (FRIA-Belgium); the Agentschap voor Innovatie door Wetenschap en Technologie (IWT-Belgium); the Ministry of Education, Youth and Sports (MEYS) of the Czech Republic; the Council of Science and Industrial Research, India; the HOMING PLUS program of the Foundation for Polish Science, cofinanced from European Union, Regional Development Fund; the OPUS program of the National Science Center (Poland); the Compagnia di San Paolo (Torino); the Consorzio per la Fisica (Trieste); MIUR project 20108T4XTM (Italy); the Thalis and Aristeia programs cofinanced by EU-ESF and the Greek NSRF; the National Priorities Research Program by Qatar National Research Fund; the Rachadapisek Sompot Fund for Postdoctoral Fellowship, Chulalongkorn University (Thailand); and the Welch Foundation, contract C-1845.

\bibliography{auto_generated}

\clearpage

\appendix

\section{Input variables to linear discriminant for event reconstruction}
\label{sec:appendix_match_vars}

\begin{table}[htb]\small
  \topcaption{Variables used to match leptons and jets to their parent particles in
           reconstructing $\ttW$, $\ttZ$, and $\ttbar$ events.  In $\ttZ$ events,
           the two leptons matched to the $\PZ$ boson decay are removed and the remaining
           $\ttbar$ system is reconstructed.  In $\ttW$ events, the $\ttbar$ system
           is reconstructed with the leptons that best match the $\ttbar$ decay,
           and the lepton from the associated $\PW$ boson (and any
           variables with \ptmiss) are not used.
           In $\ttbar$ events, $\ell_{\PQb}$ denotes a lepton from $\PQb$-hadron decay.   \label{tab:match_vars_ttW_ttZ} }
  \renewcommand{\arraystretch}{1.2}
  \centering
  \begin{tabular}{l|c|c|c|c|c}
    \hline
    \multicolumn{1}{l|}{} &
    \multicolumn{5}{c}{\textbf{Decay products of the $\ttbar$ system}} \\
    \hline
    \textbf{Reconstructed event} &
    $\PQb\PQq\PAQq$ $\PAQb\PQq\PAQq$ &
    $\PQb\ell\nu$ $\PAQb\PQq\PAQq$ &
    $\PQb\ell\nu$ $\PAQb\ell\nu$ &
    $\ell_{\PQb}\PQq\PAQq$  $\PAQb\ell\nu$ &
    $\ell_{\PQb}\ell\nu$  $\PAQb\ell\nu$ \\

    \hline

    OS dilepton $\ttZ$                                    & X &   &   &   &   \\
    SS dilepton $\ttW$                                    &   & X &   &   &   \\
    3$\ell$ $\ttZ$                                        &   & X &   &   &   \\
    3$\ell$ $\ttW$                                        &   &   & X &   &   \\
    OS dilepton $\ttbar$                                    &   &   & X &   &   \\
    SS dilepton $\ttbar$                                    &   &   &   & X &   \\
    3$\ell$ $\ttbar$                                        &   &   &   &   & X \\
    \hline
    \textbf{Input variables}                              &   &   &   &   &   \\
    \hline
    \textit{\textbf{Jet CSV (b tag) discriminator}}       &   &   &   &   &   \\
    \PQb jet CSV                                             & X & X & X & X & X \\
    Higher jet CSV from $\PW  \to \PQq\PAQq$         & X & X &   & X &   \\
    Lower jet CSV from $\PW  \to \PQq\PAQq$          & X & X &   & X &   \\
    \hline
    \textit{\textbf{Jet charge}}                          &   &   &   &   &  \\
    Charge of \PQb jet from \PQt                                & X &   & X &   &   \\
    Charge of \PQb jet from $\PAQt$                          & X &   & X &   &   \\
    Charge of \PQb jet from $\PQt  \to \PQb\ell\nu$   &   & X &   &   &   \\
    Charge of \PQb jet from $\PQt  \to \PQb\PQq\PAQq$ &   & X &   &   &   \\
    Charge of \PQb jet not decaying to a lepton              &   &   &   & X & X \\
    Sum of charges of jets from $\PW  \to \PQq\PAQq$ & X & X &   & X &   \\
    \hline
    \textit{\textbf{Invariant mass}}                                              &   &   &   &  &  \\
    Mass of lepton and \PQb jet from \PQt                                               &   &   & X &   &   \\
    Mass of lepton and \PQb jet from $\PAQt$                                         &   &   & X &   &   \\
    Mass of lepton and \PQb jet from $\PQt  \to \PQb\ell\nu$                  &   & X &   & X & X \\
    Mass of leptons from $\PQt  \to \ell_{\PQb}\ell\nu$                    &   &   &   &   & X \\
    \MT of \ptvecmiss and \ptvec of lepton and \PQb jet from \PQt                      &   & X &   & X &   \\
    Mass of two jets from $\PW  \to \PQq\PAQq$                             & X & X &   & X &   \\
    Mass of \PQb jet and quark jets from $\PQt  \to \PQb\PQq\PAQq$            & X & X &   &   &   \\
    Mass of lepton from \PQb and jets from $\PQt  \to \ell_{\PQb}\PQq\PAQq$   &   &   &   & X &   \\
    Ratio of \MT to mass for jets from \PQt or $\PW$                                     & X & X & X &   &   \\
    \hline
  \end{tabular}
\end{table}

\clearpage

\section{Input variables to final discriminants (BDTs)}
\label{sec:appendix_BDT_vars}

\begin{table}[htb]\small
	\topcaption{Input variables to the BDT that distinguishes SS $\ttW$ from $\ttbar$,
	         ranked by signal-background separation.}
  \renewcommand{\arraystretch}{1.2}
  \centering
  \begin{tabular}{lcc}
    \hline
    BDT inputs: SS $\ttW$ vs.~$\ttbar$                   & 3 jet & ${\geq}4$ jets \\
    \hline
    \MT of \ptvecmiss and \ptvec of leptons and jets                            & 1 & 1 \\
    \ptmiss                                                                   & 4 & 2 \\
    Second-highest lepton \pt                                     & 6 & 3 \\
    Match score for $\ttbar  \to \ell_{\PQb}\PQq\PAQq$ $\PAQb\ell\nu$  & 2 & 4 \\
    Highest lepton \pt                                                       & 5 & 5 \\
    Second-highest CSV value of a jet                             & 8 & 6 \\
    $\ttbar$ matched top quark \MT from $\PQb\ell\nu$                                   & 7 & 7 \\
    Match score for $\ttW  \to \PQb\ell\nu$ $\PAQb\PQq$                   & 9 & 8 \\
    Match score for $\ttW  \to \PQb\ell\nu$ $\PAQb\PQq\PAQq$            & \NA & 9 \\
    $\ttbar$ matched top quark mass from $\ell_{\PQb}\PQq\PAQq$                          & 3 & \NA \\
	\hline
	\end{tabular}
	\label{tab:BDT_inputs_SS_ttW}
\end{table}

\begin{table}[htb]\small
	\topcaption{Input variables to BDT that distinguishes 3$\ell$ $\ttW$ from $\ttbar$,
	         ranked by signal-background separation.}
  \renewcommand{\arraystretch}{1.2}
  \centering
  \begin{tabular}{lcc}
    \hline
    BDT inputs: 3$\ell$ $\ttW$ vs.~$\ttbar$                    & 1 jet & ${\geq}$2 jets \\
	\hline
	Second-highest CSV value of a jet                            & \NA & 1 \\
    \MT of \ptvecmiss and \ptvec of leptons and jets                           & 1 & 2 \\
    Match score for $\ttW  \to \ell\nu$ $\PQb\ell\nu$ $\PAQb\ell\nu$  & \NA & 3 \\
    Second-highest SS lepton \pt                          & 4 & 4 \\
    $\ttbar$ matched top quark mass from $\ell_{\PW}$ and $\ell_{\PQb}$                  & \NA & 5 \\
    Highest SS lepton \pt                                            & 3 & 6 \\
    Match score for $\ttW  \to \ell\nu$ $\PQb\ell\nu$ $\ell\nu$         & 2 & \NA \\
    \ptmiss                                                                  & 5 & \NA \\
	Jet \pt                                                         & 6 & \NA \\
	\hline
	\end{tabular}
	\label{tab:BDT_inputs_3l_ttW}
\end{table}

\begin{table}[htb]\small
	\topcaption{Input variables to BDT that distinguishes 3$\ell$ $\ttZ$ from $\PW \PZ$ and $\ttbar$,
	         ranked by signal-background separation.}
  \renewcommand{\arraystretch}{1.2}
  \centering
  \begin{tabular}{lcc}
    \hline
    BDT inputs: 3$\ell$ $\ttZ$ vs.~$\PW \PZ$ and $\ttbar$              & 3 jet & ${\geq}4$ jets \\
	\hline
    Match score for $\ttZ  \to \ell\ell$ $\PQb\ell\nu$ $\PAQb\PQq$        & 1 & 1 \\
    Match score for $\ttZ  \to \ell\ell$ $\PQb\ell\nu$ $\PAQb\PQq\PAQq$ & \NA & 2 \\
    Match score for $\ttZ  \to \ell\ell$ $\ell\nu$ $\PAQb\PQq\PAQq$  & 8 & 3 \\
    Match score for $\ttZ  \to \ell\ell$ $\PQb\ell\nu$ $\PQq\PAQq$        & 9 & 4 \\
    Number of medium $\PQb$-tagged jets                                           & 3 & 5 \\
	Mass of lepton pair matched to $\PZ$ boson                                   & 7 & 6 \\
    \MT of \ptvecmiss and \ptvec of leptons and jets                             & 4 & 7 \\
    Match score for $\ttZ  \to \ell\ell$ $\PQb\ell\nu$ $\PAQb$         & 2 & \NA \\
    Match score for $\ttZ  \to \ell\ell$ $\ell\nu$ $\PAQb\PQq$         & 5 & \NA \\
    Match score for $\ttZ  \to \ell\ell$ $\PQb\ell\nu$ $\PQq$               & 6 & \NA \\
	\hline
	\end{tabular}
	\label{tab:BDT_inputs_3l_ttZ}
\end{table}

\begin{table}[htb]\small
	\topcaption{Input variables to BDT that distinguishes OS $\ttZ$ from $\ttbar$ (used as input to the final discriminant),
	         ranked by signal-background separation.}
  \renewcommand{\arraystretch}{1.2}
  \centering
  \begin{tabular}{lcc}
    \hline
    BDT inputs: OS $\ttZ$ vs.~$\ttbar$                            & 5 jet & ${\geq}6$ jets \\
	\hline
	$\Delta$R between leptons                                                  & 1 & 1 \\
	\pt of dilepton system                                                     & 2 & 2 \\
	Dilepton invariant mass                                                    & 3 & 3 \\
	\MHT                                                                        & 4 & 4 \\
    Match score for $\ttbar  \to \PQb\ell\nu$  $\PAQb\ell\nu$            & 5 & 5 \\
    Number of jets with \pt $> 40 \GeV $                                         & 9 & 6 \\
    Match score for $\ttZ  \to \ell\ell$ $\PQb\PQq\PAQq$ $\PAQb\PQq\PAQq$  & \NA & 7 \\
    Match score for $\ttZ  \to \ell\ell$ $\PQb\PQq$ $\PAQb\PQq\PAQq$         & 8 & 8 \\
    Match score for $\ttZ  \to \ell\ell$ $\PQb\PQq\PAQq$ $\PAQb\PQq$  & 7 & 9 \\
	Ratio of \MT to mass of jets                                              & 6 & 10 \\
	CSV of jet matched to $\PQb$ from $\ttbar$                                      & 11 & 11 \\
	CSV of jet matched to $\PAQb$ from $\ttbar$                     & 10 & 12 \\
	\hline
	\end{tabular}
	\label{tab:BDT_inputs_OS_ttZ_vs_ttbar}
\end{table}

\begin{table}[htb]\small
	\topcaption{Input variables to BDT that distinguishes OS $\ttZ$ from $\PZ$ boson and $\ttbar$ (the final discriminant),
	         ranked by signal-background separation.}
  \renewcommand{\arraystretch}{1.2}
  \centering
  \begin{tabular}{lcc}
    \hline
    BDT inputs: OS $\ttZ$ vs.~$\PZ$ and $\ttbar$                             & 5 jet & ${\geq}6$ jets \\
	\hline
	OS $\ttZ$ vs.~$\ttbar$ BDT                                                          & 1 & 1 \\
    Match score for $\ttZ  \to \ell\ell$ $\PQb\PQq$ $\PAQb\PQq\PAQq$                & 3 & 2 \\
    Match score for $\ttZ  \to \ell\ell$ $\PQb\PQq\PAQq$ $\PAQb\PQq$                & 4 & 3 \\
    Match score for $\ttZ  \to \ell\ell$ $\PQb\PQq\PAQq$ $\PAQb\PQq\PAQq$         & \NA & 4 \\
    Minimum $\chi^{2}$ for $\ttZ  \to \ell\ell$ $\PQb\PQq\PAQq$ $\PAQb\PQq\PAQq$  & \NA & 5 \\
    Number of jets with $\pt > 40 \GeV $                                                & 6 & 6 \\
	Fifth-highest jet \pt                                                 & 5 & 7 \\
	Ratio of \MT to mass of jets and leptons                                          & 2 & 8 \\
	Second-highest jet CSV                                                 & 7 & 9 \\
	Highest jet CSV                                                                  & 8 & 10 \\
	\hline
	\end{tabular}
	\label{tab:BDT_inputs_OS_ttZ}
\end{table}

\cleardoublepage \section{The CMS Collaboration \label{app:collab}}\begin{sloppypar}\hyphenpenalty=5000\widowpenalty=500\clubpenalty=5000\textbf{Yerevan Physics Institute,  Yerevan,  Armenia}\\*[0pt]
V.~Khachatryan, A.M.~Sirunyan, A.~Tumasyan
\vskip\cmsinstskip
\textbf{Institut f\"{u}r Hochenergiephysik der OeAW,  Wien,  Austria}\\*[0pt]
W.~Adam, E.~Asilar, T.~Bergauer, J.~Brandstetter, E.~Brondolin, M.~Dragicevic, J.~Er\"{o}, M.~Flechl, M.~Friedl, R.~Fr\"{u}hwirth\cmsAuthorMark{1}, V.M.~Ghete, C.~Hartl, N.~H\"{o}rmann, J.~Hrubec, M.~Jeitler\cmsAuthorMark{1}, V.~Kn\"{u}nz, A.~K\"{o}nig, M.~Krammer\cmsAuthorMark{1}, I.~Kr\"{a}tschmer, D.~Liko, T.~Matsushita, I.~Mikulec, D.~Rabady\cmsAuthorMark{2}, B.~Rahbaran, H.~Rohringer, J.~Schieck\cmsAuthorMark{1}, R.~Sch\"{o}fbeck, J.~Strauss, W.~Treberer-Treberspurg, W.~Waltenberger, C.-E.~Wulz\cmsAuthorMark{1}
% ...
\vskip\cmsinstskip
% ...
1:~~Also at Vienna University of Technology, Vienna, Austria\\
2:~~Also at CERN, European Organization for Nuclear Research, Geneva, Switzerland\\
% ...

\end{sloppypar}
\end{document}